\documentclass{SciPost}
\usepackage{doi,url,enumitem}
\usepackage{mathrsfs,cite}
\usepackage{bbm,simplewick}
\usepackage{graphicx}
\usepackage{hyperref}
\allowdisplaybreaks
\def\ep{\mathrm{e}}
\def\sss{\scriptscriptstyle}
\begin{document}

\begin{center}{\Large \textbf{
Momentum correlations as signature of sonic Hawking radiation in Bose-Einstein
  condensates}}\end{center}

\begin{center}
A. Fabbri\textsuperscript{1}, N. Pavloff\textsuperscript{2}
\end{center}

\begin{center}
  {\bf 1} Centro Fermi -- Museo Storico Della Fisica e Centro Studi e Ricerche
Enrico Fermi, Piazza del Viminale 1,
  00184 Roma, Italy\\ Dipartimento di Fisica dell'Universit\`a di
  Bologna and INFN Sezione di Bologna,\\ Via Irnerio 46, 40126
  Bologna, Italy\\Departamento de F\'isica Te\'orica and IFIC, Universidad
  de
  Valencia-CSIC, \\ C. Dr. Moliner 50, 46100 Burjassot, Spain\\
  Laboratoire de Physique Th\'eorique, CNRS UMR 8627, B\^at. 210,
  Univ.  Paris-Sud, \\ Universit\'e Paris-Saclay, 91405 Orsay Cedex,
  France
\\
 {\bf 2} LPTMS, CNRS, Univ. Paris-Sud, Universit\'e Paris-Saclay,
  91405 Orsay, France\\

\end{center}

\section*{Abstract} {\bf We study the two-body momentum correlation
  signal in a quasi one dimensional Bose-Einstein condensate in the
  presence of a sonic horizon. We identify the relevant correlation
  lines in momentum space and compute the intensity of the
  corresponding signal. We consider a set of different experimental
  procedures and identify the specific issues of each measuring
  process. We show that some inter-channel correlations, in particular
  the Hawking quantum-partner one, are particularly well adapted for
  witnessing quantum non-separability, being resilient to the effects
  of temperature and/or quantum quenches.}

\vspace{10pt}
\noindent\rule{\textwidth}{1pt}
\tableofcontents\thispagestyle{fancy}
\noindent\rule{\textwidth}{1pt}
\vspace{10pt}

\section{Introduction}

While the possibility of quantizing gravitation remains elusive, some
noticeable progresses have been made in the description of the
interaction between the space-time metric and a quantum field. In
particular, the dynamical Casimir effect \cite{Moo70,Full76} and
Hawking radiation from black holes \cite{Hawking} both correspond to a
quantum creation of entangled pairs of particles induced by (strong)
space-time inhomogeneities, and have both been predicted in the framework
of quantum field theory in curved spacetime. In the case of Hawking
radiation, the prospect of an experimental study in the genuine
astrophysical context seems hopeless, because the radiation has a
thermal spectrum at a temperature $T_{\scriptscriptstyle\rm
  H}=\hbar\kappa/2\pi c_{\ell}$, where $c_{\ell}$ is the speed of
light and $\kappa$ the black hole horizon's surface gravity (we use
units such that $k_{\scriptscriptstyle\rm B}=1$) and in the standard
situation of a black hole formed after a gravitational collapse,
$T_{\scriptscriptstyle\rm H}$ is much lower than the temperature of
the microwave background radiation \cite{remark}. However, the
phenomenon of Hawking radiation has a robust kinematic origin, and
elaborating on the close analogy of a transonic flow structure with
the gravitational metric near a black hole event horizon, Unruh
proposed to observe Hawking radiation in a condensed matter context
\cite{Unr81}: this idea is often considered as the birth of the field
of analogue gravity \cite{Bar11,Fac12}.

Many physical realizations have been proposed for observing analog
Hawking radiation (see, e.g.,
\cite{Fac12,Vol03,Fou10,Wei11,Euv16,Orn18}), among which the
implementation of a sonic horizon in the flow of a Bose-Einstein
condensate (BEC) rapidly appeared as quite promising
\cite{HawkingBEC}: the low temperature of the system and its
paradigmatic quantum nature makes it an ideal playground for studying
this phenomenon. However, a direct observation of the analogous sonic
radiation in this system is still hindered by thermal effects and
difficult to identify unambiguously. The recognition of this
difficulty motivated the authors of Refs. \cite{Bal08,Car08} to
propose the detection of density correlation as an evidence for
Hawking emission of correlated pairs of particles from the horizon:
Indeed, in an analogous system, contrarily to the gravitational case,
the experimentalist is a super-observer who can make measurements from
both sides of the horizon. The correlation between the Hawking
particle and its partner were shown in Refs. \cite{Bal08,Car08} to
induce a characteristic peak in the correlator of density fluctuations
which could be used to demonstrate the existence of analogous Hawking
radiation in a BEC system. The physical interpretation of this
correlated signal is similar to the one initially given by Hawking
\cite{Hawking,Haw77}: Just at the event horizon, vacuum fluctuations
produce pairs of virtual quasi-particles, one with negative norm and
one with positive norm. The negative norm quasi-particle propagates in
the region inside the black hole where it can exist as a real
quasi-particle (and is often denoted as the ``partner''). The other
quasi-particle of the pair is denoted as the ``Hawking quantum''; it
can escape to infinity, where it constitutes a part of the Hawking
radiation. In a BEC the quasi-particles are Bogoliubov excitations
which correspond to density fluctuations: hence the emission of the
correlated pairs of particles induces density correlations. An
interesting aspect of these correlations is that they are resilient to
finite temperature effects \cite{Mac09,Rec09}.

In the same line of idea, we proposed in Ref. \cite{Boi15} to study
correlations in momentum space as evidence of Hawking radiation. The
physical idea is the same as the one behind the study of density
correlations in real space, but the specifics are different, with a
number of advantages: first, the practical implementation of this type
of experiment is well documented \cite{Hod17,Cha16,Jas12,Vas12}.
Also, it was shown in Ref. \cite{Boi15} that the momentum correlation
signal is particularly well adapted to the study of Hawking radiation,
being even less affected by the background temperature than the
real-space correlation signal, and offers a clear and robust signature
of the entangled nature of the Hawking pairs.  In the present paper we
develop and explicit the results presented in Ref. \cite{Boi15}.  We
detail the theoretical description of the quantum fluctuation of the
system and precise how a local Fourier analysis can be performed. This
leads us to underline some characteristics of the experimental
detection scheme which are crucial for the detection of entanglement
(cf. the discussions in Appendix \ref{window-Fourier} and at the end
of Sec. \ref{BHT0}).  We also extend the treatment of
Ref. \cite{Boi15} in order to include what we denote as ``non
adiabatic effects'' and ``{\it in situ} measurements''
(Sec. \ref{NA}). We show that the results presented in
Ref. \cite{Boi15} are robust with respect to this more general
treatment, and that new correlation lines appear which, at variance
with the previous ones, show no signature of non-separability.

Another important motivation of our work is the recent experimental
study of Steinhauer \cite{Ste16} who studied an acoustic BEC black
hole in one of the models discussed below (the so called ``waterfall
model'' \cite{Lar12}) and presented results on entanglement similar to
the ones discussed below.

The paper is organized as follows: Sec. \ref{BHC} presents several
black hole configurations in a quasi one dimensional BEC system. In
Sec. \ref{sec.mom.corr} we compute the corresponding theoretical
momentum correlation functions and the non-separability signals in the
different configurations in a variety of situations:
In particular we present the adiabatic and
non-adiabatic regimes and also address in both cases 
the effects of temperature.
These results are compared in
Sec. \ref{correl-noBH} with the ones obtained in the absence of sonic
horizon. In section \ref{limitations} we discuss the limitations of
our theoretical approach and finally we present our conclusion in
section \ref{sec.conclu}. Some technical points are presented in the
Appendices: in Appendix \ref{window-Fourier} we discuss a rigorous
windowed Fourier analysis which induces important
restrictions to the measurement process; in Appendix \ref{appB} we
give the form of the most general correlation functions and in
Appendix \ref{appSubSub} we discuss the specific case of a subsonic
flow in the presence of a localized external potential.

\section{Black hole configurations and their theoretical description}\label{BHC}

\subsection{Quasi one-dimensional sonic black holes}\label{q1dbh}

In this work we consider a system
where bosons are confined in one
dimension by a harmonic transverse potential of angular frequency
$\omega_\perp$. We denote by $x$ the longitudinal degree of freedom
and assume no trapping along $x$. In this configuration the theoretical
description of the system is conveniently worked out in the quasi
one-dimensional limit where the particles are described by a one
dimensional (1D) quantum field $\hat{\Psi}(x)$. According to the Bogoliubov 
prescription one writes the field operator as the sum of a 
main contribution (a classical field $\Psi_0$) and a small quantum remnant  
\begin{equation}\label{1d1}
\hat{\Psi}(x)=\Psi_0(x) +\hat{\psi}(x)\; .
\end{equation}
$\Psi_0(x)$ describes the condensate order parameter and verifies
the stationary 1D Gross-Pitaevskii equation
\begin{equation}\label{1dGP}
\mu\, \Psi_0 = -\frac{\hbar^2}{2\, m}\partial_x^2 \Psi_0 + 
\left[U(x) + g_{1{\rm d}} |\Psi_0|^2\right]\Psi_0 \; ,
\end{equation}
with $g_{1{\rm d}}=2\hbar\omega_\perp a$ \cite{Ols98}, where $a$ is
the 3D s-wave scattering length. In \eqref{1dGP} $\mu$ is the chemical
potential and $U(x)$ a possible longitudinal external potential. In
the absence of external potential, for a static homogeneous system of
constant linear density $|\Psi_0|^2=n$ one gets $\mu=g_{1{\rm d}} n$.
Useful quantities are the sound velocity in the uniform system $c =
\sqrt{\mu/m}$ and the healing length $\xi = \hbar/m c$.

In 1D a description based on Equations \eqref{1d1} and \eqref{1dGP} is
not quite legitimate, both in the high and in the low density
limit: at large density, transverse excitations of the condensate
cannot be discarded and the quasi 1D description \eqref{1dGP} fails; at
low density, phase fluctuations destroy the long range order and the
possibility of a true Bose-Einstein condensation which is at the heart
of the Bogoliubov description \eqref{1d1}. In the remaining of this
section we stick to the simple approach embodied by Equations
\eqref{1d1} and \eqref{1dGP} and we postpone the discussion of its
limitations to Sec. \ref{limitations}.

We denote as a black hole configuration a 1D configuration in which
the asymptotic upstream flow is subsonic (with constant density $n_u$)
and the asymptotic downstream one is supersonic (with constant density
$n_d$).  Typically $n_u\ne n_d$ and when a region of the flow is
denoted for instance as subsonic, this means that in this region the
density of the condensate is constant, and its velocity $V_u$ is
smaller than the asymptotic sound velocity $c_u=\sqrt{g_{1{\rm d}}
  n_u/m}$.

Several black hole configurations have been proposed in
Refs. \cite{Bal08,Car08,Mac09}. The specific form of the order parameter is
always of the type:
\begin{equation}\label{w1}
\Psi_0(x)=\left\{\begin{array}{lcl}
\sqrt{n_u}\exp({\rm i} K_u x)\,\phi_u(x) & \mbox{for} & x<0, \\[2mm]
\sqrt{n_d}\exp({\rm i} K_d x)\,\phi_d(x) & \mbox{for} & x>0,
\end{array}\right.
\end{equation}
where $K_{u,d}=m V_{u,d}/\hbar$, $V_u$ being the asymptotic upstream
flow velocity and $V_d$ the downstream one ($V_u$ and $V_d$ are both
positive). We also introduce the healing lengths $\xi_\alpha=\hbar/(m
c_\alpha)$ and the Mach numbers $M_\alpha=V_\alpha/c_\alpha$
($\alpha=u$ or $d$ depending if one considers upstream or downstream
quantities). The functions $\phi_u$ and $\phi_d$ verify $|\phi_d(x)|=1$
and $\lim_{x\to -\infty}|\phi_u(x)|=1$. The asymptotic upstream and
downstream flows are respectively subsonic and supersonic, meaning
that $M_u<1<M_d$.  A remark on the location of the event horizon is in
order here. First, as in any analogous system, its actual position is
energy-dependent: we will even see below that the horizon disappears
at large energy. This effect being taken into account, the customary
procedure is to do a semi-classical analysis and to define as the true
horizon the large wavelength one. In this case the horizon is the
point where the flow velocity is equal to the local speed of sound.
However, the flow varies rapidly around $x=0$ in the configurations we
study below, and a quantity such as the local speed of sound is ill
defined in this region. As a result, the position of the event horizon
cannot be unambiguously defined. This is not a drawback of the model:
what really matters is that the asymptotic upstream and downstream
flows are truly respectively sub- and super-sonic.

\subsubsection{The ``waterfall''
  configuration}\label{bhconfiguration}
\begin{figure}
\begin{center}
\includegraphics*[width=0.99\linewidth]{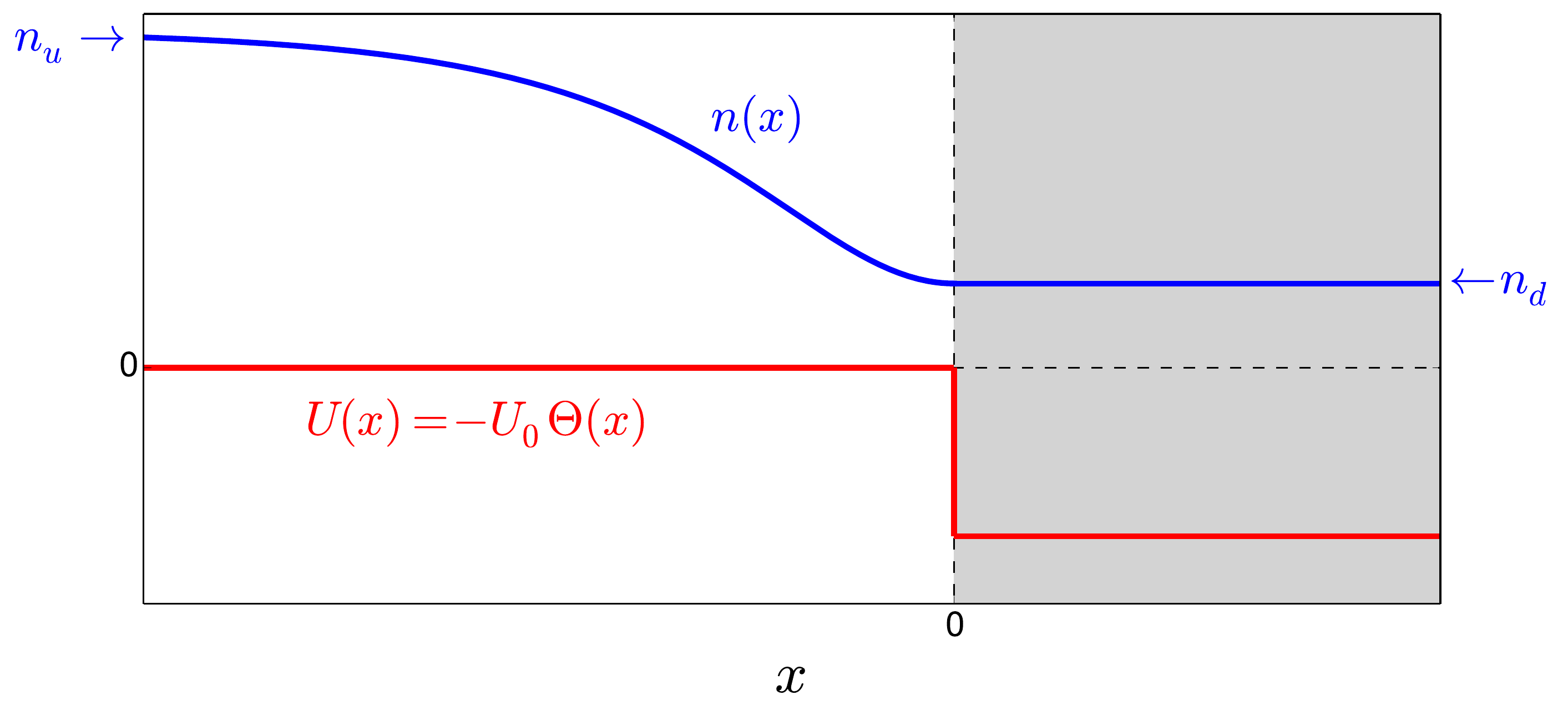}
\end{center}
\caption{Waterfall configuration. The flow is incident
  from the left with an asymptotic density $n_u$ and a (subsonic) velocity
  $V_u$. The downstream ($x>0$) velocity $V_d$ is supersonic. The
  downstream density $n_d$ is constant and lower than $n_u$. The region $x>0$ 
is shaded in order to recall that it corresponds to the interior of
the black hole.}
\label{fig-water}
\end{figure}

We first consider one of the realistic configurations introduced in
Ref. \cite{Lar12} and realized experimentally in Ref. \cite{Ste16}. In
this configuration, which we denote as ``waterfall'', the 1D flow of a
BEC is subject to an external potential which is a step function of
the form $U(x)=-U_0\, \Theta(x)$, where $\Theta$ is the Heaviside
function (and $U_0>0$). In this case, a stationary profile with a flow
which is subsonic upstream and supersonic downstream, i.e., a black
hole configuration, has been identified in Refs. \cite{Ste16,Leb03}
and is schematically represented in Fig. \ref{fig-water}. The upstream
profile is exactly one half of a dark soliton and the downstream one
corresponds to a flow with constant density and velocity (cf.
Fig. \ref{fig-water}).
In the waterfall configuration one has $\phi_d(x)=-{\rm i}$ and
$\phi_u(x)= \cos\theta\tanh(x\cos\theta/\xi_u)-{\rm i}\sin\theta$,
where $\sin\theta=M_u$.
One has also ${U_0}/{g_{1{\rm d}} n_u}=\frac{1}{2} ({M_u^2}+M_u^{-2})-1$ and
$V_u=c_d<c_u<V_d$ which indeed corresponds to a black hole type of
horizon ($M_u<1<M_d$).

\subsubsection{The ``$\delta$ peak''
  configuration}\label{delta-peak}

In this configuration the 1D flow of a BEC is subject to an external
potential which is a Dirac distribution of the form $U(x)=\kappa \,
\delta(x)$, where $\kappa>0$. In this case, a stationary profile with
a flow which is subsonic upstream and supersonic downstream, i.e., a
black hole configuration, has been identified in
Refs. \cite{Leb01,Pav02}, and it has been shown in Ref. \cite{Kam12}
how this configuration can be reached dynamically.  The downstream one
corresponds to a flow with constant density and velocity and the
upstream profile is a fraction of a dark soliton, with $\phi_u(x)=
\cos\theta\tanh[(x-x_0)\cos\theta/\xi_u]-{\rm i}\sin\theta$, where
$\sin\theta=M_u$ and $x_0$ depends on $M_u$ and $\kappa$ (see details
in \cite{Lar12}).

\subsubsection{The ``flat profile''
  configuration}\label{flatprofile}

We finally present a model configuration first introduced in
Ref. \cite{Car08}, which has been denoted as ``flat profile'' in
Ref. \cite{Lar12}. Although this configuration is not likely to be
realized experimentally, it has been demonstrated in Ref. \cite{Lar12}
that it yields a density correlation signal which is quite similar to
the one obtained in the more realistic waterfall and $\delta-$peak
configurations. We will use the flat profile configuration to present
our results below (in Sec. \ref{sec.mom.corr}) for pedagogical
reasons, because it leads to a simpler phenomenology for the momentum
correlation than the other configurations.

In the flat profile configuration one has $n_u=n_d\equiv n_0$ and
$K_u=K_d\equiv K_0$ and the $\phi_\alpha$ functions of Eq. (\ref{w1})
assume a very simple value: $\phi_u(x)=\phi_d(x)=1$. This means that
$\Psi_0(x)$ is a plane wave for all $x$. A horizon can still be
realized in this case by tuning the values of the external potential
$U(x)$ and of the non-linear constant $g_{1{\rm d}}(x)$ such that
\begin{equation}\label{gradUg}
U(x)=\left\{\begin{array}{lcl}
U_u & \mbox{for} & x<0, \\
U_d & \mbox{for} & x>0,
\end{array}\right.
\quad\mbox{and}\quad
g_{1{\rm d}}(x)=\left\{\begin{array}{lcl}
g_u & \mbox{for} & x<0, \\
g_d & \mbox{for} & x>0.
\end{array}\right.
\end{equation}
These values are
chosen so that a flow with $\Psi_0(x)=\sqrt{n_0}\exp({\rm i}K_0 x)$ 
is solution of Eq. (\ref{1dGP}) for all $x$.  This imposes
\begin{equation}\label{gradino1}
\frac{c_d}{c_u}=\frac{M_u}{M_d}=\frac{\xi_u}{\xi_d}\; ,
\quad\mbox{and}\quad
g_u n_0 + U_u = g_d n_0 + U_d\; .
\end{equation}
We finally note that in the flat profile configuration one has
$c_d<V_d=V_u<c_u$. This corresponds to a sonic black hole horizon
since the upstream and downstream Mach numbers verify $M_u<1<M_d$.

It is important to notice that, at variance with the cases of the
waterfall and of the $\delta$ peak configurations, where, once an
asymptotic flow is fixed (say, the upstream one) all the characteristics
of the flow are uniquely determined, in the case of the flat profile
configuration the values of $M_u$ and $M_d$ can be chosen
independently one from the other. As a result, one cannot directly
compare the results of, say the measure of non separability, of a
waterfall and a $\delta$ peak configuration, but each of them can be
compared with a flat profile configuration. This will be done in
Figs. \ref{fig.twobody.water}, \ref{fig.twobody.delta},
\ref{fig.CS.waterfall} and \ref{fig.CS.delta}.

\subsection{The excitation spectrum of a homogeneous
  condensate}\label{excitation-spectrum}

\begin{figure}
\begin{center}
  \includegraphics*[width=0.99\linewidth]{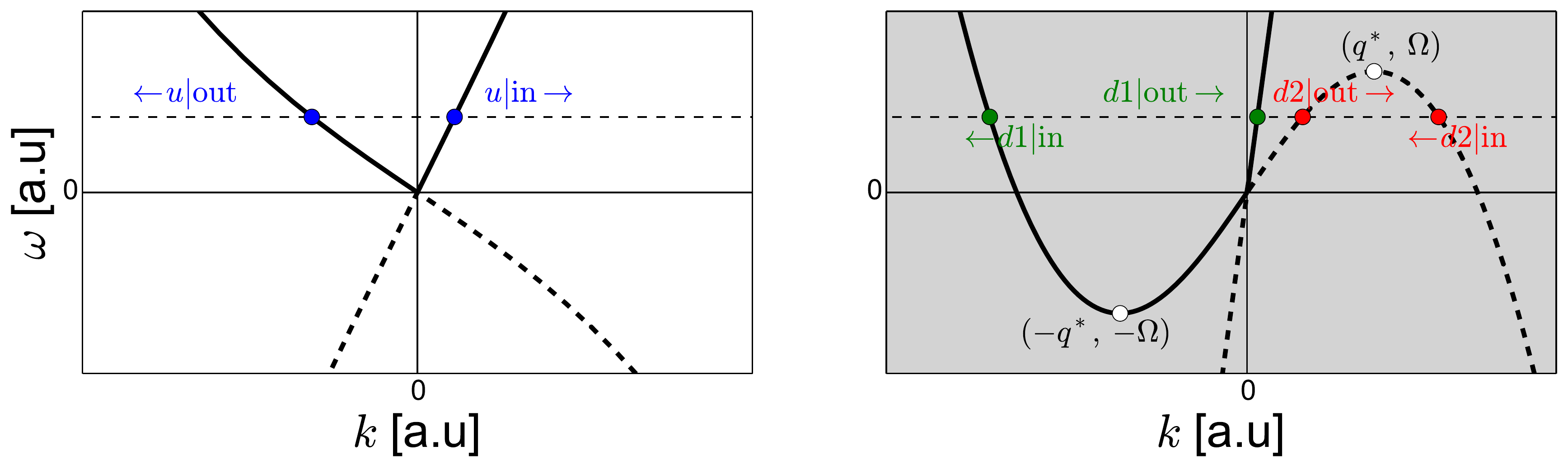}

\vspace{5mm}

\hspace{1cm}\includegraphics*[width=0.5\linewidth]{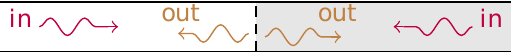}
\end{center}
\caption{Dispersion relations. The left plot
  corresponds to a subsonic flow. The right plot corresponds to a
  supersonic flow; it is shaded in order to recall that it describes
  the situation inside the black hole. In each plot the horizontal
  dashed line is fixed by the chosen value of $\omega$. The
  $q_\ell(\omega)$'s are the corresponding abscissas, with $\ell\in
  \{u|{\rm in},u|{\rm out}\}$ in the left plot and $\ell\in \{d1|{\rm
    in},d1|{\rm out},d2|{\rm in},d2|{\rm out}\}$ in the right
  plot. The direction of propagation (left or right) of the
  eigen-modes is represented by an arrow.  The lower diagram illustrates
  schematically the ``ingoing'' and ``outgoing'' terminology used in
  the text and in the two upper plots.
}\label{fig-dispersion}
\end{figure}

In the case of a static homogeneous condensate, the dispersion relation
of longitudinal excitations is the standard Bogoliubov one:
\begin{equation}\label{1d7}
  \omega = \omega_{\rm\sss B}(q) = c \, | q | 
  \left(1 + \tfrac{1}{4}\xi^2 q^2\right)^{1/2} \; .
\end{equation}
In a region where the condensate flows with a velocity $V$ this is modified
to 
\begin{equation}\label{qwe}
\left(\omega - V q\right)^2 = \omega^2_{\rm\sss B}(q)\; .
\end{equation}  
In this case $\omega$ is the energy of the elementary excitation in
the frame where the obstacle is at rest, while $\omega_{\rm\sss B}$ is
the frequency measured in the frame of the fluid. The momentum of the
excitation relative to the background flow is $\hbar q$, and its
momentum in the frame of the obstacle is $\hbar q + m V$. In a black hole
configuration, the upstream and the downstream channels are
characterized by dispersion relations of the type \eqref{qwe} (with
the appropriate values of $V$, $\xi$ and $c$), they are illustrated in
Fig. \ref{fig-dispersion}.
In this figure the upper left (upper right) plot represents the asymptotic
upstream subsonic (downstream supersonic) case. The part of the
dispersion relation represented by a dashed line correspond to
negative norm modes, as explained in the following section. In all
this work we only consider the $\omega>0$ part of the dispersion
relations, this is made possible by the $\omega \leftrightarrow
-\omega$ symmetry of the spectrum \cite{Fet74}.

One sees in the figure that, upstream, the waves of one of the
channels are directed towards the horizon (ingoing waves, denoted as
$u|{\rm in}$) whereas the waves in the other channel propagate away
from the horizon (outgoing waves: $u|{\rm out}$). The definition of
which mode is ingoing and which is outgoing depends on the side of the
horizon that is considered, this is illustrated by the lower diagram of
Fig. \ref{fig-dispersion}. Downstream, there are two ingoing waves
($d1|{\rm in}$ and $d2|{\rm in}$) and two outgoing waves ($d1|{\rm
  out}$ and $d2|{\rm out}$). Note that the two $d2$ channels disappear
at large energy, when $\omega>\Omega$, see Fig. \ref{fig-dispersion}.
$\Omega$ is the energy of an elementary excitation whose group
velocity in the frame where the condensate is at rest
($\partial\omega_{\rm\sss B}/\partial q$) is equal to the flow velocity
$V_d$ (such an equality is only possible for supersonic
flows). Excitations with momentum larger than the one of this
excitation (which is denoted as $q^*$ in Fig. \ref{fig-dispersion})
move faster than the flow and are able to escape the ``black hole''.

\subsection{The wave function in real space}\label{WFreal}
The field operator $\hat{\psi}(x)$
associated in the Schr\"odinger representation
to the particles which are out of the condensate
[as defined by Eq. (\ref{1d1})] is expanded over the scattering modes:
\begin{eqnarray}\label{q1}
\hat{\psi}(x) &=&
\ep^{{\rm i} K_{\alpha} x} \int_{0}^{\infty} \frac{{\rm d}\omega}{\sqrt{2\pi}}
\sum_{\sss L\in\{U,D1\}} \Big[ \bar{u}_{\sss L}(x,\omega) 
 \hat{b}_{\sss L}(\omega)
+ \bar{w}_{\sss L}^{*}(x,\omega)
\hat{b}_{\sss L}^{\dag}(\omega) \Big] \nonumber \\
&+& \ep^{{\rm i} K_{\alpha} x} \int_{0}^{\Omega} 
\frac{{\rm d}\omega}{\sqrt{2\pi}}
\Big[ \bar{u}_{\sss D2}(x,\omega) 
\hat{b}_{\sss D2}^{\dag}(\omega)
+ \bar{w}_{\sss D2}^{*}(x,\omega)
\hat{b}_{\sss D2}(\omega) \Big] \; .
\end{eqnarray}
$K_\alpha$ in (\ref{q1}) is equal to $K_u$ if $x<0$ and to $K_d$ if
$x>0$.  The $\hat{b}_{\sss L}^\dagger(\omega)$'s create an excitation
of energy $\hbar\omega$ in one of the three scattering modes ($L=U$,
$D1$ or $D2$), they obey the following commutation relation:
\begin{equation}\label{q2}
[\hat{b}_{\sss L}(\omega),\hat{b}^\dagger_{\sss L'}(\omega')] =
\delta_{\sss L,L'}\delta(\omega-\omega').
\end{equation}
Each of the three scattering modes ($U$, $D1$ or $D2$) is initiated by
one of the three entrance channels ($u|{\rm in}$, $d1|{\rm in}$ or
$d2|{\rm in}$). Far from the horizon, the density and the velocity of
the flow are position-independent and the corresponding wave functions
are mere plane waves of the following form:

$\bullet$ Deep in the upstream
subsonic region, i.e., for $x<0$, $x\ll -\xi_u$ :
\begin{eqnarray}\label{q3}
\begin{pmatrix}
\bar{u}_{\sss U}(x) 
\\
\bar{w}_{\sss U}(x)\end{pmatrix}
& = &
 S_{u,u}\, \begin{pmatrix}
{\cal U}_{u|{\rm out}}\\
{\cal W}_{u|{\rm out}}\end{pmatrix}
\ep^{{\rm i}q_{u|{\rm out}}x} +
\begin{pmatrix}
{\cal U}_{u|{\rm in}}\\
{\cal W}_{u|{\rm in}}\end{pmatrix}
\ep^{{\rm i}q_{u|{\rm in}}x}
\; , \nonumber \\
\begin{pmatrix}\bar{u}_{\sss D1}(x)\\
\bar{w}_{\sss D1}(x)\end{pmatrix}
& = & S_{u,d1}\, \begin{pmatrix}
{\cal U}_{u|{\rm out}}\\
{\cal W}_{u|{\rm out}}\end{pmatrix}
\ep^{{\rm i}q_{u|{\rm out}}x} \; , \\
\begin{pmatrix}
\bar{u}_{\sss D2}(x)\\
\bar{w}_{\sss D2}(x)\end{pmatrix}
& = & S_{u,d2}\, \begin{pmatrix}
{\cal U}_{u|{\rm out}}\\
{\cal W}_{u|{\rm out}}\end{pmatrix}
\ep^{{\rm i}q_{u|{\rm out}}x} \; .\nonumber
\end{eqnarray}

$\bullet$ Deep in the downstream supersonic region, i.e., when $x>0$,
$x\gg \xi_d$ :
\begin{eqnarray}\label{q4}
\begin{pmatrix}
\bar{u}_{\sss U}(x)\\
\bar{w}_{\sss U}(x)\end{pmatrix}
&\!\! = \!\!& S_{d1,u}  \begin{pmatrix}{\cal U}_{d1|{\rm out}}\\
{\cal W}_{d1|{\rm out}}\end{pmatrix}
\ep^{{\rm i}q_{d1|{\rm out}}x} + S_{d2,u} 
 \begin{pmatrix}{\cal U}_{d2|{\rm out}}\\
{\cal W}_{d2|{\rm out}}
\end{pmatrix}
\ep^{{\rm i}q_{d2|{\rm out}}x} \; , \\
\begin{pmatrix}\bar{u}_{\sss D1}(x)\\
\bar{w}_{\sss D1}(x)\end{pmatrix}
& \!\! = \!\!& 
S_{d1,d1} \begin{pmatrix} {\cal U}_{d1|{\rm out}}\\
{\cal W}_{d1|{\rm out}}\end{pmatrix}
\ep^{{\rm i}q_{d1|{\rm out}}x} + S_{d2,d1} \begin{pmatrix}
{\cal U}_{d2|{\rm out}}\\
{\cal W}_{d2|{\rm out}}\end{pmatrix}
\ep^{{\rm i}q_{d2|{\rm out}}x} 
+
\begin{pmatrix}{\cal U}_{d1|{\rm in}}\\
{\cal W}_{d1|{\rm in}}\end{pmatrix} \ep^{{\rm i}q_{d1|{\rm in}}x}
\; ,\nonumber\\
\begin{pmatrix}\bar{u}_{\sss D2}(x)\\
\bar{w}_{\sss D2}(x)\end{pmatrix}
& \!\! = \!\!& 
S_{d1,d2} \begin{pmatrix}
{\cal U}_{d1|{\rm out}}\\
{\cal W}_{d1|{\rm out}}\end{pmatrix}
\ep^{{\rm i}q_{d1|{\rm out}}x} + S_{d2,d2}
\begin{pmatrix}{\cal U}_{d2|{\rm out}}\\
{\cal W}_{d2|{\rm out}}\end{pmatrix}
\ep^{{\rm i}q_{d2|{\rm out}}x} 
+\begin{pmatrix}{\cal U}_{d2|{\rm in}}\\
{\cal W}_{d2|{\rm in}}\end{pmatrix} \ep^{{\rm i}q_{d2|{\rm in}}x}
\; .\nonumber
\end{eqnarray}
Note that the $\bar{u}_L$'s, $\bar{w}_L$'s, $q_\ell$'s,
$S_{i,j}$'s, ${\cal U}_\ell$'s and the ${\cal W}_\ell$'s in
Eqs. \eqref{q3} and \eqref{q4} all depend on $\omega$.  For instance
$q_\ell(\omega)$ is defined on Fig. \ref{fig-dispersion} for $\ell\in
\{u|{\rm out}, u|{\rm in}, d1|{\rm out}, d1|{\rm in}, d2|{\rm out},
d2|{\rm in}\}$. We chose a normalization of the the coefficients $
{\cal U}_\ell$ and ${\cal W}_\ell$ -- the so called ``Bogoliubov
amplitudes'' --  such that
\begin{equation}\label{norme}
|{\cal U}_{\ell}(\omega) |^2-|{\cal W}_{\ell}(\omega)|^2=
\frac{\pm 1}{|\partial\omega/\partial q_\ell|}.
\end{equation}
The sign $+$ ($-$) in \eqref{norme} refers to positive (negative) norm
modes. All the upstream modes ($u|{\rm in}$ and $u|{\rm out}$)
have a positive norm. In the downstream region, the $d1|{\rm in}$ and 
$d1|{\rm out}$ modes have a positive norm while the $d2|{\rm in}$ and 
$d2|{\rm out}$ ones have a negative norm.
The normalization \eqref{norme} ensures that a positive
(negative) mode carries a current $+1$ ($-1$). The explicit
expression of the
the coefficients $ {\cal U}_\ell(\omega)$ and ${\cal W}_\ell(\omega)$
corresponding to the normalization \eqref{norme} is given in
Ref. \cite{Lar12}.
Expressions \eqref{q3} are not valid close to the horizon due to (i)
the modification of the density profile which is position-dependent in
vicinity of the horizon\footnote{Note however that the density is not
  affected by the horizon in the flat profile configuration. In the
  waterfall and delta peak configurations, the corresponding explicit
  form, correct even close to the horizon, is given in
  Ref. \cite{Lar12}.} and (ii) to the occurrence of evanescent modes
(with complex momenta solutions of Eq. \eqref{qwe}) which are of
importance near the horizon.

Let us for instance discuss the physical content of the last of
Eqs. (\ref{q4}). It describes a $d2|{\rm in}$ wave incoming from
$+\infty$ (last term of the r.h.s., the corresponding group velocity
is negative, cf. Fig. \ref{fig-dispersion}) which is back scattered
into a $d1|{\rm out}$ and a $d2|{\rm out}$ wave, with respective
reflection amplitudes $S_{d1,d2}$ and $S_{d2,d2}$.  Part of this wave
is also transmitted in the $x<0$ region as a $u|{\rm out}$ wave with
transmission amplitude $S_{u,d2}$ : the corresponding expression far
from the horizon is displayed in the last of Eqs. (\ref{q3}). The
scattering amplitudes form the $S$ matrix which is $3\times 3$ for
energies $\omega$ lower than the threshold $\Omega$ defined on
Fig. \ref{fig-dispersion}:
\begin{equation}\label{Smatrix}
S=
\begin{pmatrix}
S_{u,u} & S_{u,d1} & S_{u,d2} \\
S_{d1,u} & S_{d1,d1} & S_{d1,d2} \\
S_{d2,u} & S_{d2,d1} & S_{d2,d2}
\end{pmatrix}\; .
\end{equation}
Current conservation reads
\begin{equation}\label{e19}
S^\dagger \eta S = \eta = S \eta S^\dagger, \quad \mbox{where} \quad
\eta = {\rm diag}(1,1,-1).
\end{equation}
For $\omega>\Omega$, the last row and the last column of
\eqref{Smatrix} vanish because the $d2$ mode disappears. In this
case the $S$ matrix is $2\times 2$ and satisfies $S S^\dagger =
\mathbbm{1}$.

In Eqs. (\ref{q3})
and (\ref{q4}) we did not write the contribution of the evanescent
modes since they decay exponentially and are negligible far from the
horizon (when $|x|\gg \xi_{(u,d)}$), but we fully take these modes
into account in the expression of the $\bar{u}$'s and the $\bar{w}$'s
near the horizon (around $x=0$); this is needed for an accurate
computation of the $S$ matrix.  In a given configuration (say
``waterfall''), the elements of the $S$ matrix are determined for each
value of $\omega$ by enforcing continuity of the wave functions (and
of their spatial derivatives) of the elementary excitations at
$x=0$. This represents an easy numerical task which consists in solving 
a linear $4\times 4$ system for each value of $\omega$.

\subsection{The wave function in momentum space}
 
Because of the presence of the negative norm/negative energy $d2|{\rm
  in}$ mode, stationary black hole configurations such as the ones
presented in section \ref{q1dbh} are meta-stable\footnote{In the
  gravitational context the $d2|{\rm in}$ mode is absent, and this
  metastability arises when the black hole is dynamically
  created.}. Indeed, when reaching the horizon, the $d2|{\rm in}$
waves give rise to a radiation of $u|{\rm out}$ quanta in the upstream
region, which constitutes the spontaneous Hawking radiation.  In BEC
systems however, this radiation is not easily detected. The reason is
that the occupation of the Hawking radiating modes is approximately of
thermal type, with an effective temperature much lower than the true
temperature of the system (typically by a factor 10), and the Hawking
signal is thus drowned in the thermal noise. This circumstance led the
authors of Refs. \cite{Bal08,Car08} to look for density correlations
as alternative evidence of the Hawking effect.

The idea, checked in Refs. \cite{Bal08,Car08} is that outgoing waves
generated by the same $d2|{\rm in}$ mode are all correlated. Moreover,
since the corresponding amplitudes are governed by the $S$ matrix
which describes how ingoing waves hitting the horizon generate
outgoing ones, the knowledge of the $S$ matrix makes a detailed
description of the correlation signal possible. This point has been
checked thoroughly in Refs. \cite{Mac09,Rec09,Lar12}.  In particular,
the characteristic peaks of the density correlations correspond to the
Hawking quantum ($u|{\rm out}$) - partner ($d2|{\rm out}$)
correlations (for points situated on both sides of the horizon), and
also to correlations along the $u|{\rm out}-d1|{\rm out}$ (again, for
points situated on both sides of the horizon) and $d2|{\rm
  out}-d1|{\rm out}$ (for both points inside the horizon) channels.

In a BEC, momentum correlations could be more precisely detected than
density correlations, by following a procedure used in
Ref. \cite{Jas12} in a similar context, in the case of the dynamical
Casimir effect. This is the reason why we proposed in
Ref. \cite{Boi15} to demonstrate the existence of
sonic Hawking radiation by the means of correlations in momentum
space.

\subsubsection{A local Fourier transform}\label{local-FT}
 
\begin{figure}[h]
\begin{center}
\begin{picture}(13,4.5)
\put(0.,0.25){\includegraphics[width=10.5cm]{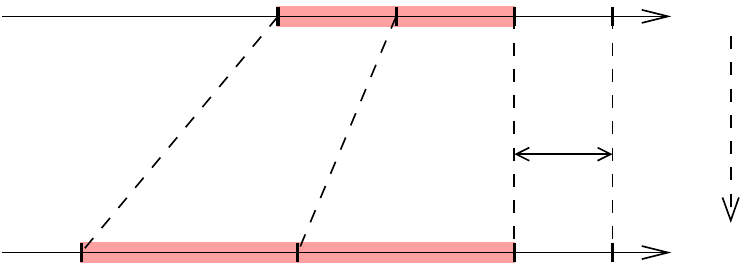}}
\put(3.2,4.0){$X_u-\sigma_u$}
\put(5.4,4.0){$X_u$}
\put(6.6,4.0){$X_u+\sigma_u$}
\put(8.55,4.0){$0$}
\put(9.6,3.7){$x$}
\put(0.5,0.0){$X_u-\sigma_u$}
\put(4.0,0.0){$X_u$}
\put(6.6,0.0){$X_u+\sigma_u$}
\put(8.55,0.0){$0$}
\put(9.6,0.35){$x$}
\put(7.55,2.){$\gg \xi_u$}
\put(10.6,2.5){increasing}
\put(10.6,2.1){$|X_u|$ and $\sigma_u$}
\end{picture}
\end{center}
\caption{Schematic representation of the behavior of the parameters
$X_u$ and $\sigma_u$ of the window function $\Pi_u(x)$ defined in
  Eq. \eqref{win-up}. The shaded zone is the region where the window
  function notably differs from zero. The spacing between
  $X_u+\sigma_u$ and the origin is large compared to $\xi_u$, and this
  ensures that the Fourier analysis, which is made local thanks to the
  window function, is performed in the deep subsonic
  region.}\label{fig.window}
\end{figure}
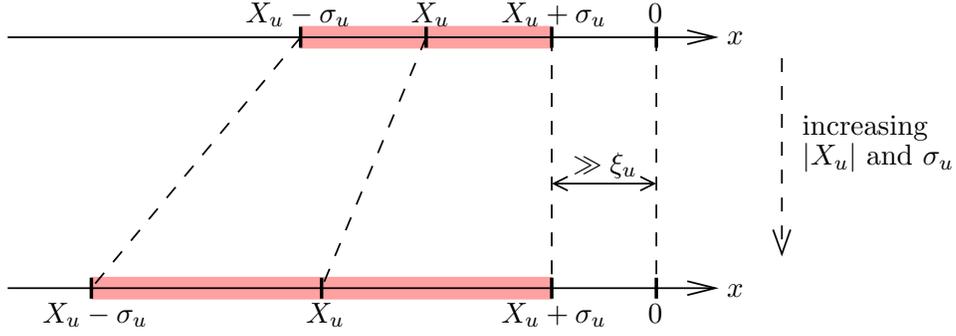

By appropriately introducing a local Fourier transform in both the
subsonic (exterior of the black hole) and supersonic (black hole
interior) regions, we shall explicitly construct the momentum
correlator and analyze, in Sec. \ref{sec.mom.corr}, its nontrivial
qualitative features, which are in correspondence with those present
in the density correlation signal.
This signal concerns the
occupation number in the momentum representation: $\hat{N}(K) =
\hat{\psi}^\dagger(K) \hat{\psi}(K)$, where $\hat{\psi}(K)$ is the
Fourier transform of $\hat{\psi}(x)$\footnote{We only consider here
  the momentum distribution of particles which are outside of the
  condensate and discard the contribution of the condensate.}:
\begin{equation}\label{defFT}
\hat{\psi}(K)=\frac{1}{\sqrt{2\pi}}\int_{\mathbb{R}} 
{\rm d}x \exp\{-{\rm i}K x\}
\hat{\psi}(x) \; .
\end{equation} 
From expression \eqref{q1} and the mode analysis presented in
Secs. \ref{excitation-spectrum} and \ref{WFreal}, it is clear that the momentum
distribution has a different form in the far-upstream and
far-downstream regions. Hence, instead of (\ref{defFT}), it is more
appropriate to perform a specific mode analysis in each of these
regions \cite{Nov14,Fin14,Nov15}. This can be done by using a window
function selecting the desired region of space. The precise form of
this window function is irrelevant, but for concreteness we will
consider a Gaussian. In the upstream region for instance, one takes a
window
\begin{equation}\label{win-up}
\Pi_u(x)=\Lambda_u \exp\left\{-\frac{(x-X_u)^2}{\sigma_u^2}\right\}\; ,
\end{equation}
and the corresponding windowed Fourier transform is
\begin{equation}\label{FT-win}
\hat{\psi}_u(K)=\frac{1}{\sqrt{2\pi}}\int_{\mathbb{R}} 
{\rm d}x \exp\{-{\rm i}K x\} \Pi_u(x)
\hat{\psi}(x) \; .
\end{equation}
This procedure is meant to select the momentum components which can be
identified from \eqref{q3}. For this purpose, the parameters of the
window function have to be chosen in order to work in the appropriate
region of space. This is done by taking $X_u<0$, $\sigma_u>0$ and
letting $X_u$ and $\sigma_u$ respectively tend to $-\infty$ and
$+\infty$, imposing for the ratio $X_u/\sigma_u\to -1$ which allows that
$X_u+\sigma_u=C^{\rm st}\ll -\xi_u$. This procedure is illustrated in
Fig. \ref{fig.window}.
It is important to take the limit of large $\sigma_u$ and $|X_u|$ for
obtaining a sharp ($\delta$-like) distribution in momentum space. In
Eq. \eqref{win-up} the normalization parameter $\Lambda_u$ is
introduced to effectively describe the finite efficiency of the
experimental detection apparatus. More precisely, it is argued in Sec.
\ref{correl-noBH} that the quantity $\lambda_u=\sigma_u
\Lambda_u^2 /\sqrt{2\pi}$ describes the rate of detection of particles
in the window $[X_u-\sigma_u,X_u+\sigma_u]$: if $\lambda_u=0$ none of
the particles are detected, if instead $\lambda_u=1$, they are all
detected. 
It is interesting to stress that 
 the explicit results given
in Sec.  \ref{sec.mom.corr} for the normalized momentum correlation
function \eqref{m2bis} do not depend on the efficiencies $\lambda_u$
and $\lambda_d$ of the detectors. 

Of course, a local Fourier transform similar to \eqref{FT-win} is
performed downstream, using a different window $\Pi_d(x)$
with parameters $\Lambda_d$, $X_d$ ($>0$) and $\sigma_d$, leading to
the downstream Fourier transform $\hat{\psi}_d(K)$. We will see in Appendix
\ref{window-Fourier} that the parameters of the upstream and of the
downstream window function have to be chosen in a specific manner when
one wants to consider a specific correlation signal. However, these
precise conditions -- which are of importance for the theoretical
analysis of the experimental 
detection scheme -- can be replaced, once met, by the following
schematic, but natural rules:

\begin{enumerate}
\item\label{R1} : A contribution to \eqref{q1}, which, for $x<0$, reads
$\ep^{{\rm i}K_u x} \int{\rm d}\omega \, C^{\rm st} \, \ep^{\pm{\rm
    i}q_\ell(\omega) x}$  
yields a contribution $\sqrt{2\pi}\int{\rm d}\omega \, C^{\rm st} \,
\delta(K-(K_u \pm q_\ell(\omega)))$ to the upstream Fourier transform
$\hat\psi_u(K)$.
 \item\label{R2} : A  contribution to \eqref{q1}, which, for $x>0$, reads
$\ep^{{\rm i}K_d x} \int{\rm d}\omega \, C^{\rm st} \, \ep^{\pm{\rm
    i}q_\ell(\omega) x}$ 
yields a contribution $\sqrt{2\pi} \int{\rm d}\omega \, C^{\rm st} \,
\delta(K-(K_d \pm q_\ell(\omega)))$ to the downstream Fourier
transform $\hat\psi_d(K)$.
\end{enumerate}

These rules are less rigorous than the correct mathematical procedure
\eqref{FT-win} which uses window functions for defining the local
Fourier transforms, but it is shown in Appendix \ref{window-Fourier}
that, provided some simple and natural redefinitions (in particular of
the singular Dirac distributions) are considered, they yield the correct
result. Hence, we shift discussion of the more rigorous results to
Appendix \ref{window-Fourier} and present our results in the main text
using these simplified rules.  Performing the schematic Fourier
transform in the $x<0$ region we get,
\begin{eqnarray}\label{q5}
\hat{\psi}_u(K)=\int_0^\infty {\rm d}\omega
& \Big\{ &\delta(K-K_u-q_{u|{\rm out}}){\cal U}_{u|{\rm out}}
(S_{u,u}\hat{b}_{\sss U} + S_{u,d1}\hat{b}_{\sss D1} 
+S_{u,d2}\hat{b}^\dagger_{\sss D2})\nonumber \\
& +&\delta(K-K_u+q_{u|{\rm out}})  
{\cal W}_{u|{\rm out}}^*(S_{u,u}^*\hat{b}_{\sss U}^\dagger + 
S_{u,d1}^*\hat{b}_{\sss D1}^\dagger +S_{u,d2}^*\hat{b}_{\sss D2})\nonumber \\
&+&\delta(K-K_u+q_{u|{\rm in}}){\cal W}^*_{u|{\rm in}} \hat{b}^\dagger_{\sss U} 
\nonumber \\
&+&\delta(K-K_u-q_{u|{\rm in}}){\cal U}_{u|{\rm in}} \hat{b}_{\sss U} \Big\}
 \; ,
\end{eqnarray}
whereas in the $x>0$ one gets 
\begin{eqnarray}\label{q6}
\hat{\psi}_d(K)=\int_0^\infty {\rm d}\omega
& \Big\{&
\delta(K-K_d-q_{d1|{\rm out}}){\cal U}_{d1|{\rm out}}
(S_{d1,u}\hat{b}_{\sss U} + 
S_{d1,d1}\hat{b}_{\sss D1} +S_{d1,d2}\hat{b}^\dagger_{\sss D2})\nonumber \\
&+&\delta(K-K_d+q_{d1|{\rm out}}){\cal W}_{d1|{\rm out}}^*
(S_{d1,u}^*\hat{b}_{\sss U}^\dagger + 
S_{d1,d1}^*\hat{b}_{\sss D1}^\dagger +S_{d1,d2}^*\hat{b}_{\sss D2})\nonumber \\
&+&\delta(K-K_d-q_{d2|{\rm out}}){\cal U}_{d2|{\rm out}}
(S_{d2,u}\hat{b}_{\sss U} + 
S_{d2,d1}\hat{b}_{\sss D1} +S_{d2,d2}\hat{b}^\dagger_{\sss D2})\nonumber \\
&+&\delta(K-K_d+q_{d2|{\rm out}}){\cal W}_{d2|{\rm out}}^*
(S_{d2,u}^*\hat{b}_{\sss U}^\dagger + 
S_{d2,d1}^*\hat{b}_{\sss D1}^\dagger +S_{d2,d2}^*\hat{b}_{\sss D2})\nonumber \\
&+&
\delta(K-K_d+q_{d1|{\rm in}}){\cal W}^*_{d1|{\rm in}} \hat{b}^\dagger_{\sss D1}
\nonumber \\
&+&
\delta(K-K_d-q_{d1|{\rm in}}){\cal U}_{d1|{\rm in}} \hat{b}_{\sss D1}\nonumber \\
&+&
\delta(K-K_d-q_{d2|{\rm in}}){\cal U}_{d2|{\rm in}} \hat{b}^\dagger_{\sss D2}\nonumber \\
&+&
\delta(K-K_d+q_{d2|{\rm in}}){\cal W}_{d2|{\rm in}}^* \hat{b}_{\sss D2}
 \Big\} \; .
\end{eqnarray}
In the two above integrals the $q_\ell$'s are functions of $\omega$
computed as schematically represented in
Fig. \ref{fig-dispersion}. The $\omega$ integration yields factors
$|\partial q_\ell/\partial \omega|$ which can be absorbed in a
re-definition of the ${\cal U}$'s and of the ${\cal W}$'s:
\begin{equation}\label{q7}
\widetilde{\cal U}_\ell(q)={\cal U}_\ell(\omega_\ell(q)) 
\left|\frac{\partial \omega_\ell}{\partial q}\right| \;\mbox{and}\;\;
\widetilde{\cal W}_\ell(q)={\cal W}_\ell(\omega_\ell(-q)) 
\left|\frac{\partial \omega_\ell}{\partial q}\right|  \; ,
\end{equation}
where $\omega_\ell(q)$ is the reciprocal function of
$q_\ell(\omega)$. The ``tilde Bogoliubov coefficients'' satisfy the
following normalization:
\begin{equation}\label{norme.tilde}
|\widetilde{\cal U}_{\ell}(q) |^2-|\widetilde{\cal W}_{\ell}(q)|^2=
\pm 
\left|\frac{\partial \omega_\ell}{\partial q}\right|
\; .
\end{equation}
Note that in the integrals defining the
expressions \eqref{q5} and \eqref{q6} of $\hat{\psi}_u(K)$ and
$\hat{\psi}_d(K)$, all the terms (Bogoliubov coefficients and
coefficients of the $S$-matrix) involving a
$d2$-subscript cancel when $\omega>\Omega$.

Defining $k_u=K-K_u$ and $k_d=K-K_d$ this yields for the upstream
Fourier transform [instead of \eqref{q5}],
\begin{eqnarray}\label{q8}
  \hat{\psi}_u(k_u<0)& = & 
  \widetilde{\cal U}_{u|{\rm out}}(k_u)
  \Big[S_{u,u}\; \hat{b}_{\sss U}
  + S_{u,d1}\; \hat{b}_{\sss D1}+
  S_{u,d2}\; \hat{b}^\dagger_{\sss D2}\Big]_{\omega=\omega_{u|{\rm out}}(k_u)}
  \nonumber \\ 
  &+&
  \widetilde{\cal W}^*_{u|{\rm in}}(k_u) \; 
  \hat{b}^\dagger_{\sss U}(\omega_{u|{\rm in}}(-k_u))
  \; ,
\end{eqnarray}
\begin{eqnarray}\label{q8bis}
\hat{\psi}_u(k_u>0)& = & 
\widetilde{\cal W}_{u|{\rm out}}^*(k_u)
\Big[S_{u,u}^*\; \hat{b}_{\sss U}^\dagger
+ S_{u,d1}^*\; \hat{b}_{\sss D1}^\dagger+
S_{u,d2}^*\; \hat{b}_{\sss D2}\Big]_{\omega=\omega_{u|{\rm out}}(-k_u)}
\nonumber \\ 
&+&
\widetilde{\cal U}_{u|{\rm in}}(k_u) \; 
\hat{b}_{\sss U}(\omega_{u|{\rm in}}(k_u))
 \; ,
\end{eqnarray}
and for the downstream Fourier transform [instead of \eqref{q6}], 
\begin{eqnarray}\label{q9}
  \hat{\psi}_d(k_d>0)& = & 
  \widetilde{\cal U}_{d1|{\rm out}}(k_d)
  \Big[S_{d1,u}\; \hat{b}_{\sss U} +
  S_{d1,d1}\; \hat{b}_{\sss D1}
  + S_{d1,d2}\; \hat{b}^\dagger_{\sss D2}\Big]_{\omega=\omega_{d1|{\rm out}}(k_d)}
  \nonumber \\ 
  &+&
  \widetilde{\cal U}_{d2|{\rm out}}(k_d)
  \Big[S_{d2,u}\; \hat{b}_{\sss U}+
  S_{d2,d1}\; \hat{b}_{\sss D1}
  + S_{d2,d2}\; \hat{b}^\dagger_{\sss D2}\Big]_{\omega=\omega_{d2|{\rm out}}(k_d)}
  \nonumber \\
  &+&\widetilde{\cal W}^*_{d1|{\rm in}}(k_d) \; 
  \hat{b}^\dagger_{\sss D1}(\omega_{d1|{\rm in}}(-k_d))
 \nonumber \\
  &+&\widetilde{\cal U}_{d2|{\rm in}}(k_d) \; 
\hat{b}^\dagger_{\sss D2}(\omega_{d2|{\rm in}}(k_d))
  \; ,\end{eqnarray}

\begin{eqnarray}\label{q9bis}
  \hat{\psi}_d(k_d<0)& = & 
  \widetilde{\cal W}_{d1|{\rm out}}^*(k_d)
  \Big[S_{d1,u}^*\; \hat{b}_{\sss U}^\dagger +
  S_{d1,d1}^*\; \hat{b}_{\sss D1}^\dagger
  + S_{d1,d2}^*\; \hat{b}_{\sss D2}\Big]_{\omega=\omega_{d1|{\rm out}}(-k_d)}
  \nonumber \\ 
  &+&
  \widetilde{\cal W}_{d2|{\rm out}}^*(k_d)
  \Big[S_{d2,u}^*\; \hat{b}_{\sss U}^\dagger+
  S_{d2,d1}^*\; \hat{b}_{\sss D1}^\dagger
  + S_{d2,d2}^*\; \hat{b}_{\sss D2}\Big]_{\omega=\omega_{d2|{\rm out}}(-k_d)}
  \nonumber \\
  &+&\widetilde{\cal U}_{d1|{\rm in}}(k_d) \; 
  \hat{b}_{\sss D1}(\omega_{d1|{\rm in}}(k_d))
  \nonumber \\
  &+&\widetilde{\cal W}_{d2|{\rm in}}^*(k_d) \; 
  \hat{b}_{\sss D2}(\omega_{d2|{\rm in}}(-k_d))
  \; .\end{eqnarray}
All the terms in the expressions \eqref{q8}, \eqref{q8bis},
\eqref{q9} and \eqref{q9bis} depend on $k_u$ ($k_d$) either directly, either 
{\it via}
$\omega_\ell(\pm k_u)$ ($\omega_m(\pm k_d)$) where $\ell \in \{
u|{\rm in}, u|{\rm out} \}$ ($ m \in \{ d1|{\rm in}, d1|{\rm out},
d2|{\rm in}, d2|{\rm out} \}$).

\subsubsection{The case of the flat profile configuration}
\label{FPFourier}
In this configuration -- presented in Sec. \ref{flatprofile} -- the
fact that $K_u=K_d\equiv K_0$ makes it possible to express \eqref{q8},
\eqref{q8bis}, \eqref{q9} and \eqref{q9bis} in terms of $k$
($=k_u=k_d=K-K_0$) only. It is then possible -- and useful, see
sec. \ref{sec.mom.corr} below -- to regroup the momentum operators in
Eqs. \eqref{q8}, \eqref{q8bis}, \eqref{q9} and \eqref{q9bis} in terms
of $k<0$ and $k>0$ contributions, and to write
\begin{eqnarray}\label{q10}
\hat{\psi}(k<0)& = & 
\widetilde{\cal U}_{u|{\rm out}}(k)
\Big[S_{u,u}\; \hat{b}_{\sss U}
+ S_{u,d1}\; \hat{b}_{\sss D1}+
S_{u,d2}\; \hat{b}^\dagger_{\sss D2}\Big]_{\omega=\omega_{u|{\rm out}}(k)}
\nonumber \\ 
&+&
\widetilde{\cal W}^*_{u|{\rm in}}(k) \; 
\hat{b}^\dagger_{\sss U}(\omega_{u|{\rm in}}(-k))
\nonumber \\ 
&+& \widetilde{\cal W}_{d1|{\rm out}}^*(k)
  \Big[S_{d1,u}^*\; \hat{b}_{\sss U}^\dagger +
  S_{d1,d1}^*\; \hat{b}_{\sss D1}^\dagger
  + S_{d1,d2}^*\; \hat{b}_{\sss D2}\Big]_{\omega=\omega_{d1|{\rm out}}(-k)}
  \nonumber \\ 
  &+&
  \widetilde{\cal W}_{d2|{\rm out}}^*(k)
  \Big[S_{d2,u}^*\; \hat{b}_{\sss U}^\dagger+
  S_{d2,d1}^*\; \hat{b}_{\sss D1}^\dagger
  + S_{d2,d2}^*\; \hat{b}_{\sss D2}\Big]_{\omega=\omega_{d2|{\rm out}}(-k)}
  \nonumber \\
  &+&\widetilde{\cal U}_{d1|{\rm in}}(k) \; 
  \hat{b}_{\sss D1}(\omega_{d1|{\rm in}}(k))
 \nonumber \\
  &+&\widetilde{\cal W}_{d2|{\rm in}}^*(k) \; 
\hat{b}_{\sss D2}(\omega_{d2|{\rm in}}(-k))
  \; ,\end{eqnarray}
  
\begin{eqnarray}\label{q10bis}
  \hat{\psi}(k>0)& = & 
  \widetilde{\cal W}_{u|{\rm out}}^*(k)
  \Big[S_{u,u}^*\; \hat{b}_{\sss U}^\dagger
  + S_{u,d1}^*\; \hat{b}_{\sss D1}^\dagger+
  S_{u,d2}^*\; \hat{b}_{\sss D2}\Big]_{\omega=\omega_{u|{\rm out}}(-k)}
  \nonumber \\ 
  &+&
  \widetilde{\cal U}_{u|{\rm in}}(k) \; 
  \hat{b}_{\sss U}(\omega_{u|{\rm in}}(k))
  \nonumber \\ 
  &+&\widetilde{\cal U}_{d1|{\rm out}}(k)
  \Big[S_{d1,u}\; \hat{b}_{\sss U} +
  S_{d1,d1}\; \hat{b}_{\sss D1}
  + S_{d1,d2}\; \hat{b}^\dagger_{\sss D2}\Big]_{\omega=\omega_{d1|{\rm out}}(k)}
  \nonumber \\ 
  &+&
  \widetilde{\cal U}_{d2|{\rm out}}(k)
  \Big[S_{d2,u}\; \hat{b}_{\sss U}+
  S_{d2,d1}\; \hat{b}_{\sss D1}
  + S_{d2,d2}\; \hat{b}^\dagger_{\sss D2}\Big]_{\omega=\omega_{d2|{\rm out}}(k)}
  \nonumber \\
  &+&\widetilde{\cal W}^*_{d1|{\rm in}}(k) \; 
  \hat{b}^\dagger_{\sss D1}(\omega_{d1|{\rm in}}(-k))
  \nonumber \\
  &+&\widetilde{\cal U}_{d2|{\rm in}}(k) \; 
  \hat{b}^\dagger_{\sss D2}(\omega_{d2|{\rm in}}(k))
  \; .
\end{eqnarray}
Each of the above expressions ($k<0$ and $k>0$) contains terms coming from
both the subsonic and the supersonic regions. This procedure can induce a
problem of normalization. As discussed in Sec. \ref{correl-noBH}
after Eq. \eqref{spurious1} this problem is easily solved by using an
appropriate overall multiplicative factor which we do not include for
readability. Furthermore, the problem disappears when one considers
normalized quantity such as $g_2(K,Q)$ defined below
[Eq. \eqref{m2bis}].

\section{Momentum correlations in the presence of 
a sonic
  horizon}\label{sec.mom.corr}
The momentum-momentum correlation signal is embodied in the 
function
\begin{equation}\label{m2}
G_2(K,Q)=\langle : \! \hat N(K)\hat N(Q) \!: \rangle -\langle \hat N(K)\rangle
\langle \hat N(Q)\rangle \; ,
\end{equation}
where $\hat{N}(K)=\hat{\psi}^\dagger(K)\hat{\psi}(K)$. We also
consider in some details the normalized quantity
\begin{equation}\label{m2bis}
g_2(K,Q)=
\frac{\langle : \! \hat N(K)\hat N(Q) \!: \rangle}{\langle \hat N(K)\rangle
\langle \hat N(Q)\rangle} \; .
\end{equation}
In the definitions (\ref{m2}) and (\ref{m2bis}) the normally ordered
product eliminates the diagonal shot noise contribution. 

The computation of the two-body momentum correlation \eqref{m2}
in the generic case is quite tedious because the upstream and
downstream Fourier transform are different, as explained in
Sec. \ref{local-FT}. Besides the formulas assume different forms
depending of the values of $K$ and $Q$ relative to $K_u$ and $K_d$,
cf. Eqs. \eqref{q8}, \eqref{q8bis}, \eqref{q9} and \eqref{q9bis}. As a
result, one has to consider nine different cases. Although we will
treat all the black-hole configurations presented in Sec. \ref{q1dbh}
(plots encompassing all the different cases for the waterfall,
$\delta$ peak and flat profile configurations are given in
Figs. \ref{fig.twobody.water} and \ref{fig.twobody.delta}), to
simplify the presentation we give here the explicit results in the
``flat profile'' configuration where the background density is a
uniform plane wave (cf. Sec. \ref{flatprofile}). In this case one has to
consider only 4 cases: the four quadrants in the $(k,q)$ plane (where
$k=K-K_0$ and $q=Q-K_0$) and one can
rewrite $G_2$ in terms of $k$ and $q$
\begin{equation}\label{m15}
G_2(k,q) = \langle \hat\psi^\dagger(k)\hat\psi^\dagger(q)
\hat\psi(k)\hat\psi(q)\rangle
- \langle \hat\psi^\dagger (k)\hat\psi (k) \rangle
\langle \hat\psi^\dagger (q)\hat\psi (q) \rangle \; ,
\end{equation}
and then perform explicit computations 
using the expressions \eqref{q10} and \eqref{q10bis}. 

The theoretical evaluation of the momentum correlation function is
performed in order to match with an ex\-pe\-ri\-men\-tal detection
scheme which consists of opening the trap and letting the
e\-le\-men\-tary excitations be converted into particles expelled from
both ends of the condensate, according to a process known as ``phonon
evaporation'' \cite{Toz04}. If this process is assumed to be gentle
and adiabatic, each elementary excitation is converted into a single
particle. More precisely, one has, in this case, for the positive norm
$u$ and $d1$ modes:
\begin{equation}\label{adia-pos}
{\cal U}_\ell(\omega) \to |\partial\omega/\partial q_\ell|^{-1/2}
\quad\mbox{and}\quad  {\cal W}_\ell(\omega) \to 0\; , \quad\mbox{for}\quad 
\ell\in\{u|{\rm in}, u|{\rm out},
d1|{\rm in}, d1|{\rm out}\}\; ;
\end{equation} 
while for the negative norm $d2$ modes:
\begin{equation}\label{adia-neg}
{\cal U}_\ell(\omega) \to 0 \quad\mbox{and}\quad {\cal W}_\ell(\omega) 
\to |\partial\omega/\partial
q_\ell|^{-1/2}\; , \quad\mbox{for}\quad 
\ell\in\{d2|{\rm in}, d2|{\rm out}\}\; .
\end{equation}
With the normalization \eqref{norme}, this corresponds, for instance
for a positive norm mode, to a perfect transmutation of an elementary
excitation into a particle state which carries a unit current. For the
``tilde Bogoliubov coefficients'' \eqref{q7}, the prescription
\eqref{adia-pos} yields
\begin{equation}\label{tadia-pos}
\widetilde{\cal U}_\ell(q) \to
|\partial\omega_\ell/\partial q|^{1/2} \quad\mbox{and}\quad 
\widetilde{\cal W}_\ell(q)
\to 0\; , 
\end{equation} 
for the positive norm states ($\ell\in\{u|{\rm in}, u|{\rm out},
d1|{\rm in}, d1|{\rm out}\}$) whereas one gets
\begin{equation}\label{tadia-neg}
\widetilde{\cal U}_\ell(q) \to
0 \quad\mbox{and}\quad 
\widetilde{\cal W}_\ell(q) \to |\partial\omega_\ell/\partial
q|^{1/2}\; , 
\end{equation} 
for the negative norm modes ($\ell\in\{d2|{\rm in}, d2|{\rm out}\}$).

As demonstrated in Ref. \cite{Jas12}, after an adiabatic opening of
the trapping potential, a measure of the velocity distribution of the
emitted particles gives access to the momentum distribution within the
condensate and to the correlators defined in Eqs. \eqref{m2} and
\eqref{m2bis}. The process can be sudden, in which case the adiabatic
hypothesis breaks down. More precisely, if $t_{\rm open}$ is the
characteristic time during which the trap is opened and an elementary
excitation is converted into a real particle, the adiabatic
approximation fails for excitations of energy $\omega$ verifying:
$\omega^{-1}\gg t_{\rm open}$: for low lying excitations (those for
which $\omega\to 0$) the opening of the trap always seems abrupt
\cite{Toz04}. 
A quantitative study of this phenomenon
will be presented in a future publication \cite{prepa-adia}.

An outline of the complications introduced by non adiabatic effects is
postponed to Section (\ref{NA}), but until then we give the results
after an adiabatic expansion, in which case, owing to
\eqref{tadia-pos} and \eqref{tadia-neg}, the expression of $\hat\psi$
given in \eqref{q10}, \eqref{q10bis} reduces to
\begin{eqnarray}\label{q11}
\hat{\psi}(k<0)& = & 
\widetilde{{\cal U}}_{u|{\rm out}}\Big[S_{u,u}\; \hat{b}_{\sss U}
+ S_{u,d1}\; \hat{b}_{\sss D1}+
S_{u,d2}\; \hat{b}^\dagger_{\sss D2}\Big]_{\omega=\omega_{u|{\rm out}}(k)}
\nonumber \\ 
  &+&
\widetilde{{\cal W}}^*_{d2|{\rm out}}  \Big[S_{d2,u}^*\; \hat{b}_{\sss U}^\dagger+
  S_{d2,d1}^*\; \hat{b}_{\sss D1}^\dagger
  + S_{d2,d2}^*\; \hat{b}_{\sss D2}\Big]_{\omega=\omega_{d2|{\rm out}}(-k)}
  \nonumber \\
  &+& \; 
\widetilde{{\cal U}}_{d1|{\rm in}}  \hat{b}_{\sss D1}(\omega_{d1|{\rm in}}(k))
 +\widetilde{{\cal W}}^*_{d2|{\rm in}} 
\hat{b}_{\sss D2}(\omega_{d2|{\rm in}}(-k))
  \; ,\end{eqnarray}
and  
\begin{eqnarray}\label{q11bis}
  \hat{\psi}(k>0)& = & 
  \widetilde{{\cal U}}_{d1|{\rm out}} \Big[S_{d1,u}\; \hat{b}_{\sss U} +
  S_{d1,d1}\; \hat{b}_{\sss D1}
  + S_{d1,d2}\; \hat{b}^\dagger_{\sss D2}\Big]_{\omega=\omega_{d1|{\rm out}}(k)}
  \nonumber \\  &+& \widetilde{{\cal U}}_{u|{\rm in}}
  \hat{b}_{\sss U}(\omega_{u|{\rm in}}(k)) \ .\ \  \end{eqnarray}
In expressions \eqref{q11} and \eqref{q11bis} the ``tilde Bogoliubov 
coefficients'' which are non zero 
assume the limiting values \eqref{tadia-pos} and 
\eqref{tadia-neg}.

\subsection{Zero temperature}\label{BHT0}
Let us first consider the case where the system is initially in its
vacuum state, i.e., the vacuum of excitations. This is the zero
temperature case. We find for the one-body momentum distribution:
\begin{equation}\label{adia.1corps.zerot.neg}
\begin{split}
\langle \hat{N}(k<0)\rangle = &
\left(|\widetilde{\cal U}_{u|{\rm out}}|^2
|S_{u,d2}|^2
\right)_{\omega_{u|{\rm out}}(k)}
\times \delta\left(\omega_{u|{\rm out}}(k)-
\omega_{u|{\rm out}}(k)\right)\\
+& 
|\widetilde{\cal W}_{d2|{\rm out}}|^2
\left(
|S_{d2,u}|^2 + |S_{d2,d1}|^2\right)_{\omega_{d2|{\rm out}}(-k)}
\times \delta\left(\omega_{d2|{\rm out}}(-k)-
\omega_{d2|{\rm out}}(-k)\right)
\; 
\end{split}
\end{equation}
and
\begin{equation}\label{adia.1corps.zerot.pos}
\begin{split}
\langle \hat{N}(k>0)\rangle = &
\left(|\widetilde{\cal U}_{d1|{\rm out}}|^2
|S_{d1,d2}|^2
\right)_{\omega_{d1|{\rm out}}(k)}
\times \delta\left(\omega_{d1|{\rm out}}(k)-
\omega_{d1|{\rm out}}(k)\right)
\; .
\end{split}
\end{equation}
These relations can be cast under the form
\begin{equation}\label{adia.1corps.zerot}
\langle \hat{N}(k)\rangle = {\cal N}(k) \times \delta(k-k)\; ,
\end{equation}
where
\begin{equation}\label{adia.1corps.zerot.posneg}
{\cal N}(k)=\begin{cases}
|S_{u,d2}|^2_{\omega_{u|{\rm out}}(k)}
+
\left(
|S_{d2,u}|^2 + |S_{d2,d1}|^2\right)_{\omega_{d2|{\rm out}}(-k)}
& \mbox{for}\;\; k<0\; ,\\
|S_{d1,d2}|^2_{\omega_{d1|{\rm out}}(k)}& \mbox{for}\;\; k>0\; .
\end{cases}
\end{equation}
Note that within the adiabatic approximation, the $T=0$ momentum
signal \eqref{adia.1corps.zerot.posneg} would cancel in the absence of
horizon, since the $d2$ mode and all the corresponding elements of the
$S$-matrix would disappear in this case. This is confirmed by the
study of Sec. \ref{correl-noBH}.

In expression \eqref{adia.1corps.zerot}, the $\delta$-peak contribution is
singular: one has a $\delta(0)$ [as in \eqref{adia.1corps.zerot.neg} and 
\eqref{adia.1corps.zerot.pos}]. This is due to the schematic nature
of the rules R\ref{R1} and R\ref{R2} defined in section
\ref{local-FT} for the Fourier transform. The more rigorous local
Fourier transform in terms of a window function --explained in the
first part of Sec. \ref{local-FT}-- yields a finite contribution, as
shown in Appendix \ref{window-Fourier}: see, e.g.,
Eqs. \eqref{adia.1corps.zerot.neg.FT} and
\eqref{adia.1corps.zerot.pos.FT} which are the rigorous versions of
Eqs.  \eqref{adia.1corps.zerot.neg} and \eqref{adia.1corps.zerot.pos}.

Our main interest in this work is the study of the correlation signal
in momentum space, that is of the two-points correlation functions
$G_2$ and $g_2$ defined in Eqs. \eqref{m2} and \eqref{m2bis}. The most
robust signals are those present even at $T=0$. These are
\begin{equation}\label{rescorr4}
\begin{split}
G_2(k<0,q<0)= 
\left[
|S_{u,d2}|^2_{\omega_{u|{\rm out}}(k)}+\left.\left(|S_{d2,u}|^2+|S_{d2,d1}|^2 
\right)\right|_{\omega_{d2|{\rm out}}(-k)}
 \right]^{2}  \delta^2(k-q) \\
+\Big[ |\widetilde{{\cal U}}_{u|{\rm out}}|^2
|\widetilde{{\cal W}}_{d2|{\rm out}}|^2
\big|S_{u,d2}^*S_{d2,d2}\big|^2 
\delta^2(\omega_{u|{\rm out}}(k)-\omega_{d2|{\rm out}}(-q))+ 
(k\leftrightarrow q) \Big],
\end{split}
\end{equation}
\begin{equation}\label{rescorr5}
G_2(k>0,q>0)=  |S_{d1,d2}|^2_{\omega_{d1|{\rm out}}(k)} \delta^2(k-q) \ ,   
\end{equation}
\begin{equation}\label{rescorr6}
\begin{split}
G_2(k<0,q>0)= &
|\widetilde{{\cal U}}_{u|{\rm out}}|^2
|\widetilde{{\cal U}}_{d1|{\rm out}}|^2
\Big|S_{u,d2}^*S_{d1,d2}\Big|^2
\delta^2(\omega_{u|{\rm out}}(k)-\omega_{d1|{\rm out}}(q)) \\
+ &|\widetilde{{\cal W}}_{d2|{\rm out}}|^2
|\widetilde{{\cal U}}_{d1|{\rm out}}|^2
\Big|S_{d2,d2}^*S_{d1,d2}\Big|^2
\delta^2(\omega_{d2|{\rm out}}(-k)-\omega_{d1|{\rm out}}(q)) .
\end{split}
\end{equation}
As already noted for the one-body signal, 
a first obvious outcome of this computation is that the $T=0$
correlations in momentum space 
disappear in the absence of sonic horizon, since in this case
the $d2$ mode does not exist and all the corresponding elements of the
$S$ matrix cancel in \eqref{rescorr4}, \eqref{rescorr5} and \eqref{rescorr6}.

Looking more in detail into the results, one sees that in the ($k<0$,
$q<0$) and in the ($k>0$, $q>0$) sectors one has terms with a
$\delta^2(k-q)$ correlation which we henceforth denote as
diagonal. These terms are simply of the form ${\cal N}^2(k)
\delta^2(k-q)$. Again, the occurrence of the highly singular squared
$\delta$ distribution in
the expressions \eqref{rescorr4}, \eqref{rescorr5} and
\eqref{rescorr6} is an artifact of our schematic Fourier
transform rules R\ref{R1} and R\ref{R2} which disappears when one
considers windowed Fourier transforms, see Appendix
\ref{window-Fourier}.

Besides the diagonal term, one has correlation lines along the $u|{\rm
  out}- d2|{\rm out}$ (Hawking quanta-partner), $d2|{\rm out}-d1|{\rm
  out}$ and $u|{\rm out}-d1|{\rm out}$ channels. All these
correlations are also present {\it mutatis mutandis} in the density
fluctuation sector \cite{Bal08,Car08,Rec09,Lar12}. As in the
interpretation of black-hole radiation first given by Hawking
\cite{Hawking}, the existence of these correlations is an indication
of the fact that the vacuum of the ingoing modes is not the same as
the vacuum of outgoing ones, and this results in spontaneous emission
of pairs of outgoing quasi-particles even in the absence of incoming
ones \cite{Fabbri:2005mw,Rob12}. This is possible even in our
stationary setting because the number of created quasiparticles of
positive energy equals the number negative energy ones, i.e. energy is
conserved \cite{Lar12,Balbinot:2012xw}.

\begin{figure}
\begin{center}
\includegraphics*[width=7cm]{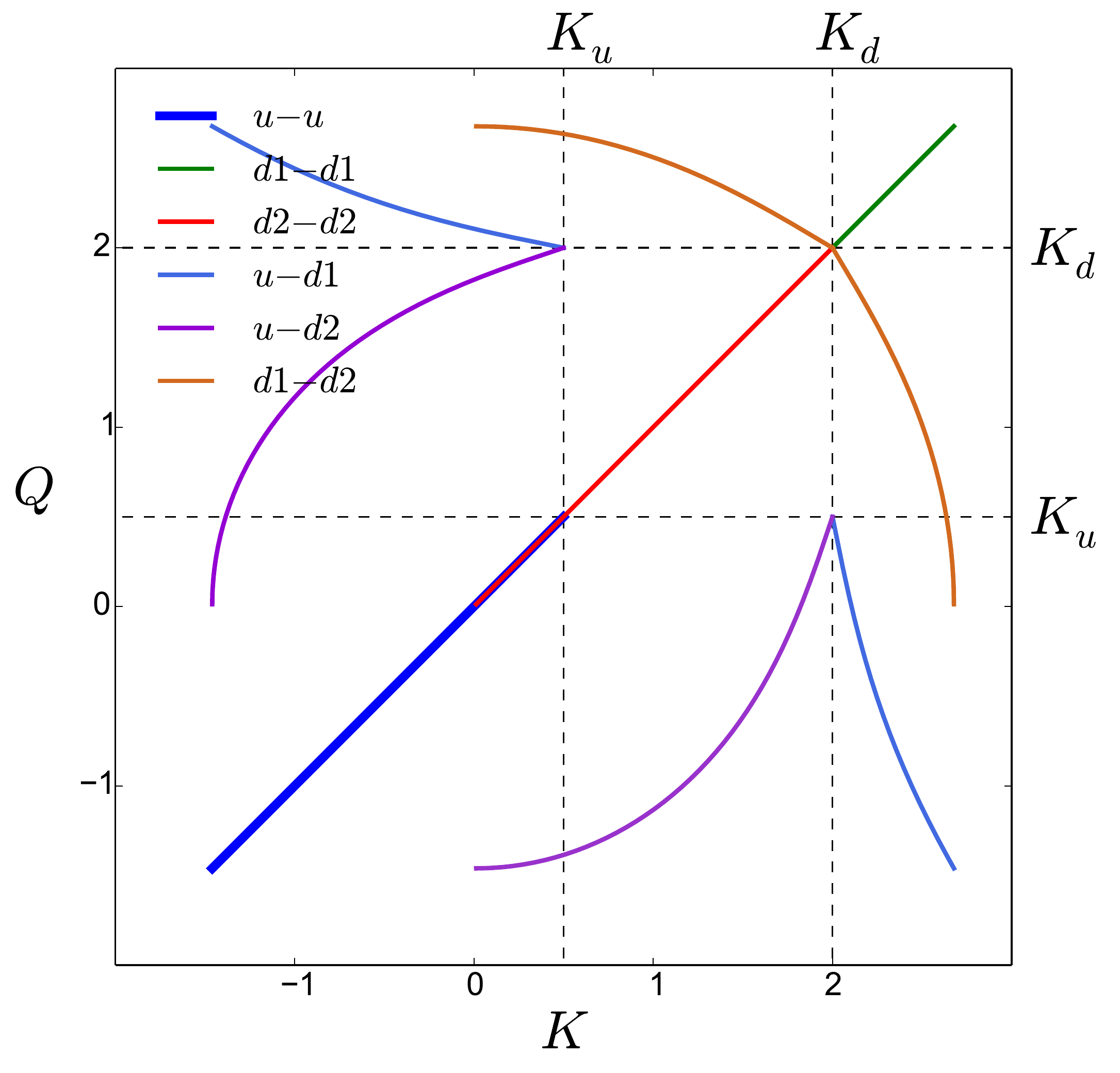}
\includegraphics*[width=7cm]{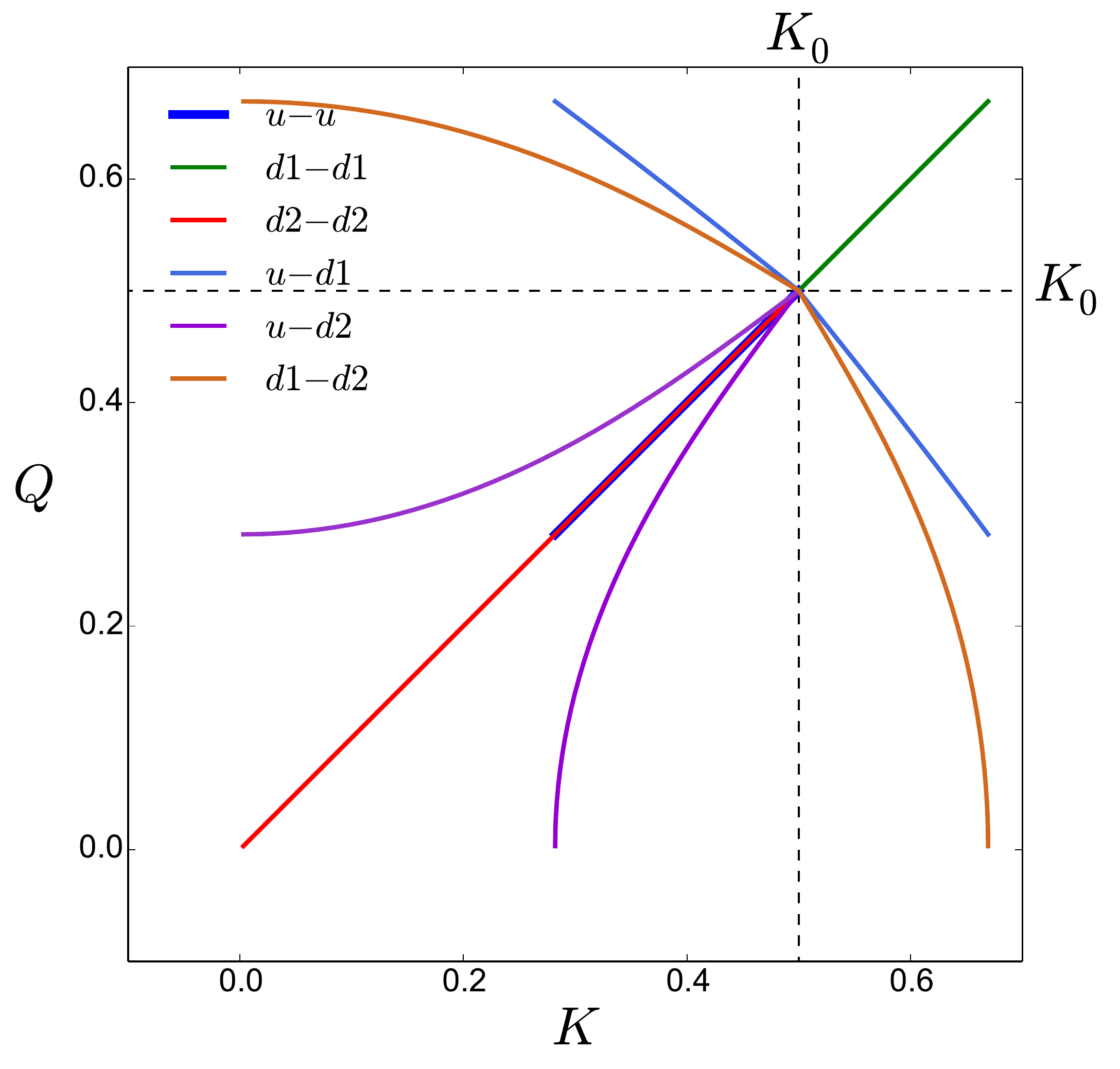}
\caption{Zero temperature momentum space correlation lines from
  Eqs. \eqref{rescorr4}, \eqref{rescorr5} and \eqref{rescorr6}. These
  curves correspond to the loci of points with finite value of the two
  body momentum correlation function $G_2(K,Q)$. They are labeled with
  the names of the modes of correlated momenta, for instance the
  ``$u-d1$'' curve corresponds to the line of correlation between the
  $u|{\rm out}$ and the $d1|{\rm out}$ modes. The left plot displays
  the results for a waterfall configuration with $M_u=0.5$ and
  $M_d=4$. The right plot displays the results for a flat profile
  configuration with the same values of $M_u$ and $M_d$. The momenta
  are expressed in units of $\xi_u^{-1}$.}
\label{fig.twobody.water}
\end{center}
\end{figure}

\begin{figure}
\begin{center}
\includegraphics*[width=7cm]{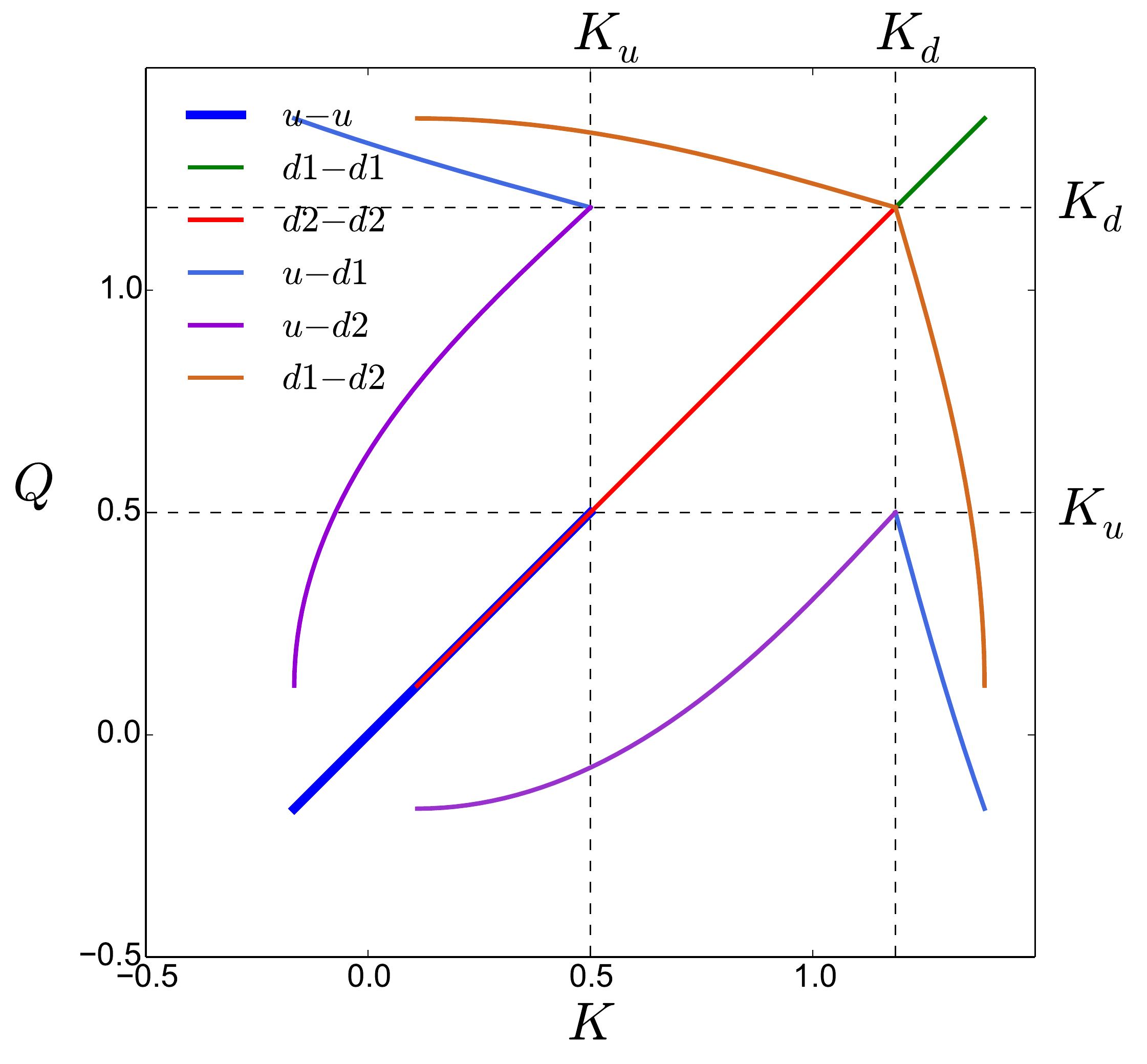}
\includegraphics*[width=7cm]{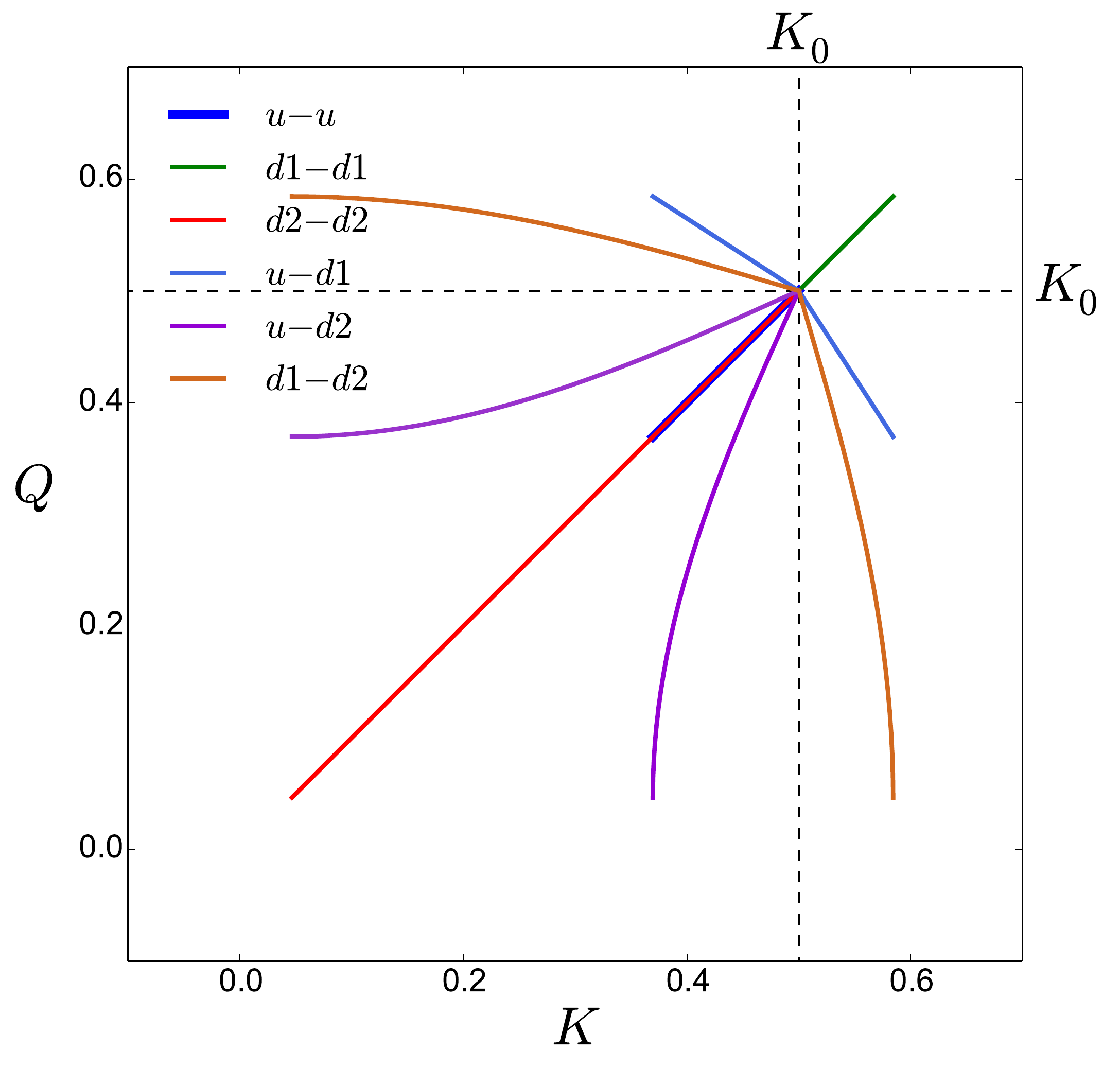}
\caption{Same as Fig. \ref{fig.twobody.water} for a $\delta$ peak
  configuration (left plot) and flat profile (right plot) having both
  the same asymptotic Mach numbers $M_u=0.5$ and
  $M_d=1.827$.}\label{fig.twobody.delta}
\end{center}
\end{figure}

The results corresponding to the zero temperature correlation signals
\eqref{rescorr4}, \eqref{rescorr5} and \eqref{rescorr6} are displayed
in Figs.  \ref{fig.twobody.water} and \ref{fig.twobody.delta}.  In
figure \ref{fig.twobody.water}, the left plot presents the results for
a waterfall configuration with $M_u=0.5$; this imposes $M_d=4$. The
results for the flat profile configuration with the same values of
$M_u$ and $M_d$ are displayed in the right plot. As noted at the end
of Sec. \ref{flatprofile}, it is not possible to realize a waterfall
and a $\delta$ peak profile having the same values of $M_u$ and $M_d$.
Hence we compare in figure \ref{fig.twobody.delta}
the results for a $\delta$ peak configuration (left plot) with
$M_u=0.5$ (this imposes $M_d=1.827$) with the ones for a flat profile
configuration having the same asymptotic Mach numbers (right plot).

The graphical rule for drawing the correlation lines in these figures
is simple. Let's consider the $u|{\rm out}- d2|{\rm out}$ correlation
line for instance. From Eq. \eqref{rescorr4} one sees that the
relative wave vectors $k$ and $q$ are correlated when $\omega_{u|{\rm
    out}}(k)=\omega_{d2|{\rm out}}(-q)$, and also along the curve
obtained by exchanging the roles of $k$ and $q$. This corresponds to
the two curves $q=- q_{d2|{\rm out}} (\omega_{u|{\rm out}} (k))$ and
$q=q_{u|{\rm out}} (\omega_{d2|{\rm out}} (-k))$. Going to the
absolute wave vectors $K$ and $Q$, these correlation lines correspond
to the curves $Q-K_d=- q_{d2|{\rm out}} (\omega_{u|{\rm out}}
(K-K_u))$ and $Q-K_u=q_{u|{\rm out}} (\omega_{d2|{\rm out}}
(-K+K_d))$. These two curves are symmetrical with respect to the
diagonal.  In the case of the flat profile configuration they meet on
the diagonal at the point of coordinate $(K_u,K_d=K_u)$.

Figures \ref{fig.twobody.water} and \ref{fig.twobody.delta} indicate
the location of the relevant correlation signal, but not its
amplitude. The theoretical evaluation of this amplitude can be done
either using the schematic rules R\ref{R1} and R\ref{R2} [but then
yields to singular expressions as in \eqref{rescorr4},
\eqref{rescorr5} and \eqref{rescorr6}] either using windowed Fourier
transforms (but then depends on the specific choice of the windows,
cf. Appendix \ref{window-Fourier}). One can circumvent these
difficulties and obtain a non-ambiguous result by working with the
rescaled $g_2$ function \eqref{m2bis}, as explained now.

Let us first consider ``diagonal correlation terms'' which are
intra-channel correlation in the $u|{\rm out}$, $d1|{\rm out}$ and
$d2|{\rm out}$ channels corresponding to the diagonal lines in
Figs. \ref{fig.twobody.water} and \ref{fig.twobody.delta}. These
diagonal terms are conveniently isolated by studying correlation
functions of the type
\begin{equation}\label{resc-g1}
g_2(K,K)_{u|{\rm out}}=
\frac{G_2(K,K)_{u|{\rm out}}}{\langle\hat{N}(K)\rangle^2_{u|{\rm out}}}+1
\; ,
\end{equation}
where $\langle\hat{N}(K)\rangle_{u|{\rm out}}$ and $G_2(K,K)_{u|{\rm
    out}}$ are the $u|$out contributions to $\langle\hat{N}(K)\rangle$
and to the diagonal part of $G_2(K,K)$. One obtains straightforwardly
\begin{equation}\label{resc-g2}
g_2(K,K)_{u|{\rm out}}= 2
\; ,
\end{equation}
a result which follows from Wick's theorem and which is also valid for
the other channels, even at finite temperature, for non-adiabatic opening
of the trap, and also when considering Fourier transforms less
schematic than in Eqs. \eqref{q5} and \eqref{q6} (cf. Appendix
\ref{window-Fourier}).

At zero temperature, in the adiabatic regime we consider in the present
subsection, the only non-diagonal contributions to $G_2$ are of the
type $u|{\rm out}$--$d2|{\rm out}$, $u|{\rm out}$--$d1|{\rm out}$ and
$d1|{\rm out}$--$d2|{\rm out}$.  One considers here for instance
correlation functions of the type (with obvious notations)
\begin{equation}\label{resc-g3}
g_2(K,Q)_{u|{\rm out}-d2|{\rm out}}=
\frac{G_2(K,Q)_{u|{\rm out}-d2|{\rm out}}}
{\langle\hat{N}(K)\rangle_{u|{\rm out}}\langle\hat{N}(Q)\rangle_{d2|{\rm out}}}
+1 \; .
\end{equation}
From expressions \eqref{adia.1corps.zerot.neg} 
and \eqref{rescorr4} one gets
\begin{equation}\label{resc-g4}
g_2(K,Q)_{u|{\rm out}-d2|{\rm out}}=
\frac{|S_{d2,d2}|^2}{|S_{d2,u}|^2+|S_{d2,d1}|^2}+1
=\frac{2|S_{d2,d2}|^2-1}{|S_{d2,d2}|^2-1}\; .
\end{equation}
For obtaining the r.h.s. of Eq. \eqref{resc-g4} we used the
pseudo-unitarity condition \eqref{e19} and also the fact that
\begin{equation}\label{resc-g5}
\frac{\delta^2(\omega_{u|{\rm out}}(k)-\omega_{d2|{\rm out}}(-q))}
{\delta\left(\omega_{u|{\rm out}}(k)-\omega_{u|{\rm out}}(k)\right)
\delta\left(\omega_{d2|{\rm out}}(-k)-
\omega_{d2|{\rm out}}(-k)\right)}=1\; .
\end{equation}
While this relation is easily verified in the schematic framework used
in the main text (which originates from the rules R\ref{R1} and
R\ref{R2} of Sec. \ref{local-FT}), it appears less straightforward
when studying the more rigorously defined local Fourier transform (see
Appendix \ref{window-Fourier}). In this case, the equivalent of
Eq. \eqref{resc-g5} is Eq. \eqref{wf18} and is equal to unity only if
Eqs. \eqref{wf20} are verified, i.e., if the window
functions fulfill specific conditions, the physical content of which
is discussed in Appendix \ref{window-Fourier}. We will assume that
these conditions are met in the following (or equivalently we keep on
using the schematic rules R\ref{R1} and R\ref{R2} of
Sec. \ref{local-FT}). For the other inter-channel correlators the
normalized two-body functions read:
\begin{equation}\label{resc-g6}
g_2(K,Q)_{u|{\rm out}-d1|{\rm out}}=2\; ,
\end{equation}
and
\begin{equation}\label{resc-g7}
g_2(K,Q)_{d1|{\rm out}-d2|{\rm out}}=\frac{2|S_{d2,d2}|^2-1}{|S_{d2,d2}|^2-1}\; .
\end{equation}
The study of $g_2$ is of interest because the occurrence of
entanglement and the quantum nature of the Hawking process can be
tested through the violation of the Cauchy-Schwarz inequality, as
recently studied in a similar context in
Refs. \cite{Khe12,Bru13,Bus14a,Bus14b,Was14,Nov14,Fin14,Wei16,Cou17}. 
For instance, the Cauchy- Schwarz inequality is violated
along the characteristic Hawking quanta–partner correlation lines
$u|{\rm out}$--$d2|{\rm out}$ of Figs. \ref{fig.twobody.water} or
\ref{fig.twobody.delta} if (see, e.g., \cite{WallsMilburn})
\begin{equation}\label{CS}
g_2(K,Q)\Big|_{u|{\rm out}-d2|{\rm out}}
>
\, \sqrt{g_2(K,K)\Big|_{u|{\rm out}}\!\!\times
\, g_2(Q,Q)\Big|_{d2|{\rm out}} } = 2\; .
\end{equation}
As already noticed after Eq. \eqref{resc-g2}, the right-hand side of
inequality \eqref{CS} is equal to 2 for all temperature. From
expression \eqref{resc-g4} one sees that the Cauchy-Schwarz inequality
is violated at $T=0$ along the $u|{\rm out}$--$d2|{\rm out}$
correlation line for those values of $K$ and $Q$ such that, at energy
$\omega=\omega_{u|{\rm out}}(k)=\omega_{d2|{\rm out}}(-q)$, one has
$|S_{d2,d2}(\omega)|>1$. $S_{d2,d2}$ is the scattering amplitude from
the $d2|{\rm in}$ mode towards the $d2|{\rm out}$ mode; its modulus
can be larger than unity without violating the pseudo-unitarity
condition \eqref{e19}. $|S_{d2,d2}(\omega)|$ diverges as
$\omega^{-1/2}$ when $\omega\to 0$ \cite{Lar12} and the Cauchy-Schwarz
inequality is thus always violated for $\omega > 0$.
The same holds true for the violation of
the Cauchy-Schwarz inequality along the $d1|{\rm out}$--$d2|{\rm out}$
channels at $T=0$ since the formula \eqref{resc-g7} yields for
$g_2(K,Q)_{d1|{\rm out}-d2|{\rm out}}$ the same result than
\eqref{resc-g4} does for $g_2(K,Q)_{u|{\rm out}-d2|{\rm out}}$. From
Eq. \eqref{resc-g6} it is clear that the Cauchy-Schwarz inequality is
not violated at $T=0$ along the $u|{\rm out}$--$d1|{\rm out}$ correlation
channel.

A remark should be made here concerning the experimental detection
process. Our theoretical analysis corresponds to windowed upstream and
downstream momentum detection. It is thus theoretically possible to
distinguish an upstream and a downstream component in the signal in
momentum space. This is the reason why it is legitimate to identify,
for instance a $u|{\rm out}-d2|{\rm out}$ component in the total
$G_2(K,Q)$, or a $u|{\rm out}$ component in the total $\langle
\hat{N}(K)\rangle$, as done in Eq. \eqref{resc-g3}. However, some
apparatuses may mix the upstream and downstream signals in their
detection scheme. In this case, the overlap in momentum space of the
domain of existence of the $u|{\rm out}$ and $d2|{\rm out}$ signals
forbids the use of a definition as precise as \eqref{resc-g3}. In this
case, a clear separation can only be done for the momenta located
outside of the overlap region. In the waterfall configuration
represented in the left plot of Fig. \ref{fig.twobody.water} for
instance, with such a detection apparatus, one cannot study the
$u|{\rm out}-d2|{\rm out}$ correlation signal in regions where the
$d2|{\rm out}$ and $u|{\rm out}$ momenta overlap, i.e, when $K$ or $Q$
lie in a segment for which the blue and red diagonal curves of the figure
overlap. In this respect, such an apparatus would not have access to
the $u|{\rm out}-d2|{\rm out}$ correlation signal at all for the flat
profile configuration, since in this case the region of momenta issued
from the $u|{\rm out}$ channel is completely contained in the region
of momenta issued from the $d2|{\rm out}$ channel. To recall this
possible experimental issue, when we plot below quantities such as
$g_2(K,Q)\Big|_{u|{\rm out}-d2|{\rm out}}$ or $g_2(K,Q)\Big|_{u|{\rm
    out}-d1|{\rm out}}$, we shade the region of momenta where an
ambiguity is possible (cf. Figs. \ref{fig.CS.waterfall} and
\ref{fig.CS.delta}). We stress however that not all experimental
detection schemes have to be subject to this drawback.

\subsection{Finite temperature}\label{finiteT}

The analog stationary black hole configuration we consider is
thermodynamically unstable, and cannot support a thermal
state. However, a thermal-like occupation of the states can be defined
by considering a time dependent process of formation of the horizon 
starting from a uniform
thermal subsonic uniform configuration, following the procedure
already considered in Ref. \cite{Rec09} (see also \cite{Mac09}): The
condensate has initially a uniform density $n_u$, a flow velocity
$V_u$ and a sound velocity $c_u$ and is at thermal equilibrium at
temperature $T$ in the moving frame. The horizon is then adiabatically
switched on, either by ramping down the scattering length in the
downstream region -- leading to a flat profile configuration --, or by
ramping up an external potential -- leading to a waterfall or a delta
peak configuration.

Then, one obtains occupation numbers $n_{\sss U}(\omega)$, $n_{\sss D1}(\omega)$
and $n_{\sss D2}(\omega)$ for each of the scattering modes. For instance
$n_{\sss U}(\omega)=n_{\rm th}[\omega_{\rm\sss B}(q_{u|{\rm in}}(\omega))]$, where
$n_{\rm th}(\Omega)=(\exp(\Omega/T)-1)^{-1}$ is the thermal Bose
occupation factor, and $\omega_{\rm\sss B}(q)$ is the Bogoliubov
dispersion relation \eqref{1d7} in the moving frame (with here $c\to
c_u$ and $\xi\to \xi_u$ in expression \eqref{1d7}). This
procedure leads for the other occupation numbers
$n_{\sss D1}(\omega)=n_{\rm th}[\omega_{\rm\sss B}(q_{d1|{\rm in}}(\omega))]$
and $n_{\sss D2}(\omega)=n_{\rm th}[\omega_{\rm\sss B}(q_{d2|{\rm
    in}}(\omega))]$, it corresponds to the experimental situation where
an external potential is swept at constant velocity through a
condensate initially at rest \cite{Ste16,defect}.

Note that one could choose other prescriptions for defining the
occupation of the scattering modes. For instance the initial state
(uniform density $n_u$ with uniform velocity $V_u$) could be in
thermal equilibrium in the frame of the obstacle (and not in the frame
where the condensate is at rest, as considered above); this would
modify the explicit expression of the $n_L$'s in \eqref{m9}. Precisely
one would have in this case: $n_{\sss U} (\omega) = n_{\rm
  th}(\omega)$, $n_{\sss D1}(\omega) = n_{\rm th} (\omega_{u|{\rm
    out}} (q_{d1|{\rm in}}(\omega)))$ and $n_{\sss D1} (\omega) =
n_{\rm th}(\omega_{u|{\rm out}} (-q_{d2|{\rm in}}(\omega)))$.  In the
following we just use the contraction rules
\begin{eqnarray}\label{m9}
\langle \hat b_{\sss L}(\omega) \hat b^\dagger_{\sss L'}(\omega')\rangle 
& =&  
[1+n_{\sss L}(\omega)]\, \delta_{\sss L,L'}\,
\delta(\omega-\omega')\; ,\nonumber \\
\langle \hat b^\dagger_{\sss L}(\omega) \hat b_{\sss L'}(\omega')\rangle 
& = & 
n_{\sss L}(\omega)\, \delta_{\sss L,L'}\,  \delta(\omega-\omega')\; ,
\end{eqnarray}
without specifying the expressions of the $n_L$'s, and
the formulas written below are thus generally valid.
 
\begin{figure}[h!]
\begin{center}
\includegraphics*[width=0.99\linewidth]{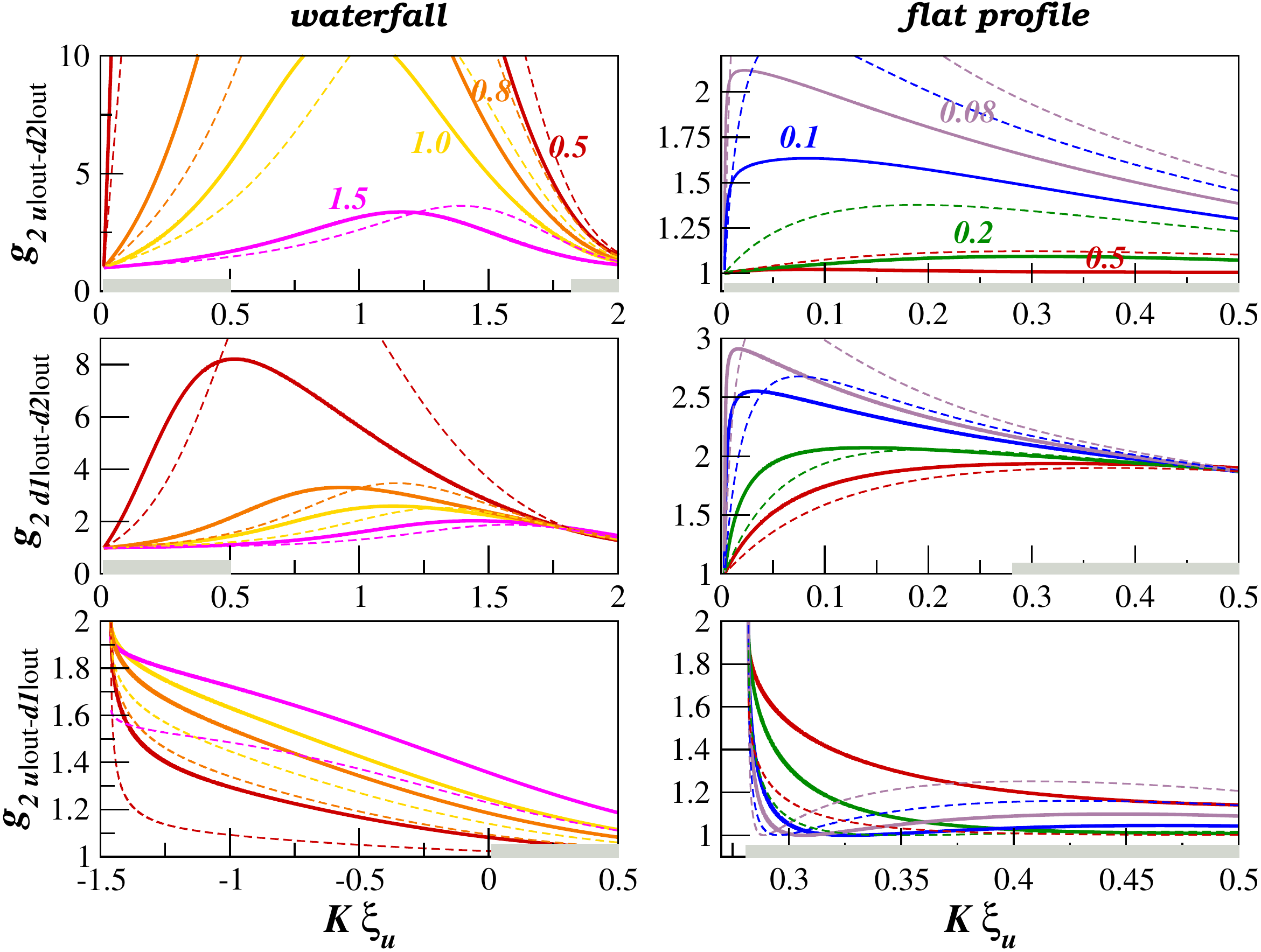}
\caption{Normalized correlation functions in the waterfall
  configuration (left plots) and flat profile (right plots). $M_u=0.5$
  and $M_d=4$ in both cases. The vertical arrangement of the different
  plots corresponds to the intensity of the correlation signal along
  the $u|{\rm out}-d2|{\rm out}$, $d1|{\rm out}-d2|{\rm out}$ and
  $u|{\rm out}-d1|{\rm out}$ correlation lines identified in
  Fig. \ref{fig.twobody.water}. The Cauchy-Schwarz inequality is
  violated when $g_2$ is larger than 2.  In each plot the different
  lines correspond to different temperatures, the values of which are
  indicated in units of $m c^2_u$. The thick solid (thin dashed) lines
  correspond to a situation where the system is at thermal equilibrium
  in the frame of the condensate (of the obstacle) before the
  formation of the sonic horizon. The shaded zone on the abscissa axis
  represent the region of overlap of the momenta of the outgoing
  channels, see the discussion at the end of
  Sec. \ref{BHT0}.}\label{fig.CS.waterfall}
\end{center}
\end{figure}

\begin{figure}[h]
\begin{center}
\includegraphics*[width=0.99\linewidth]{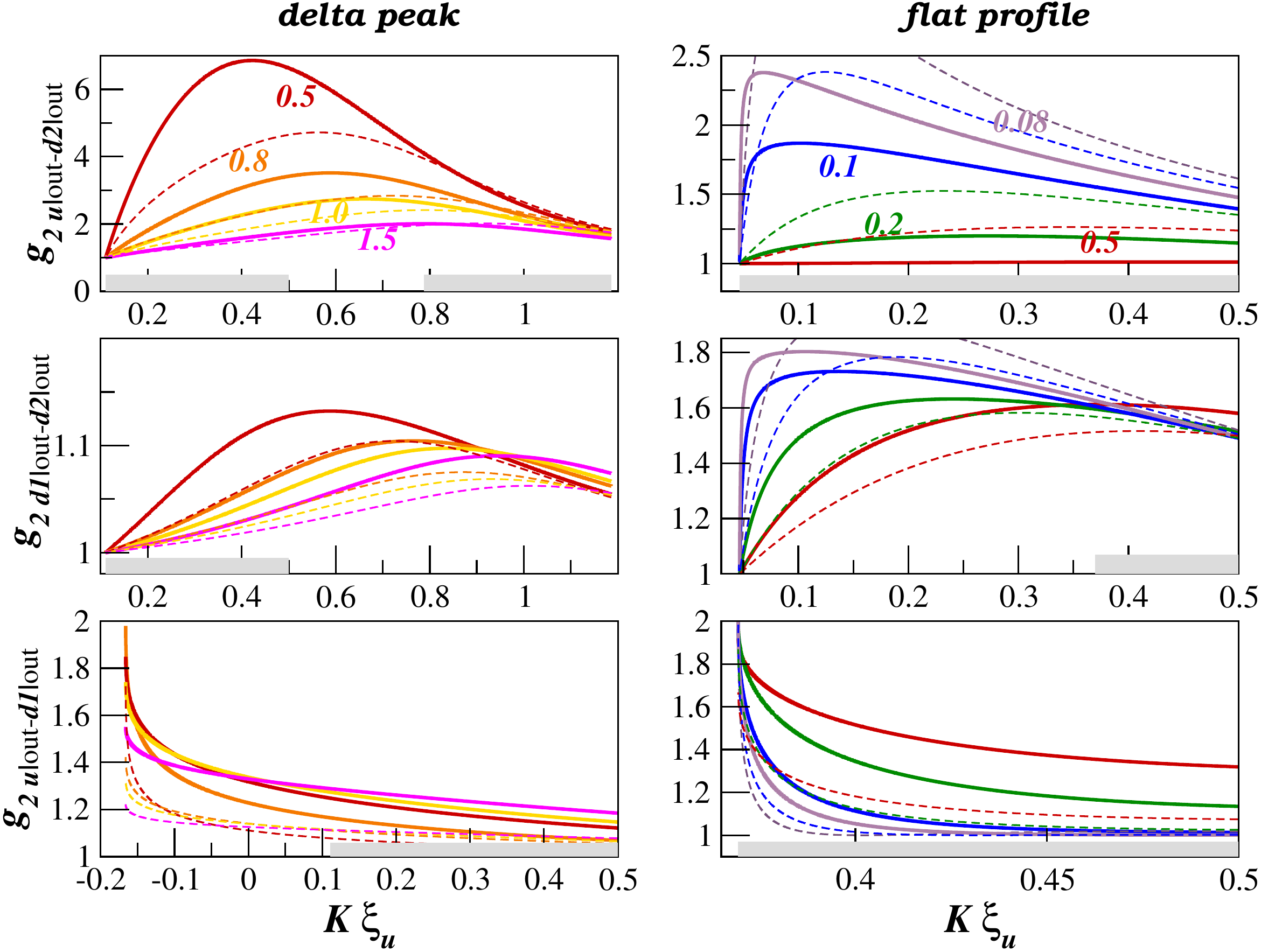}
\caption{Same as Fig. \ref{fig.CS.delta} for a $\delta$-peak
  configuration (left plot) and a flat profile one (right
  plot). $M_u=0.5$ and $M_d=1.827$ in both cases.}\label{fig.CS.delta}
\end{center}
\end{figure}

In the adiabatic regime where Eqs. \eqref{tadia-pos} and
\eqref{tadia-neg} hold, we find, for negative $k$:
\begin{equation}\label{nkn-adia}
\begin{split}
\langle \hat{N}(k<0)\rangle = &\;
|\widetilde{\cal U}_{u|{\rm out}}|^2
\left(
|S_{u,u}|^2 n_{\sss U} + |S_{u,d1}|^2 n_{\sss D1} + |S_{u,d2}|^2 (1+n_{\sss D2})
\right)_{\omega_{u|{\rm out}}(k)}\\
& \times \delta\left(\omega_{u|{\rm out}}(k)-
\omega_{u|{\rm out}}(k)\right)\\
& +
|\widetilde{\cal W}_{d2|{\rm out}}|^2
\left(
|S_{d2,u}|^2 (1+n_{\sss U}) + |S_{d2,d1}|^2 (1+n_{\sss D1}) + |S_{u,d2}|^2 n_{\sss D2}
\right)_{\omega_{d2|{\rm out}}(-k)}\\
& \times \delta\left(\omega_{d2|{\rm out}}(-k)-
\omega_{d2|{\rm out}}(-k)\right)\\
& +|\widetilde{\cal U}_{d1|{\rm in}}|^2
n_{\sss D1}(\omega_{d1|{\rm in}}(k)) \times 
 \delta\left(\omega_{d1|{\rm in}}(k)-\omega_{d1|{\rm in}}(k)\right)\\
&
+|\widetilde{\cal W}_{d2|{\rm in}}|^2
n_{\sss D2}(\omega_{d2|{\rm in}}(-k)) \times 
 \delta\left(\omega_{d2|{\rm in}}(-k)-\omega_{d2|{\rm in}}(-k)\right)\; ,
\end{split}
\end{equation}
and, for positive $k$:
\begin{equation}\label{nkp-adia}
\begin{split}
\langle \hat{N}(k>0)\rangle = & \; |\widetilde{\cal U}_{d1|{\rm out}}|^2
\left(
|S_{d1,u}|^2 n_{\sss U} + |S_{d1,d1}|^2 n_{\sss D1} + |S_{d1,d2}|^2 (1+n_{\sss D2})
\right)_{\omega_{d1|{\rm out}}(k)}\\
& \times \delta\left(\omega_{d1|{\rm out}}(k)-
\omega_{d1|{\rm out}}(k)\right)\\
& +|\widetilde{\cal U}_{u|{\rm in}}|^2
n_{\sss U}(\omega_{u|{\rm in}}(k)) \times 
 \delta\left(\omega_{u|{\rm in}}(k)-\omega_{u|{\rm in}}(k)\right)
.
\end{split}
\end{equation}
As in the zero temperature case, formulas \eqref{nkn-adia} and
\eqref{nkp-adia} can be cast under the form \eqref{adia.1corps.zerot},
with here
\begin{equation}\label{ncurln}
\begin{split}
{\cal N}(k<0) = &
\left(
|S_{u,u}|^2 n_{\sss U} + |S_{u,d1}|^2 n_{\sss D1} + |S_{u,d2}|^2 (1+n_{\sss D2})
\right)_{\omega_{u|{\rm out}}(k)} \\
& + 
\left(
|S_{d2,u}|^2 (1+n_{\sss U}) + |S_{d2,d1}|^2 (1+n_{\sss D1}) + 
|S_{d2,d2}|^2 n_{\sss D2}
\right)_{\omega_{d2|{\rm out}}(-k)}\\
&
+
n_{\sss D1}(\omega_{d1|{\rm in}}(k)) 
+n_{\sss D2}(\omega_{d2|{\rm in}}(-k))\; .
\end{split}
\end{equation}
and
\begin{equation}\label{ncurlp}
\begin{split}
{\cal N}(k>0) =
& \left(
|S_{d1,u}|^2 n_{\sss U} + |S_{d1,d1}|^2 n_{\sss D1} + |S_{d1,d2}|^2 (1+n_{\sss D2})
\right)_{\omega_{d1|{\rm out}}(k)}\\
& + n_{\sss U}(\omega_{u|{\rm in}}(k)) \; .
\end{split}
\end{equation}
Then, one obtains for the momentum correlation:
\begin{eqnarray}
& & G_2(k<0,q<0) = {\cal N}^2(k<0)\, \delta^2(k-q) \nonumber \\
&+&\Big[ 
|\widetilde{{\cal U}}_{u|{\rm out}}|^2
|\widetilde{{\cal W}}_{d2|{\rm out}}|^2
\big| S_{u,u}^*S_{d2,u}n_{\sss U}+S_{u,d1}^*S_{d2,d1}n_{\sss D1}+
S_{u,d2}^*S_{d2,d2}(1+n_{\sss D2})\big|^2 \times \nonumber \\
&& \delta^2(\omega_{u|{\rm out}}(k)-\omega_{d2|{\rm out}}(-q)) \nonumber \\ 
&+& 
|\widetilde{{\cal U}}_{u|{\rm out}}|^2
|\widetilde{{\cal U}}_{d1|{\rm in}}|^2
|S_{u,d1}|^2n_{\sss D1}^2 
\delta^2(\omega_{u|{\rm out}}(k)-\omega_{d1|{\rm in}}(q)) \nonumber \\  
&+&
|\widetilde{{\cal U}}_{u|{\rm out}}|^2
|\widetilde{{\cal W}}_{d2|{\rm in}}|^2
|S_{u,d2}|^2n_{\sss D2}(1+n_{D2}) 
\delta^2(\omega_{u|{\rm out}}(k)-\omega_{d2|{\rm in}}(-q)) \nonumber \\ 
&+ & 
|\widetilde{{\cal W}}_{d2|{\rm out}}|^2
|\widetilde{{\cal U}}_{d1|{\rm in}}|^2
|S_{d2,d1}|^2n_{\sss D1}(1+n_{\sss D1}) 
\delta^2(\omega_{d2|{\rm out}}(-k)-\omega_{d1|{\rm in}}(q)) \nonumber \\ 
&+ & 
|\widetilde{{\cal W}}_{d2|{\rm out}}|^2
|\widetilde{{\cal W}}_{d2|{\rm in}}|^2
|S_{d2,d2}|^2n_{\sss D2}^2 
\delta^2(\omega_{d2|{\rm out}}(-k)-\omega_{d2|{\rm in}}(-q)) \nonumber \\ 
&+ & (k\leftrightarrow q) \Big]\; ,
\label{rescorr1}\end{eqnarray}
\begin{eqnarray}  
&&G_2(k>0,q>0)= {\cal N}^2(k>0)\, \delta^2(k-q) \nonumber \\  
&+& \left[  |\widetilde{{\cal U}}_{u|{\rm in}}|^2
|\widetilde{{\cal U}}_{d1|{\rm out}}|^2
|S_{d1,u}|^2n_{\sss U}^2 \; 
\delta^2(\omega_{d1|{\rm out}}(k)-\omega_{u|{\rm in}}(q)) 
+ (k\leftrightarrow q) \right]\ .   \label{rescorr2} 
\end{eqnarray}
In the first term of the last line of this formula there is no ambiguity:
$|S_{d1,u}|^2$ and $n_{\sss U}$ have to be evaluated at
$\omega_{d1|{\rm out}}(k)=\omega_{u|{\rm in}}(q)$, $|\widetilde{{\cal
    U}}_{u|{\rm in}}|^2$ has to be evaluated at $q$ and
$|\widetilde{{\cal U}}_{d1|{\rm out}}|^2$ has to be evaluated at $k$.
\begin{eqnarray} 
&&G_2(k<0,q>0)= \nonumber \\ 
&& |\widetilde{{\cal U}}_{u|{\rm out}}|^2
|\widetilde{{\cal U}}_{d1|{\rm out}}|^2
\Big|S_{u,u}^*S_{d1,u}n_{\sss U}+S_{u,d1}^*S_{d1d1}n_{\sss D1} 
+ S_{u,d2}^*S_{d1,d2}(1+n_{\sss D2})\Big|^2\nonumber \\ 
&& \delta^2(\omega_{u|{\rm out}}(k)-\omega_{d1|{\rm out}}(q))
\nonumber \\ 
&+& |\widetilde{{\cal W}}_{d2|{\rm out}}|^2
|\widetilde{{\cal U}}_{d1|{\rm out}}|^2
\Big|S_{d2,u}^*S_{d1,u}n_{\sss U}+S_{d2,d1}^*S_{d1,d1}n_{\sss D1} + 
S_{d2,d2}^*S_{d1,d2}(1+n_{\sss D2})\Big|^2\nonumber \\ 
&& \delta^2(\omega_{d2|{\rm out}}(-k)-\omega_{d1|{\rm out}}(q))\nonumber \\ 
&+& |\widetilde{{\cal U}}_{u|{\rm out}}|^2
|\widetilde{{\cal U}}_{u|{\rm in}}|^2
|S_{u,u}|^2n_{\sss U}^2
\delta^2(\omega_{u|{\rm out}}(k)-\omega_{u|{\rm in}}(q))\nonumber \\ 
&+& |\widetilde{{\cal U}}_{d1|{\rm out}}|^2
|\widetilde{{\cal U}}_{d1|{\rm in}}|^2
|S_{d1,d1}|^2n_{\sss D1}^2
\delta^2(\omega_{d1|{\rm in}}(k)-\omega_{d1|{\rm out}}(q))\nonumber \\ 
&+&  |\widetilde{{\cal W}}_{d2|{\rm out}}|^2
|\widetilde{{\cal U}}_{u|{\rm in}}|^2
|S_{d2,u}|^2n_{\sss U}(1+n_{\sss U})
\delta^2(\omega_{d2|{\rm out}}(-k)-\omega_{u|{\rm in}}(q)) \nonumber \\ 
&+& |\widetilde{{\cal U}}_{d1|{\rm out}}|^2
|\widetilde{{\cal W}}_{d2|{\rm in}}|^2
|S_{d1,d2}|^2n_{\sss D2}(1+n_{\sss D2})
\delta^2(\omega_{d2|{\rm in}}(-k)-\omega_{d1|{\rm out}}(q))\ . 
\label{rescorr3}\end{eqnarray}
We remark that at $T\neq 0$ there appear in-out correlators, which
were absent at $T=0$.  We shall focus here on the most robust
correlators, those already present at $T=0$, and evaluate the
intensity of the correlation signal along these lines by using the
normalized correlation function $g_2$ which is window independent. 
The Hawking quanta-partner
correlator \eqref{resc-g4} gets modified to
\begin{equation}\label{resc-h1}
g_2(K,Q)_{u|{\rm out}-d2|{\rm out}}=
\frac{\left|S_{u,u}^* S^{\phantom{*}}_{d2,u}n_{\sss U} + 
S_{u,d1}^* S^{\phantom{*}}_{d2,d1}n_{\sss D1}+S_{u,d2}^*S^{\phantom{*}}_{d2,d2}
(1+n_{\sss D2})\right|^2}
{{\cal N}_{u|{\rm out}}(k)\, {\cal N}_{d2|{\rm out}}(q)}
+1
\; ,
\end{equation}
where [from \eqref{ncurln}]
\begin{equation}\label{resc-h2}
{\cal N}_{u|{\rm out}}(k)=\left(|S_{u,u}|^2n_{\sss U} + |S_{u,d1}|^2n_{\sss D1}
+|S_{u,d2}|^2(1+n_{\sss D2})\right)_{\omega_{u|{\rm out}}(k)}\; ,
\end{equation}
and
\begin{equation}\label{resc-h3}
\begin{split}
{\cal N}_{d2|{\rm out}}(k)& =
\left(
|S_{d2,u}|^2 (1+n_{\sss U}) + |S_{d2,d1}|^2 (1+n_{\sss D1}) + 
|S_{d2,d2}|^2 n_{\sss D2}\right)_{\omega_{d2|{\rm out}}(-k)}\\
& =\left(-1+|S_{d2,u}|^2n_{\sss U} + |S_{d2,d1}|^2n_{\sss D1}+
|S_{d2,d2}|^2(1+n_{\sss D2})\right)_{\omega_{d2|{\rm out}}(-k)}\; .
\end{split}
\end{equation}
The $u|{\rm out}-d1|{\rm out}$ correlator \eqref{resc-g6} becomes
\begin{equation}\label{resc-h5}
g_2(K,Q)_{u|{\rm out}-d1|{\rm out}}=
\frac{\left|
S_{u,u}^* S^{\phantom{*}}_{d1,u}n_{\sss U} + 
S_{u,d1}^* S^{\phantom{*}}_{d1,d1}n_{\sss D1}+
S_{u,d2}^*S^{\phantom{*}}_{d1,d2}
(1+n_{\sss D2})
\right|^2}
{{\cal N}_{u|{\rm out}}(k)\, {\cal N}_{d1|{\rm out}}(q)}
+1
\; ,
\end{equation}
and the $d1|{\rm out}-d2|{\rm out}$ correlator \eqref{resc-g7} takes the form
\begin{equation}\label{resc-h5bis}
g_2(K,Q)_{d1|{\rm out}-d2|{\rm out}}=
\frac{
\left|
S_{d1,u}^* S^{\phantom{*}}_{d2,u}n_{\sss U} + 
S_{d1,d1}^* S^{\phantom{*}}_{d2,d1}n_{\sss D1}
+S_{d1,d2}^*S^{\phantom{*}}_{d2,d2}
(1+n_{\sss D2})
\right|^2}
{{\cal N}_{d1|{\rm out}}(k)\, {\cal N}_{d2|{\rm out}}(q)}
+1
\; ,
\end{equation}
where [from \eqref{ncurlp}]
\begin{equation}\label{resc-h4}
{\cal N}_{d1|{\rm out}}(k) =
\left(
|S_{d1,u}|^2 n_{\sss U} + |S_{d1,d1}|^2 n_{\sss D1} + |S_{d1,d2}|^2 (1+n_{\sss D2})
\right)_{\omega_{d1|{\rm out}}(k)}\; .
\end{equation}
Eqs. \eqref{resc-h1}, \eqref{resc-h5} and \eqref{resc-h5bis} are
particularly important because they allow us to study how an initial
nonzero temperature affects the violation of the Cauchy-Schwarz
inequality, $g_2>2$, see e.g. Eq. (\ref{CS}). The results in the
waterfall, $\delta$-peak and flat profile configurations are presented
in Figs. \ref{fig.CS.waterfall} and \ref{fig.CS.delta}.  We saw in the
previous subsection that at $T=0$ the Cauchy-Schwarz inequality is
always violated along the $u|{\rm out}-d2|{\rm out}$ and $d1|{\rm
  out}-d2|{\rm out}$ channels for all $\omega>0$, while it is not
violated in the $u|{\rm out}-d1|{\rm out}$ one.  In a more realistic
case in which there is an initial nonzero temperature, the amount of
entanglement is, as expected, reduced with respect to the $T=0$
case. In particular, there is always a region, for small enough and
large enough momenta (when the corresponding frequency is close to 0
or to $\Omega$), where $g_2<2$. In both the $u|{\rm out}-d2|{\rm out}$
and $d1|{\rm out}-d2|{\rm out}$ channels, however, there is always an
intermediate $\omega$ region in which the Cauchy-Schwarz inequality is
violated provided the temperature is not too large.

We clearly see that the most favorable case for demonstrating quantum
entanglement -- i.e. spontaneous Hawking radiation -- corresponds to
the violation of the Cauchy-Schwarz inequality along the Hawking
quanta -- partner ($u|{\rm out}-d2|{\rm out}$) channel in the
waterfall configuration, which is exactly the situation which has been
recently studied experimentally by Steinhauer \cite{Ste16}. We remark
that this configuration corresponds to the case where the region of
overlap of the signals in momentum space is the smallest.

We also note from Fig. \ref{fig.CS.delta} that, for a $\delta$-peak
potential, the Cauchy-Schwarz inequality is violated up to a
temperature slightly larger than the chemical potential. This has been
already noticed in Ref. \cite{Nov14}. The signature of quantum
correlation is more robust for a waterfall configuration: one sees on
Fig. \ref{fig.CS.waterfall} that the Cauchy-Schwarz inequality is
violated up to a temperature of order of twice the chemical potential.

\subsection{Non adiabatic effects and {\it in situ}
  measurements.}\label{NA}

Since the adiabatic limit described by Eqs. \eqref{tadia-pos} and
\eqref{tadia-neg} is rather idealized, especially for small
frequencies (as discussed earlier, for low lying excitations the
assumption of adiabaticity is always violated), it is useful to
describe how the results
presented in the previous two subsections are modified by
non-adiabatic effects. Concretely, instead of Eqs. \eqref{tadia-pos}
and \eqref{tadia-neg}, at late times after the opening of the trap,
the Bogoliubov coefficients will behave as
\begin{equation}\label{alphabet}
\begin{split}
& \widetilde {\cal U}_{u|{\rm in}}\; , \; \widetilde {\cal U}_{u|{\rm out}}
\sim  \alpha_u\; , \quad\quad
\widetilde {\cal U}_{d1|{\rm in}}\; , \; \widetilde {\cal U}_{d1|{\rm out}} 
\sim \alpha_d \; ,\\
&\widetilde {\cal W}_{u|{\rm in}}\; , \; \widetilde {\cal W}_{u|{\rm out}}
\sim \beta_u\; , \quad 
\widetilde{\cal W}_{d1|{\rm int}}\; , \; \widetilde {\cal W}_{d1|{\rm out}} 
\sim  \beta_d\; ,\\
&\widetilde {\cal U}_{d2|{\rm in}}\; , \;\widetilde {\cal U}_{d2|{\rm out}}
\sim \beta_d\; , \quad 
\widetilde {\cal W}_{d2|{\rm in}}\; , \; \widetilde {\cal W}_{d2|{\rm out}}
\sim \alpha_d\; ,
\end{split}
\end{equation}
where the $\alpha$ and $\beta$ coefficients satisfy the same
normalization condition \eqref{norme.tilde} as the corresponding
$\widetilde {\cal U}$, $\widetilde {\cal W}$ ``tilde Bogoliubov
coefficients'' \eqref{q7}. Note that our approach encompasses also the
situation which we denote as {\it in situ} below. In this situation
the measurement process gives a direct access to the actual value of
the tilde Bogoliubov coefficients without any evaporative process
which would affect these coefficients as described in
Eqs. \eqref{tadia-pos} and \eqref{tadia-neg} for the adiabatic limit,
or possibly differently in a non adiabatic regime. In the following,
we use the generic terminology ``non adiabatic'' for describing both
the case of {\it in situ} measurements and of non-adiabatic
modification of the situation \eqref{tadia-pos}, \eqref{tadia-neg}.

To ease the presentation we have -- only temporarily -- considered in
\eqref{alphabet} a single set of coefficients for the upstream $u$
region and also a single one for the downstream $d$ region, but it is
clear that in a rigorous analysis (presented below) each mode will
have its own $\alpha$ and $\beta$ coefficients. Hence the expressions
\eqref{q10} and \eqref{q10bis} remain valid, but the $\widetilde{\cal
  U}$ and $\widetilde{\cal W}$'s are replaced by $\alpha$ and $\beta$
coefficients, following the rules \eqref{alphabet}.

The reason for the change of notation \eqref{alphabet} is
threefold. First, the expression of the non-adiabatic coefficients
maybe quite non-trivial and different from the ones of the $\widetilde
{\cal U}$'s and $\widetilde {\cal W}$'s. This will occur for instance
after a step of dynamical Casimir amplification where the system is
artificially submitted to a rapid quench. Also, this notation allows
for the possibility that, during the opening of the trap, the initial
``tilde Bogoliubov coefficients''
get modified in a manner less trivial
than \eqref{tadia-pos} and \eqref{tadia-neg}. Second, this
(momentarily) simplified notations where the mode indices are
omitted permits a simple presentation of the main features of the
$G_2$ function in the non-adiabatic case (see Appendix
\ref{appB}). Finally, in the weakly non adiabatic case all the
$\beta$'s are small compared with the $\alpha$'s, whereas keeping the
previous $\widetilde{\cal U}$'s and $\widetilde{\cal W}$'s notations
would make it difficult to identify the small terms in the expression
for $G_2$.

The results for the correlator are much more complex than those
presented in the previous two subsections in the adiabatic limit. We
give here the general structure of the correlators, the explicit
results -- with also account of finite temperature --
are given in Appendix \ref{appB}: 
\begin{equation}\label{rescorr7} 
G_2(k<0,q<0)=
  \textrm{Diag}_{<0}\delta^2(k-q)+ O(|\alpha|^4)_{<0} +
  O(|\alpha|^2|\beta|^2)_{<0} + O(|\beta|^4)_{<0} +(k\leftrightarrow
  q),
\end{equation}
\begin{equation}\label{rescorr8} 
G_2(k>0,q>0)=  
\textrm{Diag}_{>0}\delta^2(k-q)+ O(|\alpha|^4)_{>0} 
+ O(|\alpha|^2|\beta|^2)_{>0} + O(|\beta|^4)_{>0} +(k\leftrightarrow q),
\end{equation}
\begin{equation} \label{rescorr9} 
G_2(k<0,q>0)=  
A\delta^2(k+q)+ O(|\alpha|^4) + O(|\alpha|^2|\beta|^2) + O(|\beta|^4) .
\end{equation}
We recall that in the adiabatic limit, corresponding to
Eqs. \eqref{tadia-pos} and \eqref{tadia-neg} with the substitution
\eqref{alphabet}, all the $\beta$'s are zero.  In the more general
case considered here they are not; the $\alpha$'s and $\beta$'s, with
the substitution \eqref{alphabet} -- satisfy the normalization
\eqref{norme.tilde} -- implying that $\alpha$ is bigger than its
adiabatic value.  We see from the results presented in Appendix
\ref{appB} that the terms already present in the adiabatic regime are
now multiplied by a factor $\alpha^4$. In particular, the finite
temperature adiabatic terms of \eqref{rescorr1}, \eqref{rescorr2} and
\eqref{rescorr3} are now given by the $\alpha^4$ diagonal terms given
in Eqs.  \eqref{rescorr7a} and \eqref{rescorr8a} and by the
off-diagonal terms $O(|\alpha|^4)_{<0}$ in Eq. \eqref{rescorr7b},
$O(|\alpha|^4)_{>0}$ in Eq. \eqref{rescorr8b}, and $O(|\alpha|^4)$ in
Eq. \eqref{rescorr9b}. If we consider the weakly non adiabatic regime,
this means that they are now larger than the corresponding adiabatic
value. New sub-leading terms appear, of order $\alpha^2\beta^2$ terms
(among which an antidiagonal term) and also higher order $\beta^4$
contribution. This results in a very complicated pattern.

\subsubsection{Violation of the Cauchy-Schwarz inequality along the
  Hawking quantum-partner correlation lines}

Having derived the structure of the correlation lines, we now turn to
the intensity of the correlation signal, and to the possible violation
of the Cauchy-Schwarz inequality in the non adiabatic regime. 
We shall restrict our
attention to the study of the incidence of non-adiabaticity on the
Hawking quantum - partner $u|{\rm out}$-$d2|{\rm out}$ correlator; 
this will bring pieces of information valid for all the other
correlation lines. The results will be given first at
$T=0$\footnote{A similar analysis in the case of the density
  correlator has been presented in \cite{Balbinot:2014cfa}.}, then at
finite temperature for completeness.  

$\bullet$ From the results in Appendix \ref{appB}, in the case where
both $k$ and $q$ are negative, the zero temperature contribution of
the $u|{\rm out}$--$d2|{\rm out}$ modes to $G_2$ reads [see
Eq. \eqref{rescorr7b}]
\begin{equation}\label{NA70}
  G_2(k,q)_{u|{\rm out}-d2|{\rm out}}\longleftarrow |\alpha_{u|{\rm out}}|^2
  |\alpha_{d2|{\rm out}}|^2 |S_{u,d2}|^2|S_{d2,d2}|^2
  \delta^2(\omega_{u|{\rm out}}(k)-\omega_{d2|{\rm out}}(-q)), 
\end{equation}
The arrow in this equation indicates that its right hand side is not the sole
contribution to the $u|{\rm out}$--$d2|{\rm out}$ correlation signal:
not only should it be supplemented by a contribution in which the
roles of $k$ and $q$ are exchanged, but also new non-adiabatic terms
arise (see below). The contribution \eqref{NA70} corresponds to
the signal which already exists in the adiabatic case, whose intensity
is here modified by the $\alpha$ coefficients. Note that, contrarily
to the schematic presentation of Appendix \ref{appB}, we consider here
the most general case, and have explicitly written the mode-dependence
of the $\alpha$ coefficients. This will remain the case in the
rest of the section.

In the negative $k$ sector, the zero temperature contributions of the
$u|$out and $d2|$out modes to ${\cal N}(k)$ read
\begin{equation}\label{NA71}
\begin{split}
& {\cal N}_{u|{\rm out}}(k<0)=
|\alpha_{u|{\rm out}}|^2|S_{u,d2}|^2|_{\omega_{u|{\rm out}}(k)}\; ,\\
&{\cal N}_{d2|{\rm out}}(k<0)=
|\alpha_{d2|{\rm out}}|^2(|S_{d2,d2}|^2-1)|_{\omega_{d2|{\rm out}}(-k)}\; .
\end{split}
\end{equation}
In the calculation of $g_2(K,Q)_{u|{\rm out}-d2|{\rm out}}$ 
the $\alpha$ coefficients of Eqs. \eqref{NA70} and \eqref{NA71}
factorize out and we get for the normalized correlator
the same result as in 
zero temperature adiabatic case, Eq. (\ref{resc-g4}), namely
\begin{equation}
g_2(K,Q)_{u|{\rm out}-d2|{\rm out}}\longleftarrow
\frac{2|S_{d2,d2}|^2-1}{|S_{d2,d2}|^2-1}\ge 2\; .
\end{equation}
The same factorization of the $\alpha$ coefficients occurs also at
finite temperature, and the adiabatic expression \eqref{resc-h1} is
thus also valid in the non-adiabatic regime. We stress that this
important signal is thus quite robust, not being affected by possible
non adiabatic effects, and the violation of Cauchy-Schwarz inequality
is also observable from {\it in situ} measurements.

\begin{figure}
\begin{center}
\includegraphics*[width=0.6\linewidth]{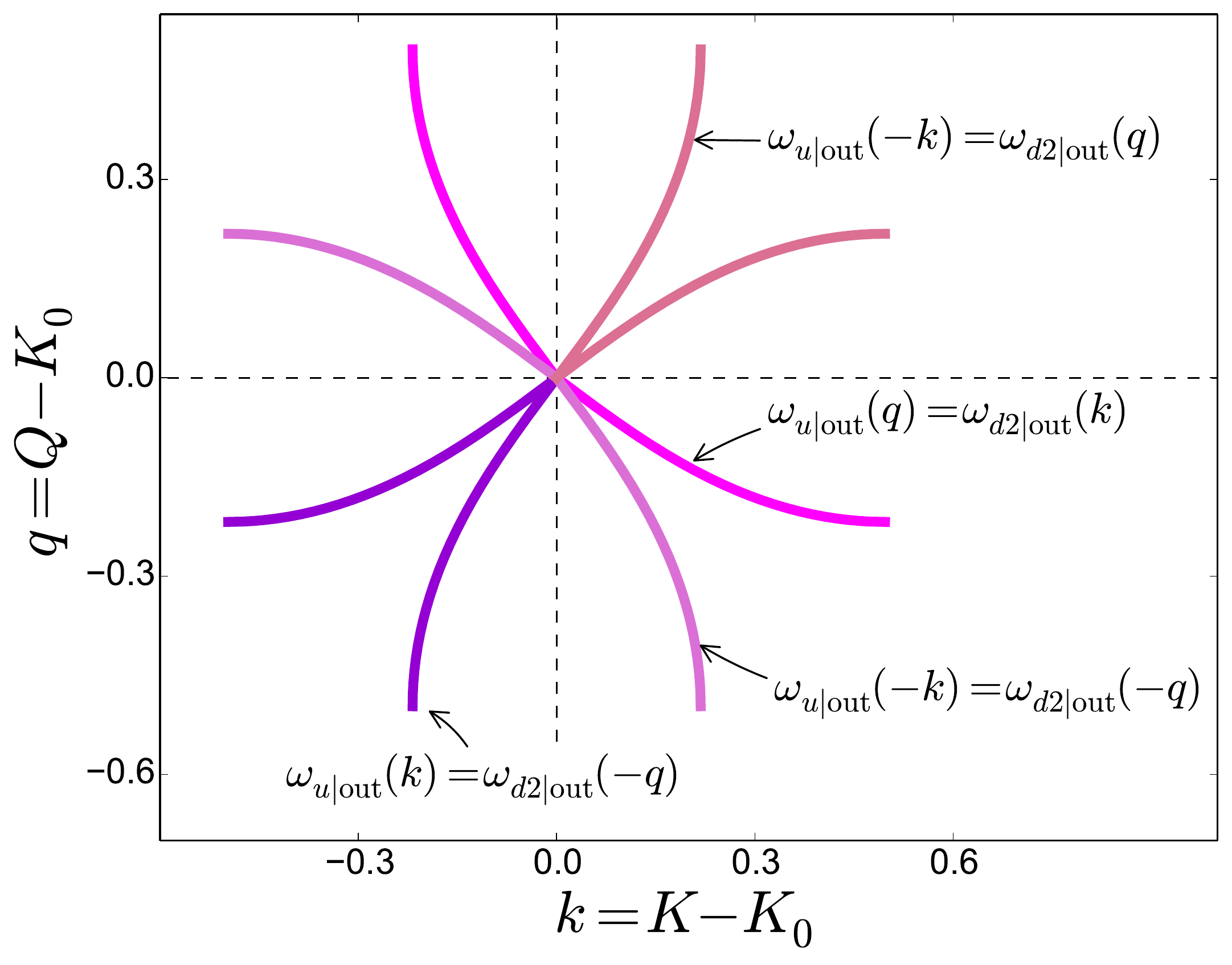}
\end{center}
\caption{Momentum space correlation lines for the $u|{\rm
    out}$-$d2|{\rm out}$ signal in a non-adiabatic case for a flat
  profile configuration with $M_u=0.5$ and $M_d=4$. The correlation line
  corresponding to $\omega_{u|{\rm out}}(k)=\omega_{d2|{\rm out}}(-q)$
  is already present in the adiabatic regime (see the right plot of
  Fig. \ref{fig.twobody.water}).  The lines of identical colors are
  obtained one from the other by an exchange of $k$ and $q$.  The same
  pattern arises {\it in situ}. The momenta are
  expressed in units of $\xi_u^{-1}$.}\label{uoutd2outNA}
\end{figure}

From the results presented in Appendix \ref{appB} one sees that in the
non-adiabatic regime three more couples of $u|{\rm out}$--$d2|{\rm
  out}$ correlation lines appear with respect to the adiabatic
situation. They correspond to the conditions
\begin{equation}
\begin{split}
& \omega_{d2|{\rm out}}(k)=\omega_{u|{\rm out}}(-q)
\quad\mbox{for}\quad k>0\;\mbox{and}\; q >0\; ,\\
& \omega_{d2|{\rm out}}(q)=\omega_{u|{\rm out}}(k)
\quad\quad\mbox{for}\quad k<0\;\mbox{and}\; q >0\; ,\\
&\omega_{d2|{\rm out}}(-q)=\omega_{u|{\rm out}}(-k)
\quad\mbox{for}\quad k>0\;\mbox{and}\; q <0\; ,
\end{split}
\end{equation}
and to the same conditions in which the roles of $k$ and $q$ are
exchanged. All these correlation lines are displayed in
Fig. \ref{uoutd2outNA}, together with the ones which already exist in
the adiabatic case\footnote{For legibility we only show in
Fig. \ref{uoutd2outNA} the $u|{\rm out}$--$d2|{\rm out}$ correlation
lines. In the non-adiabatic case there are many more similar
lines, corresponding to the correlations identified in Appendix \ref{appB}.}.
We now evaluate the intensity of the signal
corresponding to each of these lines.

$\bullet$ In the $k$ and $q>0$ sector, we get, at $T=0$
[cf. Eq. \eqref{rescorr8d}]:
\begin{equation} 
G_2(k,q)_{u|{\rm out}-d2|{\rm out}}\longleftarrow |\beta_{u|{\rm out}}|^2
|\beta_{d2|{\rm out}}|^2 |S_{u,d2}|^2|S_{d2,d2}|^2
\delta^2(\omega_{d2|{\rm out}}(k)-\omega_{u|{\rm out}}(-q)) 
\end{equation}
and
\begin{equation}
\begin{split}
& {\cal N}_{u|{\rm out}}(k>0)=|\beta_{u|{\rm out}}|^2 
(1+|S_{u,d2}|^2)|_{\omega_{u|{\rm out}}(-k)}
\; ,\\
&{\cal N}_{d2|{\rm out}}(k>0)=
|\beta_{d2|{\rm out}}|^2 |S_{d2,d2}|^2|_{\omega_{d2|{\rm out}}(k)}
\; ,
\end{split}
\end{equation}
leading to the following contribution to the normalized correlator:
\begin{equation}\label{NA76} 
g_2(K,Q)_{u|{\rm out}-d2|{\rm out}}\longleftarrow 2-\frac{1}{1+|S_{u,d2}|^2}\le 2
\; .
\end{equation}
As was previously the case for the $\alpha$ coefficients, the
$\beta$'s here also factorize out. A similar factorization will occur
in all the subsequent cases considered in this section.  At finite
temperature, expression \eqref{NA76} becomes
\begin{equation}\label{NA77}
g_2(K,Q)_{u|{\rm out}-d2|{\rm out}}\longleftarrow 
\frac{\left|S_{u,u}^* S^{\phantom{*}}_{d2,u}n_{\sss U} + 
S_{u,d1}^* S^{\phantom{*}}_{d2,d1}n_{\sss D1}+S_{u,d2}^*S^{\phantom{*}}_{d2,d2}
(1+n_{\sss D2})\right|^2}
{\mathscr{N}_{u|{\rm out}}(k>0)\, \mathscr{N}_{d2|{\rm out}}(q>0)}
+1
\; ,
\end{equation}
where here
\begin{equation}
\begin{split}
\mathscr{N}_{u|{\rm out}}(k>0) 
& \equiv \frac{1}{|\beta_{u|{\rm out}}|^2}{\cal N}_{u|{\rm out}}(k>0)\\
& =
\left(
|S_{u,u}|^2(1+n_{\sss U})+|S_{u,d1}|^2(1+n_{\sss D1})+|S_{u,d2}|^2n_{\sss D2}
\right)_{\omega_{u|{\rm out}}(-k)}
\; ,\\
\mathscr{N}_{d2|{\rm out}}(q>0)
&\equiv \frac{1}{|\beta_{d2|{\rm out}}|^2}{\cal N}_{d2|{\rm out}}(q>0)\\
& =\left(
|S_{d2,u}|^2 n_{\sss U}+|S_{d2,d1}|^2 n_{\sss D1}
+|S_{d2,d2}|^2(1+n_{\sss D2})\right)_{\omega_{d2|{\rm out}}(q)}\; .
\end{split}
\end{equation}
The important information here is that, even at zero temperature, the
new $u|{\rm out}$--$d2|{\rm out}$ correlation line which appears in
the $k$ and $q>0$ sector due to non-adiabatic effects is not
associated to a non separable signal [cf. Eq. \eqref{NA76}]. As
expected --and can be verified from expression \eqref{NA77}-- thermal
effects do not modify this situation. As we will now see, this
conclusion remains also valid for all the other correlation lines
which where not present in the adiabatic regime.

$\bullet$ For $k<0$ and $q>0$, one of the contributions to
$G_2(k,q)_{u|{\rm out}-d2|{\rm out}}$ reads  [cf. Eq. \eqref{rescorr9c}]
\begin{equation} 
G_2(k,q)_{u|{\rm out}-d2|{\rm out}}\longleftarrow 
|\alpha_{u|{\rm out}}|^2|\beta_{d2|{\rm out}}|^2 |S_{u,d2}|^2|S_{d2,d2}|^2
\delta^2(\omega_{d2|{\rm out}}(q)-\omega_{u|{\rm out}}(k))
\; , \end{equation}
and at $T=0$ this corresponds to a normalized correlation signal:
\begin{equation} \label{tre}
g_2(K,Q)_{u|{\rm out}-d2|{\rm out}}\longleftarrow 2\; . 
\end{equation}
At finite temperature this contribution modifies to:
\begin{equation}
g_2(K,Q)_{u|{\rm out}-d2|{\rm out}}\longleftarrow 
\frac{\left|S_{u,u}^* S^{\phantom{*}}_{d2,u}n_{\sss U} + 
S_{u,d1}^* S^{\phantom{*}}_{d2,d1}n_{\sss D1}+S_{u,d2}^*S^{\phantom{*}}_{d2,d2}
(1+n_{\sss D2})\right|^2}
{\mathscr{N}_{u|{\rm out}}(k<0)\, \mathscr{N}_{d2|{\rm out}}(q>0)}
+1
\; ,
\end{equation}
where
\begin{equation}
\begin{split}
\mathscr{N}_{u|{\rm out}}(k<0)& \equiv
\frac{1}{|\alpha_{u|{\rm out}}|^2}{\cal N}_{u|{\rm out}}(k<0)\\
& =
\left(
|S_{u,u}|^2n_{\sss U}+|S_{u,d1}|^2n_{\sss D1}+|S_{u,d2}|^2(1+n_{\sss D2})
\right)_{\omega_{u|{\rm out}}(-k)}
\;  .
\end{split}
\end{equation}
$\bullet$ For $k>0$ and $q<0$ another contribution to
$G_2(k,q)_{u|{\rm out}-d2|{\rm out}}$ reads [cf. Eq. \eqref{rescorr9c}]
\begin{equation} 
G_2(k,q)_{u|{\rm out}-d2|{\rm out}}\longleftarrow
|\alpha_{d2|{\rm out}}|^2|\beta_{u|{\rm out}}|^2 |S_{u,d2}|^2|S_{d2,d2}|^2
\delta^2(\omega_{d2|{\rm out}}(-q)-\omega_{u|{\rm out}}(-k))\; 
, \end{equation}
giving at $T=0$
\begin{equation} \label{quattro}
g_2(K,Q)_{u|{\rm out}-d2|{\rm out}}\longleftarrow 2-\frac{|S_{d1,d2}|^2}
{(|S_{d2,d2}|^2-1)(1+|S_{u,d2}|^2)}\le 2\; .
\end{equation} 
This form of writing the result has been obtained by using the
pseudo-unitarity of the $S$ matrix. It makes clear that no violation
of the Cauchy-Schwarz inequality occurs along the correlation line
considered here. At finite temperature one obtains
\begin{equation}
g_2(K,Q)_{u|{\rm out}-d2|{\rm out}}\longleftarrow 
\frac{\left|S_{u,u}^* S^{\phantom{*}}_{d2,u}n_{\sss U} + 
S_{u,d1}^* S^{\phantom{*}}_{d2,d1}n_{\sss D1}+S_{u,d2}^*S^{\phantom{*}}_{d2,d2}
(1+n_{\sss D2})\right|^2}
{\mathscr{N}_{u|{\rm out}}(k>0)\, \mathscr{N}_{d2|{\rm out}}(q<0)}
+1
\; ,
\end{equation}
where here
\begin{equation}
\begin{split}
\mathscr{N}_{d2|{\rm out}}(q<0)& \equiv
\frac{1}{|\alpha_{d2|{\rm out}}|^2}{\cal N}_{d2|{\rm out}}(q<0)\\
& =\left(
|S_{d2,u}|^2 (1+n_{\sss U})+|S_{d2,d1}|^2 (1+n_{\sss D1})
+|S_{d2,d2}|^2 n_{\sss D2}\right)_{\omega_{d2|{\rm out}}(-q)}\; .
\end{split}
\end{equation}

To sum up, we recall that the above study is a partial focus on a
subpart of the whole correlation pattern, concerning only the most
important Hawking quantum-partner signal (in our terminology, the
$u|{\rm out}-d2|{\rm out}$ signal). We used it to demonstrate that the
most interesting correlation is the one already present in the $T=0$
adiabatic case: the Cauchy-Schwarz inequality can only be violated
along this line. The same is true for the $d1|{\rm out}-d2|{\rm out}$
signal.

\section{Momentum correlations in the absence of sonic horizon}\label{correl-noBH}
In this section we consider a configuration where the upstream and the
downstream regions are both subsonic. In this case there is no
horizon, but one can still be in a configuration where the upstream
and downstream non linear coefficients are different (as in the flat
profile configuration), also the system can be affected by the
presence of an external potential (as in the waterfall and delta peak
configuration). If this external potential is localized (i.e., tends
rapidly enough to zero at infinity), and if the nonlinear coefficient
keeps the same value in all the system, then the type of flow
considered is rather simple. More precisely, as demonstrated in
Appendix \ref{appSubSub}, the upstream flow velocity and density at
$-\infty$ are the same as the flow velocity and density at $+\infty$:
$V_u=V_d$ and $n_u=n_d$.  As a result, the general formulas given in
the present section simplify, this is explained in Appendix
\ref{appSubSub}.

In the situation we consider in the present section,
the $d2$ negative-norm mode disappears since the downstream region is
subsonic: there is now a single downstream mode which we simply 
denote by ``$d$''.  The obstacle is thus
characterized by a $S$-matrix which is $2\times 2$ and unitary:
\begin{equation}\label{nah1}
S=
\begin{pmatrix}
S_{u,u} & S_{u,d}\\ S_{d,u} & S_{d,d}
\end{pmatrix}
\quad\mbox{with}\quad
S S^\dagger = \mathbbm{1} \, .
\end{equation}
In this case, instead of Eq. \eqref{q1} one has
\begin{equation}\label{nah2}
\hat{\psi}(x) =
\ep^{{\rm i} K_{\alpha} x} \int_{0}^{\infty} \frac{{\rm d}\omega}{\sqrt{2\pi}}
\sum_{\sss L\in\{U,D\}} \Big[ \bar{u}_{\sss L}(x,\omega) 
 \hat{b}_{\sss L}(\omega)
+ \bar{w}_{\sss L}^{*}(x,\omega)
\hat{b}_{\sss L}^{\dag}(\omega) \Big]\; ,
\end{equation}
where $K_\alpha=m V_\alpha/\hbar$ as in \eqref{q1}, $V_\alpha$ being the 
value of the upstream ($\alpha=u$)
or downstream ($\alpha=d$) asymptotic flow velocity.

We perform the usual (fake) Fourier transform on $\hat \psi$, using
the rules R\ref{R1} and R\ref{R2} of section
(\ref{local-FT}). Collecting separately the $k<0$ and $k>0$
contributions we get
\begin{eqnarray}\label{nexpin}
\hat \psi(k<0) &=& 
\widetilde{\cal U}_{u|{\rm out}}
\Big( S_{u,u}\,\hat b_{\sss U} +S_{u,d}\,\hat b_{\sss D} 
\Big)_{\omega_{u|{\rm out}}(k)}
+
\widetilde{\cal W}^*_{d|{\rm out}}
\Big( S_{d,u}^*\, \hat b_{\sss U}^\dagger +
S_{d,d}^*\,\hat b_{\sss D}^\dagger \Big)_{\omega_{d|{\rm out}}(-k)}
\nonumber \\ 
&+&  \widetilde{\cal U}_{d|{\rm in}}\,\hat b_{\sss D}|_{\omega_{d|{\rm in}}(k)}
+ \widetilde{\cal W}^*_{u|{\rm in}} \,\hat b_{\sss U}^\dagger|_{\omega_{u|{\rm in}}(-k)}
\ , \\
\hat \psi (k>0) &=& \widetilde{\cal U}_{d|{\rm out}}
\Big(S_{d,u}\,\hat b_{\sss U} + S_{d,d}\,\hat b_{\sss D}
\Big)_{\omega_{d|{\rm out}}(k)} + 
\widetilde{\cal W}_{u|{\rm out}}^* \Big(S_{u,u}^*\,\hat b_{\sss U}^\dagger 
+ S_{u,d}^*\,\hat b_{\sss D}^\dagger\Big)_{\omega_{u|{\rm out}}(-k)}
\nonumber \\
&+& \widetilde{\cal U}_{u|{\rm in}}\,\hat b_{\sss U}|_{\omega_{u|{\rm in}}(k)}
+ \widetilde{\cal W}_{d|{\rm in}}^* \,\hat b_{\sss D}^\dagger|_{\omega_{d|{\rm in}}(-k)}
\ .\nonumber
\end{eqnarray}
This expression corresponds to Eqs. \eqref{q10} and \eqref{q10bis} in
which the negative norm $d2$ mode has been suppressed.  From it one
can compute the density distribution in momentum space and also the
momentum correlation function.  In particular, in
  the adiabatic limit one gets $\langle\hat N(k<0)\rangle = {\cal
  N}(k)\times \delta(k-k)$ where
\begin{equation}\label{nnd}
{\cal N}(k<0) = 
\Big( |S_{u,u}|^2 n_{\sss U} + |S_{u,d}|^2 n_{\sss D} \Big)_{\omega_{u|{\rm out}(k)}}
+  \left.
n_{\sss D}\right|_{\omega_{d|{\rm in}(k)}}
\; ,
\end{equation}
and
\begin{equation}\label{nnu}
{\cal N}(k>0) =
\Big( |S_{d,d}|^2 n_{\sss D} + |S_{d,u}|^2 n_{\sss U}\Big)_{\omega_{d|{\rm out}(k)}}
+
\left.
n_{\sss U}\right|_{\omega_{u|{\rm in}(k)}}
\; .
\end{equation}
Since one is in a situation where the flow is everywhere subsonic, one
can define a {\it bona fide} temperature state where, for all the
modes, the occupation number of a state of energy $\omega$ is $n_{\sss
  U}=n_{\sss D}=n_{\rm{th}}(\omega)$.  In the idealized case of
perfect transmission one has $S_{u,u}=0=S_{d,d}$ and
$S_{u,d}=1=S_{d,u}$, then Eqs.  \eqref{nnd} and \eqref{nnu} reduce to
\begin{equation}\label{spurious1}
{\cal N}(k) = 
2\, n_{\rm{th}}(\omega(k))  \; .
\end{equation}
The factor 2 is spurious. It comes from the fact that one does two
Fourier transforms, one upstream and one downstream, and that there is
a kind of built-in double counting in this approach. The problem is
suppressed if one considers upstream and downstream windowed Fourier
transforms, as done in Appendix \ref{window-Fourier}. In this case, in
the absence of the obstacle, instead of Eqs. \eqref{nnd} and \eqref{nnu}
one gets
\begin{equation}\label{non-spurious}
\langle\hat N(k)\rangle = n_{\sss\rm{th}}(\omega(k)) \left(
\frac{\sigma_u \Lambda_u^2}{\sqrt{8\pi}} +
\frac{\sigma_d \Lambda_d^2}{\sqrt{8\pi}} \right)\; .
\end{equation}
How the correct treatment of the windowed upstream and downstream
Fourier transforms leads to the precise form of the two terms in the
big parenthesis of the r.h.s. of \eqref{non-spurious} is explained in
Appendix \ref{window-Fourier}. 

From the present analysis one is thus led to define finite
efficiencies of the upstream and downstream particle detectors. A
finite efficiency corresponds to a measurement process in which a
fraction of particles are missed in the detection of the momentum
signal. This is described theoretically by the normalization of the
windowed Fourier transforms \eqref{win-up}: defining the efficiencies
as $\lambda_\alpha=\sigma_\alpha \Lambda_\alpha^2/\sqrt{2\pi}$
($\alpha=u$ or $d$) with $\lambda_\alpha\in[0,1]$ one casts formula
\eqref{non-spurious} under the form $\langle\hat N(k)\rangle =\tfrac12
(\lambda_u +\lambda_d) n_{\sss\rm{th}}(\omega(k))$. So, when both
efficiencies are zero, there is no particle detected and no momentum
signal, and when, on the contrary, the detection efficiencies are both
unity one gets in the absence of the obstacle $\langle\hat N(k)\rangle
= n_{\sss\rm{th}}(\omega(k))$, as it should be.

Note that the double counting which has been easily identified in
formula \eqref{spurious1} is present in all the previous formulas of
main text (Eqs. \eqref{nkn-adia} to \eqref{ncurlp} and also Eqs.
\eqref{rescorr1} to \eqref{rescorr3}). It can be cured in the same
simple way. For instance, considering perfect detectors, one should
have added a factor $1/\sqrt{2}$ in the definition of rules R\ref{R1}
and R\ref{R2} of section (\ref{local-FT}) for the schematic Fourier
transforms. We did not do that for avoiding an overall multiplicative
factor in all the formulas, and also because, as demonstrated in
Appendix \ref{window-Fourier} this double counting (and also the
finite efficiency of the detector) disappears in the formulas for the
normalized correlation signal $g_2$ which is our main interest in the
present work.

Finally, we give the expression for the momentum correlator, for
simplicity in the adiabatic regime
\begin{eqnarray}
& & G_2(k<0,q<0)=  {\cal N}^2(k<0)  \delta^2(k-q) \nonumber \\
& & + \Big( |\widetilde{\cal U}_{u|{\rm out}}|^2|
\widetilde{\cal U}_{d|{\rm out}}|^2 |S_{u,d}|^2n_D^2 
\delta^2(\omega_{u|{\rm out}}(k)-\omega_{d|{\rm in}}(q)) + 
(k\leftrightarrow q) \Big) \label{dieci} \ ,
\end{eqnarray}
\begin{eqnarray}
& & G_2(k>0,q>0)= 
 {\cal N}^2(k>0)  \delta^2(k-q) \nonumber \\
& & + \Big( |\widetilde{\cal U}_{u|{\rm in}}|^2|
\widetilde{\cal U}_{d|{\rm out}}|^2|S_{d,u}|^2n_{\sss U}^2 
\delta^2(\omega_{d|{\rm out}}(k)-\omega_{u|{\rm in}}(q)) + 
(k\leftrightarrow q) \Big) \ ,
\end{eqnarray}
\begin{eqnarray}
G_2(k<0,q>0)&=&|\widetilde{\cal U}_{u|{\rm out}}|^2|
\widetilde{\cal U}_{d|{\rm out}}|^2\Big|S_{u,u}^*S_{d,u}n_{\sss U}
+S_{u,d}^*S_{d,d}n_{\sss D}\Big|^2
\delta^2(\omega_{u|{\rm out}}(k)-\omega_{d|{\rm out}}(q))  \nonumber\\
&+&
|\widetilde{\cal U}_{u|{\rm out}}|^2|\widetilde{\cal U}_{u|{\rm in}}|^2
|S_{u,u}|^2n_{\sss U}^2
\delta^2(\omega_{u|{\rm out}}(k)-\omega_{u|{\rm in}}(q))
\\
&+&|\widetilde{\cal U}_{d|{\rm out}}|^2|\widetilde{\cal U}_{d|{\rm in}}|^2
|S_{d,d}|^2n_{\sss D}^2
\delta^2(\omega_{d|{\rm in}}(k)-\omega_{d|{\rm out}}(q))\; .
\nonumber 
\end{eqnarray} 
These expressions can be further 
simplified by considering, for this configuration
which is everywhere subsonic, a common occupation number
$n_{\rm{th}}(\omega)$ of a state of energy $\omega$. As expected the
corresponding expressions in the black hole case,
Eqs. (\ref{rescorr1}-\ref{rescorr3}), reduce to the above in the
absence of the negative norm $d2$ modes. Also, unlike in the black
hole case all contributions disappear in the case where
the initial state is
the vacuum.

\section{Limitations of the theoretical
  description}\label{limitations}

In this section we properly define the domain of applicability of our
approach.  A first limitation concerns the low density regime. As well
known, in one dimension phase fluctuations prevent a true
Bose-Einstein condensation. As a result, a description based on the
simple separation \eqref{1d1} between a classical field and a small
quantum correction does not allow to properly estimate the large $x$
behavior of the one body density matrix $\langle\Psi^\dagger(x)
\Psi(0)\rangle$ of a homogeneous 1D system (see, e.g.,
Ref. \cite{Mor03}) and, at low density, phase fluctuations blur the
sharp correlations of Figs. \ref{fig.twobody.water} and
\ref{fig.twobody.delta}, cf. Refs. \cite{Mat09} and \cite{Bou12}.
We nonetheless argue that \eqref{1d1} it is still useful for
understanding the qualitative behavior of some observables important
for analyzing correlations in the system: for instance \eqref{1d1}
yields the correct two-body correlation $\langle \hat\Psi^\dagger(0)
\hat\Psi^\dagger(x) \hat\Psi(x) \hat\Psi(0) \rangle$ of a homogeneous
system \cite{Deu09}.

The relevance of \eqref{1d1} depends on the characteristic length
involved in the spatial correlation considered. If the correlation
characteristic length is smaller than the phase coherence length
$L_\phi=\xi\exp\left[\pi\sqrt{\frac{\hbar n}{2 m a
      \omega_\perp}}\,\right]$ then \eqref{1d1} may be used. This is
what happens for the two-body correlation : this quantity is non
trivial (i.e., different from the square $n^2$ of the linear density
$n= \langle \hat\Psi^\dagger(0) \hat\Psi(0)\rangle$) only in a range
of distances $x< \xi$, typically much smaller than $L_\phi$. More
precisely, $L_\phi$ is exponentially larger than $\xi$, and the separation  
\eqref{1d1} is thus valid, when
\begin{equation}\label{nophasrfluct}
\left(\frac{a}{a_\perp}\right)^2\ll n \, a\; ,
\end{equation}
where $a_\perp=\sqrt{\hbar/m\omega_\perp}$ is the transverse harmonic
oscillator length.

In the large density limit the 1D description also fails, not because
of lack of BEC as just discussed, but because the transverse degrees
of freedom of the system are not completely frozen. A realistic 3D
black-hole configuration has been first considered in
Ref. \cite{Wus07}, with also account for 3 body losses. In the present
discussion we focus on the treatment of transverse
excitations. Assuming that Bose condensation is total, but not
disregarding the transverse degrees of freedom, one considers the
dynamics of the system as described by a Gross-Pitaevskii equation for
the classical field $\Psi_0(\vec{r},t)$:
\begin{equation}\label{1d2}
i\hbar\partial_t \Psi_0=\left(-\frac{\hbar^2 \vec{\nabla}^{\,2}}{2 m} + 
V_\perp(\vec{r}_\perp) + g\, |\Psi_0|^2\right)\Psi_0 \; ,
\end{equation}
where $\vec{r}=x\,\vec{e}_x +\vec{r}_\perp$, $\vec{r}_\perp$ denoting
the transverse coordinate, $V_\perp(\vec{r}_\perp)=\frac{1}{2}
m\omega_\perp^2 r_\perp^2$ being the transverse potential and
$g=4\pi\hbar^2 a/m$. In \eqref{1d2} the normalization is chosen so
that $\rho_0(\vec{r},t)=|\Psi_0(\vec{r},t)|^2$ is the density of
particles. At equilibrium, in the so-called Thomas-Fermi limit
\cite{Bay96}, the Laplacian term in \eqref{1d2} can be omitted and the
density has a cylindrical symmetry with
\begin{equation}\label{1d3}
\rho_0({r}_\perp) = \left\{
\begin{array} {cl}
\frac{1}{g}\left[\mu-V_\perp(\vec{r}_\perp)\right] & \mbox{if}\;\;
\mu\ge V_\perp(\vec{r}_\perp)\; ,\\[2mm]
0 & \mbox{elsewhere}\; .
\end{array}\right.
\end{equation}
Here $\mu$ is the chemical potential fixed by the normalization: $\int
d^2r_\perp \rho_0({r}_\perp) = n$~; $\mu=2\hbar\,\omega_\perp\sqrt{a
  \, n}$ \cite{Zar98,Str98}. The Thomas-Fermi approximation holds in
the large density limit $a \, n\gg 1$ \cite{Jac98,Leb01}. In this
limit, which has been denoted as ``3D cigar'' in Ref. \cite{Men02}, the
classical field description is accurate, that is, the quantum
fluctuations around $\Psi_0$ are small.

In the cylindrical geometry we consider here, the excitation spectrum
has se\-ve\-ral bran\-ches corresponding to density fluctuations of the form
$\delta\rho_0(\vec{r},t)= \delta\rho(r_\perp) e^{i m \theta} e^{i(q
  x-\omega t)}$, where $\theta$ is a polar angle in the transverse
plane. For each branch the lowest state is obtained for $m=0$ and
$q=0$ and its energy reads $\hbar\,\omega_n = \hbar\,\omega_\perp
\sqrt{2 n (n+1)}$, with $n\in\mathbb{N}$. Taking into account possible
longitudinal excitations one gets in the long wave-length limit
\cite{Zar98,Str98}
\begin{eqnarray}
\omega^2_0(q)& = & c^2_{\sss\rm{TF}} \, q^2 \left(1 -\frac{1}{48}(qR_\perp)^2 
+ \dots \right) \; ,\label{1d4}\\
\omega^2_{n\ge 1}(q) & = & 
2 n (n+1)\,\omega^2_\perp +  
c^2_{\sss\rm{TF}} \, q^2
+ \dots \; ,
\label{1d5}
\end{eqnarray}
where $c_{\sss\rm{TF}}=\sqrt{\mu/2\,m}$ is the sound velocity in the
Thomas-Fermi limit (which has been measured by the MIT group
\cite{And97}) and $R_\perp=2 \, c_{\sss\rm{TF}} / \omega_\perp$ is the
transverse extension of the condensate (in the same
limit). Eqs. \eqref{1d4} and \eqref{1d5} describe a lower mode with
sonic-like dispersion relation and gaped transverse excited states
which behave quadratically at low $q$.  Note that the low $q$
expansion displayed in Eq. \eqref{1d4} does not correspond to what is
expected from the usual Bogoliubov dispersion relation
\eqref{1d7}. This is a hint that the longitudinal dynamics of the
system is modified in the Thomas-Fermi limit. Of course, the
hydrodynamic result \eqref{1d4} is limited to the region $q\ll
R^{-1}_\perp$ and cannot provide a reliable description of the whole
excitation spectrum. But the departure from the usual Bogoliubov
dispersion is confirmed by numerical solutions of Bogoliubov-de
Gennes equations \cite{Fed01,Toz02} which are valid for the whole
range of wave vectors and for a range of densities larger than those
based on the Thomas-Fermi approximation. These computations show that
when increasing the linear density starting from a value $n\sim
a^{-1}$ (i.e., when one goes deeper in the Thomas-Fermi regime) the
dispersion relation develops a plateau in the region $q\sim
1/R_\perp$. This is interpreted as a tendency of the excitations to
explore the radial parts of the condensate where the density is lower
and where the local sound velocity accordingly decreases. This effect
is not taken into account in the theoretical analysis presented in the
main text where we work in a regime which has been denoted as ``1D
mean field'' in Ref. \cite{Men02}.  At zero temperature this
corresponds to the regime where the condition \eqref{nophasrfluct} is
supplemented by
\begin{equation}\label{1d6}
 n \, a \ll 1 \; .
\end{equation}
In this regime $\mu=2\hbar\,\omega_\perp a n$ \cite{Ols98} which is
much smaller that the energy $2\, \hbar \omega_\perp$ of the first
transverse excited state\footnote{The value $2\, \hbar \omega_\perp$
  is the same as $\hbar\omega_{n=1}(q=0)$ in \eqref{1d5} : it is model
  independent as a result of a scaling property of the
  Gross-Pitaevskii equation in two dimensions \cite{Pit97}.}, one can
thus safely neglect transverse excitations and the transverse density
profile is not of the type \eqref{1d3}, but has rather a Gaussian
shape.

It is interesting to evaluate the actual range of parameters
corresponding to the fulfillment of conditions \eqref{nophasrfluct}
and \eqref{1d6}, which is the regime of validity of our approach. For a
transverse trap of frequency of 1 kHz, one gets for $^{23}$Na
$(a_\perp/a)^2=1.7\times 10^{-5}$, for $^{87}$Rb
$(a_\perp/a)^2=2.6\times 10^{-4}$ and for He$^*$
$(a_\perp/a)^2=2.2\times 10^{-5}$ \cite{scattering}. Hence the domain
of validity of the 1D mean field approximation used in the present
work ranges over four orders of magnitudes in density.

In present time experiments, when the 1D mean field regime fails, this
is mostly due to the fact that the the linear density is large, and in
this case \eqref{1d6} may be violated. Then, the transverse density
profile has the Thomas-Fermi shape \eqref{1d3}. It is thus of interest
to briefly and qualitatively discuss the features appearing in
momentum correlators in acoustic black holes due to the transverse
modes (\ref{1d5})\footnote{Transverse modes can be incorporated in our
  analysis following the approach of Ref. \cite{Coutant:2012zh}.}.

\begin{figure}
\begin{center}
\includegraphics*[width=\linewidth]{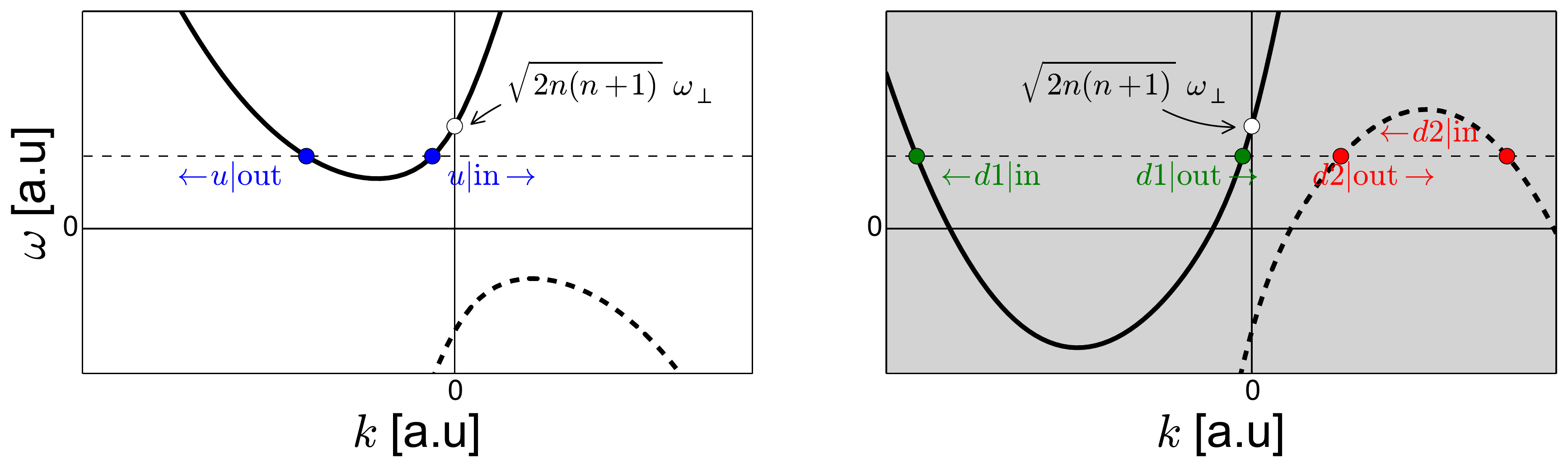}
\end{center}
\caption{Sketch of the typical dispersion relations of a transverse
  mode in the subsonic (left plot) and supersonic (right plot)
  case. The horizontal dashed line is fixed by the chosen value of
  $\omega$.}
\label{fig.tmodes}
\end{figure}

A first remark is in order here: the new transverse modes are
typically not coupled to the modes studied in the present work. The
reason for this is that the potential $V(x)$ used to implement the
sonic horizon does not couple modes with different transverse quantum
numbers. Only small imperfections and nonlinear effect would induce
such a coupling, and the results presented in the present work would
remain almost unaffected.
If the dispersion relation were limited to expression \eqref{1d5},
i.e., were of the Klein-Gordon type, new outgoing modes would appear
which would be populated by the time dependent formation process of
the horizon, as in the gravitational context. In the present case
however, the transverse dispersion relation \eqref{1d5} encompasses
terms of higher order in $q$, and, as a result, new transverse
incoming modes appear in the supersonic region, of the $d2|{\rm in}$
type, as illustrated in Fig. \ref{fig.tmodes}.
As a result of the existence of these new incoming modes, the Hawking
radiation process would occur also in the transverse sector even in the stationary context, and
consequently new correlations lines appear which should add to the ones
studied in the main text. However, in the regime \eqref{1d6} they
should correspond to a very weak signal.

\section{Conclusions}\label{sec.conclu}

In this work we have investigated in detail the two-body momentum
correlation in a quasi 1D BEC in the presence of a sonic horizon. Our
modeling of the measurement process shows that the measurements have
to be performed with some care: (1) the spatial windows selected for
the Fourier analysis must be chosen carefully in order not to damp the
expected signal (see Appendix \ref{window-Fourier} and also
Ref. \cite{Nov15}) and (2) a separation of the upstream and downstream
signals in the detection scheme favors the highlighting of quantum
non separability (end of Sec. \ref{BHT0}). Once the appropriate
requirements are met, the normalized correlator \eqref{m2bis} appears
to be a robust quantity, making it possible to test quantum
 entanglement in a rich variety of situations, namely, {\it in situ}, or
after artificially inducing a quench in the system or at the end of an
adiabatic expansion after opening of the trap. Among the possible
implementations of a sonic horizon we have studied, the largest
quantum correlation signal, observed between the Hawking quantum and
its partner, is realized in the so-called ``waterfall
configuration''. In this configuration the Cauchy-Schwarz inequality
is violated up to a temperature larger than the chemical potential and
therefore should be experimentally testable in a finite temperature
setting [the violation is still present at $T=1.5\times m c_u^2$, see
Fig. \ref{fig.CS.waterfall}, upper left panel].

\paragraph{Acknowledgements}  
Inspiring exchanges with P.-\'E. Larr\'e are gratefully acknowledged.
We also thank D. Boiron, M. Isoard, G. Martone, C. Westbrook and
P. Zi\'n for fruitful discussions.

\paragraph{Funding information}
This work was supported by the French ANR under grant n$^\circ$
ANR-15-CE30-0017 (Haralab project) and by the Spanish Mineco grant
FIS2014-57387-C3-1-P, the Generalitat Valenciana project SEJI/2017/042
and the Severo Ochoa Excellence Center Project SEV-2014-0398.
\begin{appendix}

  \section{Rigorous local Fourier transform}\label{window-Fourier}

  In this appendix we present the precise form of the Fourier
  transforms performed by using the window functions which, in the
  upstream region, are of the form \eqref{win-up}.

We first note that the Fourier transform of the window function is:
\begin{equation}\label{wf1}
\Pi_u(K)=\int_{\mathbb{R}} 
\frac{{\rm d}x}{\sqrt{2\pi}} \, \mathrm{e}^{-{\rm i}K x}\,\Pi_u(x)=
\frac{\Lambda_u \sigma_u}{\sqrt{2}}
\,\mathrm{e}^{-\tfrac14 K^2\sigma_u^2-{\rm i}K X_u}
=
\sqrt{2 \pi}\, \Lambda_u  \,\mathrm{e}^{-{\rm i}K X_u}
\delta_{u}^{(1)}(K)
\; ,
\end{equation}
where $\delta^{(1)}_{u}(K)=(\sigma_u/2\sqrt{\pi})
\exp\{- K^2\sigma_u^2/4\}$ is an approximation of the Dirac distribution
(tending towards the $\delta$ function when $\sigma_u\to\infty$). One
also has
\begin{equation}\label{wf2}
|\Pi_u(K)|^2
=\sqrt{\frac{\pi}{2}}\, \sigma_u \,  \Lambda_u^2 \, 
\delta^{(2)}_{u}(K)
\; ,
\end{equation}
where $\delta_{u}^{(2)}(K)=(\sigma_u/\sqrt{2\pi})
\exp\{- K^2\sigma_u^2/2\}$ is another approximation of the Dirac
$\delta$-distribution  defined by \eqref{wf2} and
verifying $\delta_{u}^{(2)}(0)=\sigma_u/\sqrt{2\pi}$.
So, instead of the approximate formula \eqref{q5} one gets
\begin{equation}\label{wf3}
\begin{split}
\hat{\psi}_u(K)=\int_0^\infty \frac{{\rm d}\omega}{\sqrt{2\pi}}
& \Big\{ \Pi_u(K-K_u-q_{u|{\rm out}}){\cal U}_{u|{\rm out}}
(S_{u,u}\hat{b}_{\sss U} + S_{u,d1}\hat{b}_{\sss D1} 
+S_{u,d2}\hat{b}^\dagger_{\sss D2})\\
& +\Pi_u(K-K_u+q_{u|{\rm out}})  
{\cal W}_{u|{\rm out}}^*(S_{u,u}^*\hat{b}_{\sss U}^\dagger + 
S_{u,d1}^*\hat{b}_{\sss D1}^\dagger +S_{u,d2}^*\hat{b}_{\sss D2})\\
&+\Pi_u(K-K_u+q_{u|{\rm in}}){\cal W}^*_{u|{\rm in}} \hat{b}^\dagger_{\sss U} 
+\Pi_u(K-K_u-q_{u|{\rm in}}){\cal U}_{u|{\rm in}} \hat{b}_{\sss U} \Big\}
 \; .
\end{split}
\end{equation}
Formula \eqref{q6} is modified in a similar way:
\begin{equation}\label{wf3-down}
\begin{split}
\hat{\psi}_d(K)=\int_0^\infty \frac{{\rm d}\omega}{\sqrt{2 \pi}}
& \Big\{
\Pi_d(K-K_d-q_{d1|{\rm out}}){\cal U}_{d1|{\rm out}}
(S_{d1,u}\hat{b}_{\sss U} + 
S_{d1,d1}\hat{b}_{\sss D1} +S_{d1,d2}\hat{b}^\dagger_{\sss D2})\\
&+\Pi_d(K-K_d+q_{d1|{\rm out}}){\cal W}_{d1|{\rm out}}^*
(S_{d1,u}^*\hat{b}_{\sss U}^\dagger + 
S_{d1,d1}^*\hat{b}_{\sss D1}^\dagger +S_{d1,d2}^*\hat{b}_{\sss D2})\\
&+\Pi_d(K-K_d-q_{d2|{\rm out}}){\cal U}_{d2|{\rm out}}
(S_{d2,u}\hat{b}_{\sss U} + 
S_{d2,d1}\hat{b}_{\sss D1} +S_{d2,d2}\hat{b}^\dagger_{\sss D2}) \\
&+\Pi_d(K-K_d+q_{d2|{\rm out}}){\cal W}_{d2|{\rm out}}^*
(S_{d2,u}^*\hat{b}_{\sss U}^\dagger + 
S_{d2,d1}^*\hat{b}_{\sss D1}^\dagger +S_{d2,d2}^*\hat{b}_{\sss D2}) \\
&+
\Pi_d(K-K_d+q_{d1|{\rm in}}){\cal W}^*_{d1|{\rm in}} \hat{b}^\dagger_{\sss D1}
+
\Pi_d(K-K_d-q_{d1|{\rm in}}){\cal U}_{d1|{\rm in}} \hat{b}_{\sss D1} \\
&+
\Pi_d(K-K_d-q_{d2|{\rm in}}){\cal U}_{d2|{\rm in}} \hat{b}^\dagger_{\sss D2}
+ \Pi_d(K-K_d+q_{d2|{\rm in}}){\cal W}_{d2|{\rm in}}^* \hat{b}_{\sss D2}
 \Big\} \; .
\end{split}
\end{equation}
In the following we present the results for the flat profile
configuration. In this case $K_u=K_d\equiv K_0$ and we note $k=K-K_0$,
$q=Q-K_0$.

We first evaluate the one-body term $\langle \hat{N}(K)\rangle$ which
has contributions coming from both $\langle
\hat{\psi}^\dagger_u(K)\hat{\psi}_u(K)\rangle$ and $\langle
\hat{\psi}^\dagger_d(K)\hat{\psi}_d(K)\rangle$. The double integral
over $\omega$ and $\omega'$ defining these terms is reduced to a
single integral by means of the contractions \eqref{m9}. In this
integral one can safely discard overlap terms such as
$\Pi_u(k-q_{u|{\rm out}}(\omega)) \Pi_u(k-q_{u|{\rm in}}(\omega))$
when $\sigma_u\to\infty$ since $\Pi_u(K)$ is proportional to
$\delta_{u}^{(1)}(K)$. One thus gets terms generically of the form of
the one resulting from the contraction of the first term of the
integral in the r.h.s. of \eqref{wf3} with its hermitian conjugate,
which reads:
\begin{equation}\label{wf4}
\begin{split}
& \int_0^\infty\frac{{\rm d}\omega}{2\pi}
\; |S_{u,u}(\omega)\, {\cal U}_{u|{\rm out}}(\omega)|^2  
\, n_{\sss U}(\omega)  \,
|\Pi_u(k-q_{u|\rm out}(\omega)|^2=\\
& \int_{-\infty}^0\frac{{\rm d}p}{2\pi} 
\;
\left|\frac{\partial\omega_{u|{\rm out}}}{\partial p}\right|
|S_{u,u} \, {\cal U}_{u|{\rm out}}|^2 \, n_{\sss U} \, |\Pi_u(k-p)|^2
=
\frac{\sigma_u \Lambda_u^2}{\sqrt{8\pi}}
\left|\frac{\partial\omega_{u|{\rm out}}}{\partial k}\right|
|S_{u,u} \, {\cal U}_{u|{\rm out}}|^2 \, n_{\sss U}
\; .
\end{split}
\end{equation}
In the integral of the second term of \eqref{wf4} one has made the
change of variable $p=q_{u|{\rm out}}(\omega)$ and all the
$\omega$-dependent terms have to be evaluated at $\omega_{u|{\rm
    out}}(p)$. In the last term of \eqref{wf4} one has used the fact
that $|\Pi_u(k-p)|^2$ is proportional to $\delta_{u}^{(2)}(k-p)$ and
all the $\omega$-dependent terms have to be evaluated at
$\omega_{u|{\rm out}}(k)$. As can be checked for instance by
comparison with the similar contribution to $\langle
\hat{N}(k)\rangle$ in Sec. \ref{finiteT}, using the correct windowing
for the Fourier transform, instead of the singular term $\delta(k-k)$
obtained with the schematic rules R\ref{R1} and R\ref{R2}, one gets now a
factor $\sigma_u \Lambda_u^2/\sqrt{8\pi}$ for the terms issued from
the upstream window and a factor $\sigma_d \Lambda_d^2/\sqrt{8\pi}$
for the terms issued from the downstream window. As an illustration,
the formulas equivalent to \eqref{adia.1corps.zerot.neg} and
\eqref{adia.1corps.zerot.pos} (which, we recall, correspond to the
adiabatic and zero temperature situation) read here
\begin{equation}\label{adia.1corps.zerot.neg.FT}
\langle \hat{N}(k<0)\rangle
=  |S_{u,d2}|^2|_{\omega_{u|{\rm out}}(k)} \times 
\frac{\sigma_u \Lambda_u^2}{\sqrt{8\pi}}
+ \left(
|S_{d2,u}|^2 + |S_{d2,d1}|^2\right)_{\omega_{d2|{\rm out}}(k)}
\times \frac{\sigma_d \Lambda_d^2}{\sqrt{8\pi}}.
\end{equation}
and 
\begin{equation}\label{adia.1corps.zerot.pos.FT}
\langle \hat{N}(k>0)\rangle =
|S_{d1,d2}|^2|_{\omega_{d1|{\rm out}}(k)}\times 
\frac{\sigma_d \Lambda_d^2}{\sqrt{8\pi}}
\; .
\end{equation}
It now remains to evaluate the two-body function $G_2(K,Q)$. This involves
four integrations over $\omega$, two of which disappear when using the
contraction rules \eqref{m9}. In the contributions to $G_2(K,Q)$ one
has to distinguish the diagonal terms --- i.e., intra-channel
correlations --- and the crossed ones --- inter-channel. The
evaluation of the diagonal terms is simpler, and we only state the
results: Instead of the singular term $\delta^2(k-q)$ (such as
obtained for instance in the diagonal terms of \eqref{rescorr1} and
\eqref{rescorr2}) one gets a term $\Lambda_u^4(\sigma_u/4 \sqrt{\pi})
\delta_{u}^{(1)}(k-q)$ for the contributions from the upstream
windowing and a term $\Lambda_d^4(\sigma_d/4 \sqrt{\pi})
\delta_{d}^{(1)} (k-q)$ for the contributions from the downstream
windowing. 

One now has all the tools for determining the effect of the
windowing on the evaluation of the intra-channel correlation signals,
of the type $g_2(K,K)_{u|{\rm out}}$ for instance:
\begin{equation}\label{wf5}
g_2(K,K)_{u|{\rm out}}=
\frac{G_2(K,K)_{u|{\rm out}}}{\langle\hat{N}(K)\rangle^2_{u|{\rm out}}}+1
\; ,
\end{equation}
where $\langle\hat{N}(K)\rangle_{u|{\rm out}}$ and $G_2(K,K)_{u|{\rm out}}$
are the $u|$out contributions to $\langle\hat{N}(K)\rangle$ and to
$G_2(K,K)$. With the correct rules presented
above, one gets $g_2(K,K)_{u|{\rm out}}=2$, as in the main text. The
same result holds true for all the intra-channel correlation terms.

We present the evaluation of the inter-channel terms in more detail,
because it is less straightforward than the one of the diagonal terms,
and also because it involves considerations relevant to the
experimental detection scheme. Let us focus on the $u|$out-$d2|$out
contribution for instance.  As done above [Eq. \eqref{wf4}] we
illustrate the general case by studying one of the many contributions
to $G_2|_{u|{\rm out}-d2|{\rm out}}$. In the four field quantity
\eqref{m15}, one has a product of four integrals of the type 
\eqref{wf3} and \eqref{wf3-down}.
For instance, one 
of the double contractions of terms issued from these integrals is
\begin{equation}\label{contrac}
\begin{split}
& \left[{\cal U}_{u|{\rm out}} S_{u,d2}\right]^*_{\omega_1}
\left[{\cal W}_{d2|{\rm out}} S_{d2,d2}\right]_{\omega_2}
\left[{\cal U}_{u|{\rm out}} S_{u,d2}\right]_{\omega_3}
\left[{\cal W}_{d2|{\rm out}} S_{d2,d2}\right]^*_{\omega_4}\times\\
& \big\langle 
\contraction[1ex]{}{\hat{b}}{_{\sss D2}(\omega_1)}{\hat{b}}
\contraction[1ex]{\hat{b}_{\sss D2}(\omega_1)\hat{b}^\dagger_{\sss D2}(\omega_2)}
{\hat{b}}{^\dagger_{\sss D2}(\omega_3)}{\hat{b}}
\hat{b}_{\sss D2}(\omega_1)\hat{b}^\dagger_{\sss D2}(\omega_2)
\hat{b}^\dagger_{\sss D2}(\omega_3)
\hat{b}_{\sss D2}(\omega_4)
\big\rangle\; .
\end{split}
\end{equation}
The contractions are evaluated using the finite temperature rules
\eqref{m9}, and the contribution of the term corresponding to
\eqref{contrac} can be written as the products of two independent 
integrals, $I$ and $J$:
\begin{equation}\label{wf6}
\begin{split}
I(k,q)=\mathrm{e}^{{\rm i}(k X_u+q X_d)} 
\int_0^{\infty}\frac{{\rm d}\omega}{2\pi} & (1+n_{\sss D2})
S_{d2,d2}S_{u,d2}^* {\cal U}^*_{u|{\rm out}}{\cal W}_{d2|{\rm out}} \\
& \Pi_u(k-q_{u|{\rm out}}) 
\Pi_d(q+q_{d2|{\rm out}})
\mathrm{e}^{-{\rm i}q_{u|{\rm out}}X_u} 
\mathrm{e}^{{\rm i}q_{d2|{\rm out}}X_d} \; ,
\end{split}
\end{equation}
\begin{equation}\label{wf7}
\begin{split}
J(k,q)=\mathrm{e}^{-{\rm i}(k X_u+q X_d)}
\int_0^{\infty}\frac{{\rm d}\omega}{2\pi} & n_{\sss D2}
S_{d2,d2}^*S_{u,d2} {\cal U}_{u|{\rm out}}{\cal W}^*_{d2|{\rm out}} \\
& \Pi_u(k-q_{u|{\rm out}}) 
\Pi_d(q+q_{d2|{\rm out}})
\mathrm{e}^{{\rm i}q_{u|{\rm out}}X_u} 
\mathrm{e}^{-{\rm i}q_{d2|{\rm out}}X_d} \; .
\end{split}
\end{equation}
The two integrals have similar forms. Each appears with a prefactor
which disappears when the product $I\times J$ is performed: we thus drop
this prefactor in the following and denote $\tilde{I}$ and $\tilde{J}$
the integrals where this prefactor is removed. We now focus on the
evaluation of $\tilde{I}$; after a change of variable 
$p=q_{u|{\rm out}}(\omega)$ it reads
\begin{equation}\label{wf8}
\tilde{I}(k,q)=\int_0^{\infty}\frac{{\rm d}p}{2\pi}
A(\omega_{u|{\rm out}}(p))\Pi_u(k-p) 
\Pi_d(q+q_{d2|{\rm out}}(\omega_{u|{\rm out}}(p))
\mathrm{e}^{-{\rm i}q_{u|{\rm out}}X_u} 
\mathrm{e}^{{\rm i}q_{d2|{\rm out}}(\omega_{u|{\rm out}}(p))X_d}\; ,
\end{equation}
where $A(\omega)=(1+n_{\sss D2}) S_{d2,d2}S_{u,d2}^* {\cal
  U}^*_{u|{\rm out}} {\cal W}_{d2|{\rm out}}|\partial\omega_{u|{\rm
    out}}/\partial p|$.  

\begin{figure}[h]
\begin{center}
\begin{picture}(8,4)
\put(1.6,0.3){\includegraphics[width=5cm]{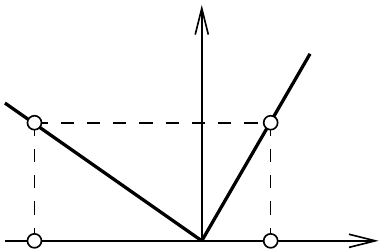}}
\put(3.9,3.4){\large$\omega$}
\put(6.3,0.6){\large$k$}
\put(1.9,0){\large$p$}
\put(4.6,0){\large$q_{d2|{\rm out}}(\omega_{u|{\rm out}}(p))=-\gamma\,p$}
\put(4.8,3){$\omega_{d2|{\rm out}}(k)=V_{d2|{\rm out}}\cdot k$}
\put(0.,2.3){$\omega_{u|{\rm out}}(k)=V_{u|{\rm out}}\cdot k$}
\end{picture}
\end{center}
\caption{Dispersion relations in the long wavelength (dispersionless)
  limit.  $V_{u|{\rm out}}=V_u-c_u(<0)$, $V_{d2|{\rm out}}=V_d+c_d$
  and $\gamma\equiv -V_{u|{\rm out}}/V_{d2|{\rm out}}$
  ($>0$).}\label{fig.nodisp}
\end{figure}
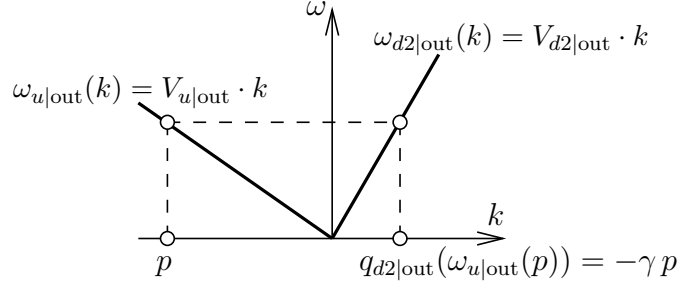

For presenting the results it is easier to work in a simple regime
where the dispersion relations are dispersionless; this will be
assumed in the remaining of this appendix, the general result being
given in the final formula \eqref{wf20}. In this case
$\partial\omega_{u|{\rm out}}/\partial k=V_u-c_u\equiv V_{u|{\rm
    out}}$ and $\partial\omega_{d2|{\rm out}}/\partial k=V_d+c_d\equiv
V_{d2|{\rm out}}$ and\footnote{We will use indifferently the notations
  $\partial\omega_\ell/\partial k$ or $V_\ell$ (with $\ell= u|$out or
  $d2|$out) in the following of this appendix.} one can write
$q_{d2|{\rm out}}(\omega_{u|{\rm out}}(p))=-\gamma\, p$, where
$\gamma\equiv -V_{u|{\rm out}}/V_{d2|{\rm out}} >0$,
cf. Fig. \ref{fig.nodisp} (the notation $\gamma$ is temporarily
introduced to make the computations easier to follow).
Then \eqref{wf8} reads
\begin{equation}\label{wf9}
\tilde{I}(k,q)=\Lambda_u\Lambda_d\frac{\sigma_u\sigma_d}{2}
\int_0^{\infty}\frac{{\rm d}p}{2\pi}
\, A(\omega_{u|{\rm out}}(p))\, \exp\{T(p,k,q)\}
\end{equation}
where
\begin{equation}\label{wf10}
\begin{split}
T(p,k,q)= &
-\frac{\sigma_u^2}{4}(k-p)^2-\frac{\sigma_d^2}{4}(q-\gamma\, p)^2
-{\rm i}p(X_u+\gamma X_d)\\
= & -\frac{\sigma_u^2+\gamma^2\sigma_d^2}{4}
\left(p-P(k,q)\right)^2
-\frac{\sigma_u^2\sigma_d^2(\gamma k - q)^2}{4(\sigma_u^2+\gamma^2\sigma_d^2)}
-Z(k,q)\; ,
\end{split}
\end{equation}
with
\begin{equation}\label{wf11}
P(k,q)=
\frac{\sigma_u^2k+\sigma_d^2\gamma q -2{\rm i}(X_u+\gamma X_d)}
{\sigma_u^2+\gamma^2\sigma_d^2}\; ,
\end{equation}
and
\begin{equation}\label{wf12}
Z(k,q)=\frac{(X_u+\gamma X_d)^2}{\sigma_u^2+\gamma^2\sigma_d^2}
 +{\rm i}\,
\frac{(X_u+\gamma X_d)(\sigma_u^2 k +\sigma_d^2\gamma q)}
{4(\sigma_u^2+\gamma^2\sigma_d^2)}\; .
\end{equation}
It suffices to evaluate the integral \eqref{wf9} for $\sigma_u$ and
$\sigma_d\to\infty$, which is the relevant limit as explained in the
main text (cf. Sec. \ref{local-FT}). In this case $A(\omega_{u|{\rm
    out}}(p))$ is a weakly dependent function of $p$ compared with the
rapidly varying exponent, and the integral in \eqref{wf9} can be
computed by means of the steepest descent method. This amounts to
evaluate $A(\omega_{u|{\rm out}}(p))$ for $p=P(k,q)$ and to compute
the remaining Gaussian integral. The result reads
\begin{equation}\label{wf13}
\tilde{I}(k,q)=\Lambda_u\Lambda_d
A(\omega_{u|{\rm out}}(P)) V_{d2|{\rm out}}\, \delta_{u.d}^{(3)}(V_{u|{\rm out}}k+
V_{d2|{\rm out}}q)\, 
\exp\{-Z(k,q)\}\; ,
\end{equation}
where
\begin{equation}\label{wf14}
  \delta_{u.d}^{(3)}(K) =
  \sqrt{\frac{\sigma_u^2\sigma_d^2/4\pi}
    {V_{d2|{\rm out}}^2\sigma_u^2+V_{u|{\rm out}}^2\sigma_d^2}}\,
  \exp\left\{
    \frac{- \sigma_u^2\sigma_d^2\, K^2/4}
{V_{d2|{\rm out}}^2\sigma_u^2+V_{u|{\rm out}}^2\sigma_d^2}  \right\}\; ,
\end{equation}
is again an approximation of the Dirac $\delta$-function. The term
$\exp\{-Z\}$ in \eqref{wf13} induces a damping which is not present in
the schematic approach presented in the main text. This term can be
removed if one imposes $X_u=-\gamma X_d$, i.e.,
\begin{equation}\label{wf15}
\frac{X_u}{V_{u|{\rm out}}} = \frac{X_d}{V_{d2|{\rm out}}} \; .
\end{equation}
This relation has a simple physical interpretation: the time taken by
an elementary excitation pertaining to the $u|$out channel to go from
the horizon to the center ($X_u<0$) of the upstream detection zone has
to be the same as the time taken by its partner ($d2|$out channel) to
go from the horizon to the center ($X_d>0$) of the downstream
detection zone.  Note that this relation depends on the signal one is
interested in (here the $u|{\rm out}-d2|{\rm out}$ channel): For other
channels (say the $u|{\rm out}-d1|{\rm out}$ channel) the condition
\eqref{wf15} will be modified and the centers of the window functions
have to be shifted accordingly.

If the condition \eqref{wf15} is not fulfilled, the measured correlation
will be damped compared to the perfect result [presented in the main
text]. Note also that when
this condition is fulfilled, since $q=\gamma\, k$ from the
$\delta_{u.d}^{(3)}$ contribution in \eqref{wf13} one has $P(k,q)=k$
and, in this equation, $A(\omega)$ is evaluated at $\omega_{u|{\rm out}}(k)$ as
expected.  Once condition \eqref{wf15} is realized, the product
$I\times J=\tilde{I}\times \tilde{J}$ is found to be equal to
\begin{equation}\label{wf16}
  I\times J=\Lambda_u^2\Lambda_d^2
  (1+n_{\sss D2})n_{\sss D2} \left| S_{d2,d2}S_{u,d2}^* 
\widetilde{\cal U}_{u|{\rm out}} \widetilde{\cal W}_{d2|{\rm out}} 
\right|^2
  \left[\delta_{u.d}^{(3)}(V_{u|{\rm out}}k+
V_{d2|{\rm out}}q) \right]^2\; .
\end{equation}
The same contribution evaluated with the less rigorous approach
presented in the main text yields to a very similar expression, where
the $\Lambda_u^2\Lambda_d^2$ prefactor is missing and where the term
$\left[\delta_{u.d}^{(3)}(V_{u|{\rm out}}k+ V_{d2|{\rm out}}q)
\right]^2$ is replaced by $\delta^2(\omega_{u|{\rm
    out}}(k)-\omega_{d2|{\rm out}}(-q))$.

Finally, we consider the evaluation of the normalized inter-channel correlator
\begin{equation}\label{wf17}
g_2(K,Q)_{u|{\rm out}-d2|{\rm out}}=
\frac{G_2(K,Q)_{u|{\rm out}-d2|{\rm out}}}
{\langle\hat{N}(K)\rangle_{u|{\rm out}}\langle\hat{N}(Q)\rangle_{d2|{\rm out}}}
+1 \; .
\end{equation}
When evaluating the fraction appearing in the r.h.s. of \eqref{wf17}
along the line $\omega_{u|{\rm out}}(k)=\omega_{d2|{\rm out}}(-q)$ one
obtains a ratio identical to the one obtained in Eq. \eqref{resc-h1} of
the main text, multiplied by a factor
\begin{equation}\label{wf18}
\left|\frac{\partial \omega_{u|{\rm out}}}{\partial k}\cdot
\frac{\partial \omega_{d2|{\rm out}}}{\partial q}\right|
\frac{\left[\delta_{u.d}^{(3)}(0)\right]^2}
{\Lambda^2_u\sigma_u\Lambda^2_d\sigma_d/8\pi}=
2\,
\frac{\sigma_u\sigma_d |V_{u|{\rm out}}| V_{d2|{\rm out}}}
{\sigma_u^2 V_{d2|{\rm out}}^2+\sigma_d^2 V^2_{u|{\rm out}}}\; .
\end{equation}
This term is equal to unity, as it should, only if
\begin{equation}\label{wf19}
\frac{\sigma_u}{|V_{u|{\rm out}}|} = \frac{\sigma_d}{ V_{d2|{\rm out}}}\; ,
\end{equation}
i.e., if the width of the window functions $\Pi_u(x)$ and $\Pi_d(x)$
are in the same ratio \eqref{wf15} as their center.

We recall that we have used a simplified linear dispersion relation
for deriving the relations \eqref{wf15} and \eqref{wf19}. However,
there is dispersion in the system; this means that these
relations have to be adapted for each $k$ and $q$ along a specific
correlation line: they should read in the $u|{\rm out}-d2|{\rm out}$
case considered here
\begin{equation}\label{wf20}
\frac{X_u}{\partial\omega_{u|{\rm out}}/\partial k} = 
\frac{X_d}{\partial\omega_{d2|{\rm out}}/\partial q} 
\; ,
\quad\mbox{and}\quad
\frac{\sigma_u}{|\partial\omega_{u|{\rm out}}/\partial k|} = 
\frac{\sigma_d}{\partial\omega_{d2|{\rm out}}/\partial q}\; .
\end{equation}
The same condition has been already derived by de Nova, Sols
and Zapata in Ref. \cite{Nov15}.

\section{Non adiabatic effects (explicit results)}\label{appB}

In this appendix we give the explicit form of the different
contributions to the correlators \eqref{rescorr7}, \eqref{rescorr8} and
\eqref{rescorr9} discussed in subsection \ref{NA}. The results are
valid at finite temperature, and also {\it in situ}. The expressions
are simplified as much as possible to ease readability:
\begin{equation}\label{rescorr7a}
\begin{split}
{\rm Diag}_{<0}=
&\Big[|\alpha_u|^2\Big(|S_{u,u}|^2n_{\sss U}+|S_{u,d1}|^2n_{\sss D1}+
|S_{u,d2}|^2(1+n_{\sss D2})\Big) \\ 
&+ |\alpha_d|^2\Big(|S_{d2,u}|^2(1+n_{\sss U})+|S_{d2,d1}|^2(1+n_{\sss D1})+
|S_{d2,d2}|^2n_{\sss D2}\Big) \\ 
&
+|\alpha_d|^2n_{\sss D1} + |\alpha_d|^2 n_{\sss D2} + 
|\beta_u|^2(1+n_{\sss U}) \\ 
&+|\beta_d|^2\Big( |S_{d1,u}|^2(1+n_{\sss U})+|S_{d1,d1}|^2(1+n_{\sss D1})
+|S_{d1,d2}|^2 n_{\sss D2}\Big)  \Big]^2  \ ,
\end{split}
\end{equation}
\begin{eqnarray}\label{rescorr7b}
&&O(|\alpha|^4)_{<0}= \\ &&
 |\alpha_u|^2|\alpha_d|^2\big| S_{u,u}^*S_{d2,u}n_{\sss U}+S_{u,d1}^*S_{d2,d1}n_{\sss D1}
+S_{u,d2}^*S_{d2,d2}(1+n_{\sss D2})\big|^2 
\delta^2(\omega_{u|{\rm out}}(k)-\omega_{d2|{\rm out}}(-q))\nonumber \\  
&&+|\alpha_u|^2|\alpha_d|^2 |S_{u,d1}|^2n_{\sss D1}^2 
\delta^2(\omega_{u|{\rm out}}(k)-\omega_{d1|{\rm in}}(q))\nonumber \\ 
&& + |\alpha_u|^2|\alpha_d|^2|S_{u,d2}|^2n_{\sss D2}(1+n_{\sss D2}) 
\delta^2(\omega_{u|{\rm out}}(k)-\omega_{d2|{\rm in}}(-q))\nonumber \\ 
&&+ 
|\alpha_d|^4|S_{d2,d1}|^2n_{\sss D1}(1+n_{\sss D1}) 
\delta^2(\omega_{d2|{\rm out}}(-k)-\omega_{d1|{\rm in}}(q))\nonumber \\ 
&&+ |\alpha_d|^4|S_{d2,d2}|^2n_{\sss D2}^2 
\delta^2(\omega_{d2|{\rm out}}(-k)-\omega_{d2|{\rm in}}(-q))\ ,
\nonumber
\end{eqnarray}
\begin{eqnarray}\label{rescorr7c}
&& O(|\alpha|^2|\beta|^2)_{<0}=\\
&&  |\alpha_u|^2|\beta_u|^2|S_{u,u}|^2n_{\sss U}(1+n_{\sss U})
\delta^2(\omega_{u|{\rm out}}(k)-\omega_{u|{\rm in}}(-q)) 
\nonumber 
\\ &&+ |\alpha_d|^2|\beta_u|^2|S_{d2,u}|^2(1+n_{\sss U})^2
\delta^2(\omega_{d2|{\rm out}}(-k)-\omega_{u|{\rm in}}(-q)) \nonumber \\ &&
+|\alpha_u|^2|\beta_d|^2|S_{u,u}^*S_{d_1u}n_{\sss U}+S_{u,d1}^*S_{d1,d1}n_{\sss D1}
+S_{u,d2}^*S_{d1,d2}(1+n_{\sss D2})|^2
\delta^2(\omega_{u|{\rm out}}(k)-\omega_{d1|{\rm out}}(-q)) \nonumber \\ 
&&  +  |\alpha_d|^2|\beta_d|^2|S_{d2,u}S_{d1,u}^*n_{\sss U}+
S_{d2,d1}S_{d1,d1}^*n_{\sss D1} +S_{d2,d2}S_{d1,d2}^*(1+n_{\sss D2})|^2
\delta^2(\omega_{d2|{\rm out}}(-k)-\omega_{d1|{\rm out}}(-q))  
\nonumber \\ 
&&+ |\alpha_d|^2|\beta_d|^2|S_{d1,d1}|^2n_{\sss D1}(1+n_{\sss D1})
\delta^2(\omega_{d1|{\rm in}}(k)-\omega_{d1|{\rm out}}(-q)) \nonumber \\ 
&&+ |\alpha_d|^2|\beta_d|^2|S_{d1,d2}|^2n_{\sss D2}^2\delta^2(\omega_{d2|{\rm in}}(-k)
-\omega_{d1|{\rm out}}(-q))  \; ,
\nonumber 
\end{eqnarray}
\begin{equation}\label{rescorr7d}
O(|\beta|^4)_{<0}=
|\beta_u|^2|\beta_d|^2|S_{d1,u}|^2(1+n_{\sss U})^2
\delta^2(\omega_{u|{\rm in}}(-k)-\omega_{d1|{\rm out}}(-k)) \ , 
\end{equation}
\begin{eqnarray}\label{rescorr8a}
{\rm Diag}_{>0}=
\Big[ \!\!\!\!\!\!\!\!\!\!\!\!
& |\alpha_d|^2\Big(|S_{d1,u}|^2n_{\sss U}+|S_{d1,d1}|^2n_{\sss D1} + 
|S_{d1,d2}|^2(1+n_{\sss D2})\Big) \\ 
&+|\alpha_u|^2n_{\sss U} + |\beta_d|^2(1+n_{\sss D1})
+|\beta_d|^2(1+n_{\sss D2})\nonumber \\ 
&+|\beta_d|^2\Big(|S_{d2,u}|^2 n_{\sss U}+|S_{d2,d1}|^2n_{\sss D1}
+|S_{d2,d2}|^2(1+n_{\sss D2})\Big) \nonumber \\ 
&
+ |\beta_u|^2\Big( |S_{u,u}|^2(1+n_{\sss U})+|S_{u,d1}|^2(1+n_{\sss D1})+
|S_{u,d2}|^2n_{\sss D2}\Big)  
\Big]^2 \; ,
\nonumber 
\end{eqnarray}
\begin{equation}\label{rescorr8b}
O(|\alpha|^4)_{>0}=
|\alpha_d|^2|\alpha_u|^2|S_{d1,u}|^2n_{\sss U}^2 
\delta^2(\omega_{d1|{\rm out}}(k)-\omega_{u|{\rm in}}(q))\; ,
\end{equation}
\begin{eqnarray}\label{rescorr8c}
&&O(|\alpha|^2|\beta|^2)_{>0}= \\ 
&& |\alpha_d|^2|\beta_d|^2|S_{d1,u}^*S_{d2,u}n_{\sss U}+S_{d1,d1}^*S_{d2,d1}n_{\sss D1}+
S_{d1,d2}^*S_{d2,d2}(1+n_{\sss D2})|^2
\delta^2(\omega_{d1|{\rm out}}(k)-\omega_{d2|{\rm out}}(q))   \nonumber \\ 
&&
 + |\alpha_d|^2|\beta_d|^2|S_{d1,d2}|^2(1+n_{\sss D2})^2
\delta^2(\omega_{d1|{\rm out}}(k)-\omega_{d2|{\rm in}}(q))\nonumber \\ 
&& +|\beta_d|^2|\alpha_u|^2|S_{d2,u}|^2n_{\sss U}^2
\delta^2(\omega_{d2|{\rm out}}(k)-\omega_{u|{\rm in}}(q))   \nonumber \\ 
&&+ |\alpha_d|^2|\beta_u|^2|S_{d1,u}^*S_{u,u}n_{\sss U}+
S_{d1,d1}^*S_{u,d1}n_{\sss D1}+S_{d1,d2}^*S_{u,d2}(1+n_{\sss D2})|^2
\delta^2(\omega_{d1|{\rm out}}(k)-\omega_{u|{\rm out}}(-q)) \nonumber \\ 
&&+ |\beta_u|^2|\alpha_u|^2|S_{u,u}|^2n_{\sss U}(1+n_{\sss U})
\delta^2(\omega_{u|{\rm out}}(-k)-\omega_{u|{\rm in}}(q))\nonumber \\
&&+ |\alpha_d|^2|\beta_d|^2|S_{d1,d1}|^2n_{\sss D1}(1+n_{\sss D1})
\delta^2(\omega_{d1|{\rm out}}(k)-\omega_{d1|{\rm in}}(-q))
\; ,\nonumber
\end{eqnarray}
\begin{eqnarray}\label{rescorr8d}
&&O(|\beta|^4)_{>0}= \\ 
&&  |\beta_d|^4|S_{d2,d1}|^2n_{\sss D1}(1+n_{\sss D1})
\delta^2(\omega_{d2|{\rm out}}(k)-\omega_{d1|{\rm in}}(-q))\nonumber \\ 
&& +|\beta_d|^4|S_{d2,d2}|^2(1+n_{\sss D2})^2
\delta^2(\omega_{d2|{\rm out}}(k)-\omega_{d2|{\rm in}}(q))   
\nonumber \\
&&+  |\beta_d|^2|\beta_u|^2|S_{d2,u}^*S_{u,u}n_{\sss U}+S_{d2,d1}^*S_{u,d1}n_{\sss D1}+
S_{d2,d2}^*S_{u,d2}(1+n_{\sss D2})|^2
\delta^2(\omega_{d2|{\rm out}}(k)-\omega_{u|{\rm out}}(-q))  \nonumber \\ 
&&+  |\beta_d|^2|\beta_u|^2|S_{u,d1}|^2n_{\sss D1}^2
\delta^2(\omega_{d1|{\rm in}}(-k)-\omega_{u|{\rm out}}(-q))\nonumber \\ 
&& + |\beta_d|^2|\beta_u|^2|S_{u,d2}|^2(1+n_{\sss D2})n_{\sss D2}
\delta^2(\omega_{d2|{\rm in}}(k)-\omega_{u|{\rm out}}(-q))\nonumber \ , 
\end{eqnarray}
\begin{eqnarray}\label{rescorr9a}
A= \!\!\!\!\!&& Re \Big[ \alpha_u\beta_u\Big(|S_{u,u}|^2(1+n_{\sss U})+
|S_{u,d1}|^2(1+n_{\sss D1})+|S_{u,d2}|^2n_{\sss D2}\Big) \\ 
&&+ \alpha_d\beta_d\Big( |S_{d2,u}|^2n_{\sss U}+|S_{d2,d1}|^2n_{\sss D1}+
|S_{d2,d2}|^2(1+n_{\sss D2})\Big)  \nonumber \\ 
&&
+ \beta_d\alpha_d\Big( |S_{d1,u}|^2n_{\sss U}+|S_{d1,d1}|^2n_{\sss D1}
+|S_{d1,d2}|^2(1+n_{\sss D2})\Big)\nonumber \\ 
&&+ \alpha_d\beta_d(1+n_{\sss D1})+\alpha_d\beta_d(1+n_{\sss D2})+
\beta_u\alpha_un_{\sss U}\Big]     \nonumber\\ 
&&
\Big[ \alpha_u\beta_u\Big(|S_{u,u}|^2n_{\sss U}+|S_{u,d1}|^2 n_{\sss D1}+
|S_{u,d2}|^2(1+n_{\sss D2})\Big)\nonumber \\ 
&& + \alpha_d\beta_d\Big( |S_{d2,u}|^2(1+n_{\sss U})+|S_{d2,d1}|^2(1+n_{\sss D1})+
|S_{d2,d2}|^2 n_{\sss D2} \Big)  \nonumber \\
&&+ \beta_d\alpha_d\Big( |S_{d1,u}|^2(1+n_{\sss U})+|S_{d1,d1}|^2(1+n_{\sss D1})+
|S_{d1,d2}|^2 n_{\sss D2} \Big)\nonumber \\ 
&& + \alpha_d\beta_d n_{\sss D1}+\alpha_d\beta_d n_{\sss D2}+
\beta_u\alpha_u(1+n_{\sss U})\Big] \; ,
\nonumber \end{eqnarray} 
\begin{eqnarray}\label{rescorr9b}
&&O(|\alpha|^4)= \\ 
&& |\alpha_u|^2|\alpha_d|^2 
\Big|S_{u,u}^*S_{d1,u}n_{\sss U}+S_{u,d1}^*S_{d1,d1}n_{\sss D1} 
+ S_{u,d2}^*S_{d1,d2}(1+n_{\sss D2})\Big|^2
\delta^2(\omega_{u|{\rm out}}(k)-\omega_{d1|{\rm out}}(q)) \nonumber \\  
&&+ |\alpha_d|^4\Big|S_{d2,u}^*S_{d1,u}n_{\sss U}+S_{d2,d1}^*S_{d1,d1}n_{\sss D1} 
+ S_{d2,d2}^*S_{d1,d2}(1+n_{\sss D2})\Big|^2
\delta^2(\omega_{d2|{\rm out}}(-k)-\omega_{d1|{\rm out}}(q)) \nonumber \\ 
&&
+ |\alpha_u|^4 |S_{u,u}|^2n_{U}^2
\delta^2(\omega_{u|{\rm out}}(k)-\omega_{u|{\rm in}}(q))\nonumber \\ 
&& + |\alpha_d|^4|S_{d1,d1}|^2n_{\sss D1}^2
\delta^2(\omega_{d1|{\rm in}}(k)-\omega_{d1|{\rm out}}(q))\nonumber \\ 
&&
+|\alpha_d|^2|\alpha_u|^2|S_{d2,u}|^2n_{\sss U}(1+n_{U})
\delta^2(\omega_{d2|{\rm out}}(-k)-\omega_{u|{\rm in}}(q))\nonumber \\ 
&& +|\alpha_d|^4|S_{d1,d2}|^2n_D(1+n_{\sss D2})
\delta^2(\omega_{d2|{\rm in}}(-k)-\omega_{d1|{\rm out}}(q))\; ,
\nonumber 
\end{eqnarray}
\begin{eqnarray}\label{rescorr9c}
&&O(|\alpha|^2|\beta|^2)= \\ 
&&  | \alpha_u|^2 
|\beta_d|^2|S_{u,u}^*S_{d2,u}n_{\sss U}+S_{u,d1}^*S_{d2,d1}n_{\sss D1}+
S_{u,d2}^*S_{d2,d2}(1+n_{\sss D2})|^2
\delta^2(\omega_{u|{\rm out}}(k)-\omega_{d2|{\rm out}}(q))   \nonumber \\ 
&&
+  |\alpha_u|^2 |\beta_d|^2 |S_{u,d1}|^2n_{\sss D1}(1+n_{\sss D1})
\delta^2(\omega_{u|{\rm out}}(k)-\omega_{d1|{\rm in}}(-q))\nonumber\\ 
&&+ |\alpha_u|^2 |\beta_d|^2 |S_{u,d2}|^2(1+n_{\sss D2})^2
\delta^2(\omega_{u|{\rm out}}(k)-\omega_{d2|{\rm in}}(q))  \nonumber \\ 
&& +|\alpha_d|^2 |\beta_d|^2|S_{d2,d1}|^2(1+n_{\sss D1})^2
\delta^2(\omega_{d2|{\rm out}}(-k)-\omega_{d1|{\rm in}}(-q))
\nonumber \\ 
&&+|\alpha_d|^2 |\beta_d|^2|S_{d2,d2}|^2n_{\sss D2}(1+n_{\sss D2})
\delta^2(\omega_{d2|{\rm out}}(-k)-\omega_{d2|{\rm in}}(q))\nonumber \\
&& + |\alpha_d|^2 |\beta_u|^2 |S_{d2,u}S_{u,u}^*n_{\sss U}+S_{d2,d1}S_{u,d1}^*n_{\sss D1}
+S_{d2,d2}S_{u,d2}^*(1+n_{\sss D2})|^2
\delta^2(\omega_{d2|{\rm out}}(-k)-\omega_{u|{\rm out}}(-q))\nonumber \\ 
&&
+  |\alpha_ d|^2 |\beta_d|^2 |S_{d2,d1}|^2n_{\sss D1}^2
\delta^2(\omega_{d1|{\rm in}}(k)-\omega_{d2|{\rm out}}(q))\nonumber \\ 
&& + |\alpha_d|^2 |\beta_u|^2 |S_{u,d1}|^2n_{\sss D1}(1+n_{\sss D1})
\delta^2(\omega_{d1|{\rm in}}(k)-\omega_{u|{\rm out}}(-q))   \nonumber \\ 
&&
+  |\alpha_d|^2 |\beta_d|^2 |S_{d2,d2}|^2n_{\sss D2}(1+n_{\sss D2})
\delta^2(\omega_{d2|{\rm in}}(-k)-\omega_{d2|{\rm out}}(q))\nonumber \\ 
&& +| \alpha_d|^2 |\beta_u|^2 |S_{u,d2}|^2n_{\sss D2}^2
\delta^2(\omega_{d2|{\rm in}}(-k)-\omega_{u|{\rm out}}(-q))    \nonumber \\ 
&&
+ |\beta_u|^2 |\alpha_d|^2|S_{d1,u}|^2n_{\sss U}(1+n_{\sss U})
\delta^2(\omega_{u|{\rm in}}(-k)-\omega_{d1|{\rm out}}(q))\nonumber \\ 
&&+|\beta_d|^2 |\alpha_u|^2|S_{d1,u}|^2n_{\sss U}(1+n_{\sss U})
\delta^2(\omega_{d1|{\rm out}}(-k)-\omega_{u|{\rm in}}(q)) \; ,  \nonumber 
\end{eqnarray}
\begin{eqnarray}\label{rescorr9d}
&&O(|\beta|^4)= \\ 
&& |\beta_d|^2 |\beta_u|^2 |S_{d2,u}|^2n_{\sss U}(1+n_{\sss U})
\delta^2(\omega_{u|{\rm in}}(-k)-\omega_{d2|{\rm out}}(q))\nonumber \\ 
&&+|\beta_u|^4|S_{u,u}|^2(1+n_{\sss U})^2
\delta^2(\omega_{u|{\rm in}}(-k)-\omega_{u|{\rm out}}(-q))    \nonumber \\ 
&&
+  |\beta_d|^4|S_{d1,u}^*S_{d2,u}n_{\sss U}+S_{d1,d1}^*S_{d2,d1}n_{\sss D1}+
S_{d1,d2}^*S_{d2,d2}(1+n_{\sss D2})|^2
\delta^2(\omega_{d1|{\rm out}}(-k)-\omega_{d2|{\rm out}}(q))     \nonumber \\ 
&&
+  |\beta_d|^4|S_{d1,d1}|^2(1+n_{\sss D1})^2
\delta^2(\omega_{d1|{\rm out}}(-k)-\omega_{d1|{\rm in}}(-q))\nonumber \\ 
&& +| \beta_d|^4|S_{d1,d2}|^2n_{\sss D2}(1+n_{\sss D2})
\delta^2(\omega_{d1|{\rm out}}(-k)-\omega_{d2|{\rm in}}(q))    
\nonumber \\ 
&&+ |\beta_d|^2 |\beta_u|^2 |S_{d1,u}^*S_{u,u}n_{\sss U}+S_{d1,d1}^*S_{u,d1}n_{\sss D1}
+S_{d1,d2}^*S_{u,d2}(1+n_{\sss D2})|^2
\delta^2(\omega_{d1|{\rm out}}(-k)-\omega_{u|{\rm out}}(-q))   
\nonumber \ .    
\end{eqnarray}

\section{Subsonic flow in the presence of a localized obstacle}\label{appSubSub}

\begin{figure}[h]
\begin{center}
\begin{picture}(10,3)
\put(0,0.25){\includegraphics[width=10.5cm]{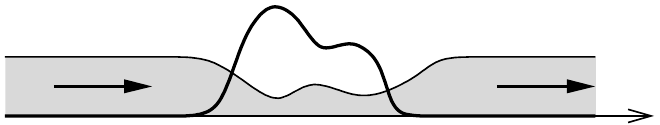}}
\put(0,1.5){\large$n_u$}
\put(2,1.6){\large$n(x)$}
\put(5.,2.){\large$\boldsymbol{U_{\rm ext}(x)}$}
\put(10.,0.){\large$x$}
\put(0.3,0.8){\large$V_u$}
\put(7.4,0.8){\large$V_d$}
\put(9,1.5){\large$n_d$}
\end{picture}
\end{center}
\caption{Sketch of the situation considered in the present
  appendix. The far upstream and far downstream asymptotic flows are
  both subsonic. The obstacle is represented by a localized external
  potential $U_{\rm ext}(x)$.}\label{fig.subsub}
\end{figure}
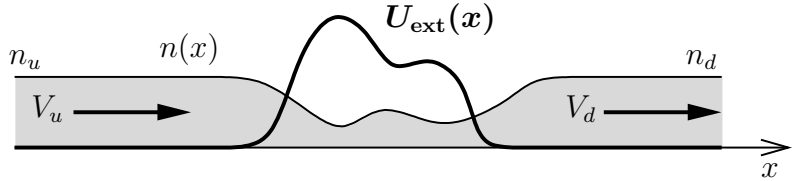

In this appendix we consider the scattering of a stationary subsonic
flow onto a {\it localized external potential} and we assume that the
downstream flow is also subsonic. This is a special case of the
situation considered in section \ref{correl-noBH}. The configuration
is illustrated in Fig. \ref{fig.subsub}. We show in this case that
-- when the non-linearity coefficient $g$ is $x$-independent -- the
far-upstream velocity and density are equal to the far-downstream
velocity and density: $V_u=V_d$ and $n_u=n_d$.

Let's initially assume that the far upstream Mach
number ($M_u=V_u/c_u$) and the far downstream one ($M_d=V_d/c_d$) are
both less than unity, possibly different, and also 
that $V_u$ ($n_u$) may be different from $V_d$ ($n_d$).
In our stationary setting, from 
the conservation of current and the definition of the speed of sound one gets
\begin{equation}\label{sst1}
\frac{V_d}{V_u}=\frac{n_u}{n_d}=\left(\frac{c_u}{c_d}\right)^2\equiv X\; ,
\end{equation}
where the last equality defines the quantity $X$.

The equality of the upstream and downstream chemical potentials reads
\begin{equation}\label{sst2}
\tfrac12 m
V_u^2 + g n_u = \tfrac12 m V_d^2 + g n_d\; . 
\end{equation}
Plugging \eqref{sst1} into \eqref{sst2} yields a third order equation
for the quantity $X$. This equation can be cast under the form
\begin{equation}
(X-1)(X^2+X-2\,M_u^{-2})=0\; .
\end{equation}
If, for the time being, one discards the trivial solution
$X=1$, the only other positive solution is $X= \tfrac12 (-1+\sqrt{
1 + 8 M_u^{-2}\,}\,)$. Then, the far downstream Mach number is
$M_d=V_d/c_d=M_u X^{3/2}$. It can easily be checked that for $M_u<1$
(which has been assumed above) this expression yields for $M_d$ a value
larger than 1: the downstream flow is supersonic, which
contradicts our hypothesis. This means that the trivial
solution $X=1$ is the only acceptable one. From \eqref{sst1} one then gets
the desired result: $n_u=n_d$ and $V_u=V_d$.

Note that the same result also holds true when the flow is supersonic both 
upstream and downstream: also in this case one has  $n_u=n_d$ and $V_u=V_d$.
An important outcome of this remark is that, in the presence of a
localized obstacle, as soon as one is able to prove that the asymptotic
upstream and downstream flow velocities are different, one can be sure that a
sonic horizon has been realized.

Since for a localized obstacle the far upstream and far downstream
characteristics of a subsonic flows are identical, the general
formulas given in Sec. \ref{correl-noBH} simplify due to the
following remarks:
\begin{itemize}
\item[(i)] Since the far upstream and far downstream flows have the
  same density and velocity, $\omega_{u|{\rm out}}(k) = \omega_{d|{\rm
      in}}(k)=\omega(k<0)$ and $\omega_{d|{\rm out}}(k) =
  \omega_{u|{\rm in}}(k) = \omega(k>0)$.
\item[(ii)] ${\cal U_\ell}$ and ${\cal W_\ell}$ are functions of
  $\omega$ and $q_\ell(\omega)$ only (cf. their expression in
  \cite{Lar12}), whereas $\widetilde{\cal U_\ell}$ and
$\widetilde{\cal W_\ell}$ are functions of $k$ and $\omega_\ell(k)$
only.
\item[(iii)] As a result of points (i) and (ii) above, in
  \eqref{nexpin} all the $\widetilde{\cal U}_\ell$'s  can be written as
$\widetilde{\cal
    U}(k)$ and all the $\widetilde{\cal W}_\ell$'s can be written as
  $\widetilde{\cal W}(-k)$.
\end{itemize}

Note that these simplifications are only possible for a
  barrier of finite extent. If, for instance, one considers a flat
  profile configuration where the upstream and the downstream regions
  are both subsonic, the upstream and the downstream speeds of sound
  are not the same (because the upstream and downstream nonlinear
  coefficient are not the same) and point (i) above is not valid.

\end{appendix}


\begin{thebibliography}{99}
\bibitem{remark} From the surface gravity formula, one obtains $T_{\rm H}\simeq
  10^{-7} M^{-1}$ K, where $M$ is the mass of the black hole expressed in
  units of the mass of the sun.
\bibitem{scattering} The values of the $s$-wave scattering length are
  extracted from Ref. \cite{Ties96} for $^{23}$Na, 
from Ref. \cite{Ver09} for $^{87}$Rb $|F= 1, m_F=-1\rangle$, and from
  Ref. \cite{Moa06} for He$^*$ spin-polarized helium atoms in the
$2^3 S_1$ metastable state.
\end{thebibliography}
\end{document}